# Are cookie banners indeed compliant with the law?

## Deciphering EU legal requirements on consent and technical means to verify compliance of cookie banners

Cristiana Santos, Nataliia Bielova, Célestin Matte
Inria, France
cristianasantos@protonmail.com
nataliia.bielova@inria.fr
celestin.matte@cmatte.me

**Abstract**

In this paper, we describe how cookie banners, as a consent mechanism in web applications, should be designed and implemented to be compliant with the ePrivacy Directive and the GDPR, defining 22 legal requirements. While some are provided by legal sources, others result from the domain expertise of computer scientists. We perform a technical assessment of whether technical (with computer science tools), manual (with a human operator) or user studies verification is needed. We show that it is not possible to assess legal compliance for the majority of requirements because of the current architecture of the web. With this approach, we aim to support policy makers assessing compliance in cookie banners, especially under the current revision of the EU ePrivacy framework.

## 1 Introduction

The ePrivacy Directive[1] 2002/58/EC, as amended by Directive 2009/136/EC, stipulates the need for consent for the storage of or access to cookies (and any tracking technology, e.g. device fingerprinting) on the user's terminal equipment, as the lawfulness ground, pursuant to Article 5(3) thereof. The rationale behind this obligation aims to give users control of their data. Hence, website publishers processing personal data are duty-bound to collect consent. Consequently, an increasing number of websites now display (cookie) consent banners.[2]

However, there is no established canonical form for the consent request. It is clear from Recital 17 of the ePrivacy Directive (hereinafter named ePD) that a user's consent may be given by any appropriate method. Website operators are free to use or develop consent flows that suit their organization, as long as this consent can be deemed valid under EU legislation (Article 29 Working Party, WP259 rev.01).[3,4] As such, excessive focus is being placed on the manufacturing of consent, taken up by consent management platforms and tools. The most well-known way to collect consent is through "cookie banners", also often referred to as *prompts*, *overlays*, *cookie bars*, or *cookie pop-up-boxes* that pop up or slide atop websites prominently.[5]

[1] In this paper we will only regard to the recent amended version of the ePrivacy Directive, the Directive 2009/136/EC of the European Parliament and of the Council of 25 November 2009 amending Directive 2002/22/EC on universal service and users' rights relating to electronic communications networks and services, Directive 2002/58/EC concerning the processing of personal data and the protection of privacy in the electronic communications sector and Regulation (EC) No 2006/2004 on cooperation between national authorities responsible for the enforcement of consumer protection laws (Text with EEA relevance) OJ L 337, 11–36 (hereinafter named "ePD") .
[2] Jannick Sørensen, Sokol Kosta, "Before and After GDPR: The Changes in Third Party Presence at Public and Private European Websites" (Proceedings of the World Wide Web Conference, ACM, NY, USA, 2019)1590–1600.
[3] In this paper, we provide many excerpts of the opinions and guidelines of the Article 29 Working Party. For readability and presentation purposes, we convey in the text of the article the abbreviation "29WP", followed by the reference number of each opinion. Even if the European Data Protection Board has endorsed the endorsed the GDPR related WP29 Guidelines, for simplicity purposes, we only mention Article 29 Working Party.
[4] Article 29 Working Party, "Guidelines on consent under Regulation 2016/679" (WP259 rev.01, 10 April 2018).
[5] For example, the French DPA (henceforth named CNIL) decided to remove its cookie banner and to leave no tracer until the user has consented by going actively to the cookie management menu or directly through the content pages. This choice not to use a banner is neither an obligation nor a recommendation for other websites that are free to adopt solutions tailored to their situation, in compliance with Regulations, CNIL, "The legal framework relating to consent has evolved, and so does the website of the CNIL" (2019) <www.cnil.fr/en/legal-framework-relating-consent-has-evolved-and-so-does-website-cnil> accessed 7 May 2020.

Their design and functionality differ – the simplest banners merely state that the website uses cookies without any option, whereas the most complex ones allow users to individually (de)select each third-party service used by the website.

Amid information overload and the development of manipulative dark patterns[6,7,8] that lead to nudging users to consent, data subjects are not always able to easily understand the outcomes of data collection, and the use of their data.

The assessment as to whether or not cookie banner designs implemented by website operators fulfil all the requirements for valid consent, as stipulated by the General Data Protection Regulation[9] (hereinafter named GDPR), is considered in the guidelines of both the Article 29 Working Party and Data Protection Authorities (hereinafter named 29WP and DPAs). These guidelines provide a useful framework of *what* is a valid consent for cookie banners, but they do not define *how to assess, in practice, their legal compliance*. Though these guidelines have an important interpretative value, in concrete settings, they offer still vague guidance on the consent implementation. Even though Recital 66 of the ePD disposes that "the enforcement of these requirements should be made more effective by way of enhanced powers granted to the relevant National Authorities", this point is still under work, despite the recent guidelines issued by various DPAs. The legislative provisions in the GDPR are purposefully general to cover a range of different scenarios, including unanticipated future developments. The ePD does not sketch procedures to guide the enforcement of its principles, nor provides guidelines to perform systematic audits. Moreover, the lack of automatic tools which can verify whether a website violates the legislative instruments possibly makes it complicated for the deputed agencies to plan systematic audits.

The consequence of not complying with the requirements for a valid consent renders the consent invalid and the controller may be in breach of Article 6 of the GDPR. Hence, the controller may be subject to fines (Article 83).[10]

We consider in this work that there is a need for a technical perspective in the analysis of a valid consent for browser-based tracking technologies (including cookies), as processing operations of web services are *technology intensive*. This means that the use of the technology underlying processing operations is such, that specific guidance on the use of that technology is needed to adequately protect personal data while managing cookies on the server side, the third-party side, and also on the side of designers and/or developers of websites. We state that a *privacy by design* approach, as posited in Article 25 of the GDPR, advocates good technical design which embeds privacy into IT systems and business practices from the outset (and does not just add privacy measures ex-post).

Our ***aim*** is to identify the requirements for a valid consent to assess compliance of cookie banners, preparing the ground for compliance of cookie consent banners to be automated. Therefore, our final goal is to evaluate to which extent Web Privacy Measurements (WPM) are capable of assessing compliance automatically. To ensure automatic WPM, we need to rely on a combination of law, policy and technology areas to operationalize requirements for consent. Hence, our intention is to contribute to closing the gap between existing legal guidelines and interpretations and technical solutions for consent banners to discern compliant banner designs and to spot invalid ones. This analysis can be useful to compliance officers, regulators, privacy NGOs, law and computer science researchers, web services business owners and other services concerned with the design or operation of web services.

This paper makes the following contributions:

[6] Harry Brignull, "What are Dark Patterns?" (2018) <https://darkpatterns.org> accessed 7 May 2020.

[7] Colin M. Gray, Yubo Kou, Bryan Battles, Joseph Hoggatt, and Austin L. Toombs, "The Dark (Patterns) Side of UX Design" (Proceedings of the CHI Conference on Human Factors in Computing Systems ACM, New York, USA, 2018).

[8] CNIL"s 6th Innovation and Foresight Report "Shaping Choices in the Digital World, "From dark patterns to data protection: the influence of UX/UI design on user empowerment" (2019) <https://linc.cnil.fr/fr/ip-report-shaping-choices-digital-world> accessed 7 May 2020.

[9] Regulation (EU) 2016/679 of the European Parliament and of the Council of 27 April 2016 on the protection of natural persons with regard to the processing of personal data and on the free movement of such data, and repealing Directive 95/46/EC (General Data Protection Regulation), OJ 2016 L 119/1 (hereafter, "GDPR").

[10] The German DPA acknowledges that if consent is required but not effectively granted, the setting or reading of a cookie is unlawful and data controllers face both the prohibition of data processing and fines, LfDI Baden-Württemberg, "On the use of cookies and cookie banners - what must be done with consent (ECJ ruling "Planet49")?" (2019) <www.baden-wuerttemberg.datenschutz.de/zum-einsatz-von-cookies-und-cookie-bannern-was-gilt-es-bei-einwilligungen-zu-tun-eugh-urteil-planet49/>, and "Guidelines for Telemedia Providers", (2019) <www.baden-wuerttemberg.datenschutz.de/wp-content/uploads/2019/04/Orientierungshilfe-der-Aufsichtsbeh%C3%B6rden-f%C3%BCr-Anbieter-von-Telemedien.pdf > accessed 21 November 2019.

- We identify 22 legal-technical requirements for a valid consent of cookie;
- We show how the 22 requirements for valid consent can be used in practice when performing a compliance audit of consent request (i.e. the banner design).
- We explore to what extent automated consent verification is possible.

We conclude that a fully automatic consent verification by technical means is not possible because the majority of the low-level requirements either require manual inspection, can be evaluated with technical tools only partially, or must be evaluated with user studies to assess users' perceptions and experience with the website's consent implementation.

The remainder of the paper is as follows. Section 2 describes the methodology adopted to construe the requirements for a valid consent for consent banners. Section 3 provides the background knowledge of the paper. Section 4 discusses the scope of browser tracking technologies and analyzes which purposes are subject to the legal basis of consent. Section 5 expounds on each of the requirements and low-level requirements for a valid consent for consent banners, providing compliant and non-compliant examples and the means to verify compliance. Section 6 summarizes different technical solutions that could be applied to detect violations of requirements that depend on natural-language processing and user perception. Section 7 discusses scenarios and consequences of a shared consent. Section 7 discusses a recent draft of the ePrivacy Regulation. Section 8 opens a discussion on the upcoming ePrivacy Regulation. Section 9 compares our work with related work in the area of consent to browser tracking technologies. Section 10 concludes the paper.

## 2 Methodology

This section presents the methodology used in our work. We propose a methodology based on two steps: a legal analysis (Section 2.1), and a technical analysis (Section 2.2). The definition of requirements emerged from joint interdisciplinary work composed of law and computer science experts in the domains of Data Protection Law and Web Tracking Technologies. The combined expertise was conducive to inspect legal and technical effects and the practical implementation of each requirement.

### 2.1 On the legal analysis

**Bottom-up approach.** In our work, we follow a bottom-up approach, using granular content from the elicited legal sources to build the devised requirements. First, we analyzed consent elements separately for general consent, and afterwards, we delved into the specificities of consent dedicated to browser-based tracking technologies (henceforth named BTT), including cookies.

**Legal sources.** We have included in our analysis the following sources:

- legislation: GDPR and ePrivacy Directive (ePD)
- regulatory overview of decisions issued by the European Court of Justice of the EU (CJEU)
- DPA decisions on the use of cookies
- DPA guidelines on the use of cookies
- related works by legal scholars regarding some requirements

**Criterion to elicit requirements based on legal effect.** By analyzing the legal sources listed above, we have defined the requirements for a valid consent considering their *legal effect*. First, we extract requirements from legal sources with a *binding effect*, which can render legal certainty and predictability which happens with the GDPR and case-law from the CJEU. Table 1 depicts the legal source according to its legal effect.

**Table 1** Legal sources according to its legal effect

| Legal source | Type | Legal effect |
|---|---|---|
| Legislation | GDPR | Binding |
| Case law | CJEU case law | Binding |
| Guidelines | EDPB | Non-binding, interpretative effect. These contain persuasive authority, which means that the court is not required to follow the analysis |
| | DPA guidelines | |

We now present requirements with binding and non-binding legal effect that we rely on in this work.

### 2.1.1 Requirements coming from binding sources.

**Standard requirements from GDPR, ePrivacy Directive and CJEU.** We include the four cumulative elements for a valid consent prescribed by Articles 4(11), 7(4) of the GDPR which amount to: freely given, specific, informed and unambiguous consent. We also include the requirement of an informed consent mandated in Article 5(3) of the ePD. We consider the Planet 49 ruling of the CJEU.[11] Table 2 shows the binding requirements coming from the GDPR, ePD and CJEU.

**Table 2** Standard requirements for a valid consent from binding sources: GDPR Article 4(11) and 7(4), ePD Article 5(3) and CJEU

| High-level requirements | Provenance |
|---|---|
| Freely given | • Article 4(11) of GDPR<br>• Article 7(4) of GDPR |
| Specific | • Article 4(11) of GDPR<br>• CJEU Planet 49 |
| Informed | • Article 4(11) of GDPR<br>• CJEU Planet 49<br>• Article 5(3) of ePD |
| Unambiguous | • Article 4(11) of GDPR<br>• CJEU Planet 49 |

**GDPR Articles 6 and 7.** Besides these mentioned elements, we make salient and autonomous three other requirements:

- Prior
- Readable and accessible
- Revocable

These elements are mentioned in the GDPR, though they are not part of the definitional elements of Article 4(11). However, these three additional requirements, depicted in Article 7, are called as *conditions for validity* of consent, and are also meaningful to be considered for their practical effects in the online environment of consent banners (as explained throughout the paper). In this line, the 29WP (WP 259 rev.01) refers that the GDPR introduces requirements (beyond Article 4(11)) for controllers to make *additional* arrangements to ensure they obtain and are able to demonstrate valid consent. It refers to Article 7 which sets out these additional conditions of validity for valid consent, with specific provisions on keeping records of consent and the right to easily withdraw consent. In this regard, the Advocate General Spuznar[12] contends that the purpose of Article 7(1) of Regulation 2016/679 requires a *broad interpretation* in that the controller must not only prove that the data subject has given his or her consent but must also prove that *all the conditions for effectiveness* have been met, hence, expressing the practical side of the conditions of Article 7. Table 3 extends Table 2 and depicts these three added high-level requirements and their respective provenance.

**Table 3** Additional requirements for valid consent from binding sources: GDPR Articles 6 and 7

| High-level requirement | Provenance in the GDPR |
|---|---|
| Prior | • Article 6, by the wording "has given" consent |
| Readable and accessible | • Article 7 (2) "conditions for consent"<br>• Recitals 32, 42 |
| Revocable | • Article 7 (3) "conditions for consent" |

### 2.1.2 Requirements coming from non-binding sources

**EDPB guidelines.** Whenever there is no binding rule, we resort to the agreed-upon and harmonized guidelines coming from the EDPB – as a new EU decision-making body building on the work of the 29 Working Party (29WP). The EDPB has adopted various opinions and guidelines to clarify fundamental provisions of the GDPR and to ensure consistency in the application of the GDPR by DPAs.

[11] cf. Planet49 Judgment (n 87) Verbraucherzentrale Bundesverband v. Planet49, Case C-673/17, [2019] OJ C 112 (ECLI:EU:C:2019:801) para 75.
[12] Case C-61/19 Orange România SA v ANSPDCP, 4 March 2020, (ECLI:EU:C:2020:158)

**DPA guidelines**. We also resort to the guidelines of DPAs on consent for browser-based tracking technologies. We give a comparative analysis of the DPA guidelines. The usefulness of these guidelines is twofold:

- they connect to the legal requirements implemented at national level in the light of the GDPR standard requirements for consent, and
- they incorporate the recent binding requirements coming from the CJEU.

In the comparative analysis of the existing DPA guidelines on the use of BTT, we discuss to what extent our 22 low-level requirements (see Table 6) are reflected in these guidelines, and where there are divergences. We only assess the guidelines under the specific criterion of being comprehensive. For example, the Italian DPA simply presents in its website Frequently Asked Questions (FAQs) on cookies along with basic information and does not specify the requirements for trackers. Even though the Finnish DPA[13] issued guidelines on cookies, it was not possible to analyze even the high-level requirements, while the CNIL, the ICO, the Irish and Greek guidelines provide a more detailed description on each of the GDPR requirements. We have selected these eight guidelines that offer a wide coverage and reasoning upon the requirements for a valid consent: UK, French, Irish, German, Belgium, Danish, Greek, Spanish.

Table 4 extends Table 3, now depicting non-binding sources, thereby consolidating the assumed requirements.

**Table 4** Additional requirements for a valid consent from non-binding sources: EDPB (WP29) and DPA guidelines

| High-level requirements | Provenance from other sources |
|---|---|
| Prior | •29WP on Consent<br>•Article 2 of CNIL Guidance for cookies, 2019<br>•DPAs: Finnish, German Guidelines |
| Readable and accessible | •29WP Guidelines on Transparency; DPAs: Belgium, Spanish, French, ICO |
| Revocable | •29WP on Consent<br>•Article 2 CNIL Guidance for cookies, 2019<br>•CNIL Recommendation for cookies, 2020<br>•DPAs: French, Greek, Irish, Danish, Spanish, German, Belgium |

### 2.1.3 Requirements coming from our own legal interpretation

We have pursued with our own interpretation regarding some requirements, demarking our explicit positioning. Concretely, we have proposed three new requirements.

- “configurable consent banner” (R12),
- “balanced choice” (explained in R 13),
- “no consent wall” (explained in R 20).

Additionally, we propose five other low-level technical requirements, as a result of our technical analysis of consent banners. See Section 2.2 for details.

**List of requirements.** We assert that the complete list of 22 low-level requirements (see Table 6 of Section 5) derived from the high-level requirements presented in this section is exhaustive from a legal perspective. However, given that technologies are constantly evolving, we do not guarantee the exhaustiveness of the low-level technical requirements. To ensure legal compliance today, a consent banner implementation must meet all low-level requirements derived from binding legal sources. However, we strongly encourage that such implementations also comply with other, non-binding requirements presented in this paper. We believe all requirements should become mandatory and binding in the near future.

**Presentation of requirements.** Whilst deciphering each high-level requirement and the respective low-level requirements, we propose a description thereof, consisting of a concise designation of a requirement (e.g. “Prior to setting cookies”), and followed by its concrete and objective explanation (e.g. consent must be obtained before cookies requiring consent are set). For readability purposes, we additionally extend this description, whenever possible, with further observations.

[13]Finish DPA, “Guidance on Confidential Communications” (2019) <www.kyberturvallisuuskeskus.fi/fi/toimintamme/saantely-ja-valvonta/luottamuksellinen-viestinta, accessed 7 May 2020 (our translation).

### 2.1.4 On the national implementation of the ePrivacy Directive

**Implementation of the ePD at national level.** Although the ePD stipulates the need for consent for the storage of and/or access to cookies, the practical implementations of the legal requirements vary among website operators across EU Member States (MSs)[14]. Fragmented transpositions of the ePD at national levels create problems and legal uncertainty for European citizens as well as for the digital single market. Interestingly, 25 MSs have not fully updated ePD since the GDPR came into force. On the 30 January 2020, the Commission[15] reported that only three Member States "seem to have properly adapted the provisions" of the ePD following the entry into application of the GDPR. On the 5th of May, the Commission mentioned[16] that it asked MSs for information regarding the implementation of certain provisions. It stated it is still assessing the situation and will take a decision on appropriate measures in accordance with the Treaties. In this line, we do not study the differences of the ePD implementation in each MS due to lack of linguistic skills, volume constraints and also due to the fact that some legislations are quite old and do not contemplate neither the GDPR standard of consent, nor the recent CJEU jurisprudence.

## 2.2 On the technical analysis

**Technical assessment of legal requirements.** A technical analysis is performed of the legal requirements by computer scientists – co-authors of the paper, experts in Web Tracking Technologies. They evaluate how each requirement translates into practice for the technology of consent banners. They notably reflect on whether the GDPR requirements are compatible with existing web technologies, as it is not always the case, and technologies need to be adapted. For instance, the "prior to sending cookies" requirement cannot be enforced with the current state of technologies, because unless configured otherwise, cookies are sent automatically by browsers in every request.

**Checking consent implementation on websites.** We proceed with an empirical step of visiting example websites where a given low-level requirement is respected or violated and investigating the consent banner implementation.

**Requirements coming from technical computer science analysis.** Other requirements resulted from the domain-expertise of computer scientists. The four technical requirements are the following:

- R2 prior to sending an identifier,
- R14 post-consent registration,
- R15 correct consent registration,
- R22 delete "consent cookie" and communicate to third parties.

For example, we explain how the GDPR's requirement for "revocable consent" could be implemented in practice: when consent is revoked, the publisher should delete the consent cookie and communicate the withdrawal to all third parties who have previously received consent. This operation implies the emergence of a new technical requirement: R22 "Delete "consent cookie" and communicate to third parties".

**Procedure to verify compliance**. For each requirement, we analyzed whether compliance thereto can be verified with technical means, using existing computer science tools or where feasible with state-of-the art technologies. If no such tools exist or seem feasible, we analyzed how each requirement can be manually verified. As a result, for each requirement, we have identified whether its violation can be detected:

- Technically, by an expert using computer tools
- Manually, relying only on a human operator, or
- Performing user studies to evaluate perception of end users.

[14] For example, Germany has not implemented the ePrivacy Directive, though it has formulated guidelines on the use of trackers.

[15] European Parliament, "Implementation of the ePrivacy Directive following the entry into application of the GDPR" (2020) <https://www.europarl.europa.eu/doceo/document/E-9-2020-000790_EN.html> accessed on the 18th June 18, 2020.

[16] European Parliament, "Implementation of the ePrivacy Directive following the entry into application of the GDPR" (2020) <https://www.europarl.europa.eu/doceo/document/E-9-2020-000790-ASW_EN.html> accessed on the 18th June 18, 2020.

Additionally, for each requirement where technical means are not possible today, we analyzed which upcoming technologies and possible technical solutions could be implemented.

### 2.3 Exclusions from this work

**Explicit consent.** We have excluded the requirement of explicit consent which is required whenever websites deal with: i) special categories of data (listed in Article 9 of the GDPR); ii) data transfers to third countries; and iii) automated decision-making (including profiling). As this requirement should contain a double-layer verification approach – following the recommendation by the 29WP (since ticking one box or pressing one button is not enough to ensure an affirmative and explicit act) – we decided not to contemplate this added layer verification effort.

**Freely given.** In the analysis of the element of a freely given consent, we did not consider the cases of *imbalance of power* (Recital 43 of the GDPR) for the same motive as above. This is mostly observed in the context of a public authority, employer, medical service relationship, or wherever there is a dominant position in relation to the data subject. In such contexts, the data subject fearing adverse consequences has no realistic alternative to accept the processing terms.

**Informed consent.** While considering the information necessary for an informed consent, we excluded the analysis of the purposes of an informed consent, meaning that we do not analyze the meaning of the purposes presented in the cookie banners. We state that in the information page, each purpose should be sufficiently unambiguous and clearly expressed, specific and clear. We nevertheless in general address intelligible and clear expression of information in the "Readable and Accessible" high-level requirement.

**Browser settings.** This paper does not analyze consent expressed through browser settings. We think that browser settings, as they exist today, do not correspond to the requirements of a valid consent for the following reasons: (a) no purposes are specified; (b) they do not reflect an informed decision; and (c) browser settings do not express an unambiguous consent. The 29WP[17] mentions that browser settings may be considered as a mechanism for expressing consent if they are clearly presented to the user. We do not agree with this statement for the reason that many browser vendors expose cookie settings in browser preferences that are hard to find. Moreover, the location and user interface of such cookie settings changes significantly from one version of the browser to another. Even though cookie settings work in some browsers, this does not generally apply to all tracking technologies. For example, since there is no precise way to detect browser fingerprinting and moreover, the purpose of such fingerprinting is not known, browser preferences are not a meaningful control mechanism for this tracking technology. Due to the complexity of this topic, we have excluded it from this paper.

**Children consent.** We do not address the specific concerns related to children's consent.

**Exceptions.** We left exceptions specified in the GDPR out of our study, e.g. cases of medical research conducted in the public interest or for compliance with legal obligations (Recital 51).

## 3 Background

In this section, we outline a summary of the legal fabric mostly related to cookies and other browser-based technologies, personal data collection and consent as reflected in a consent banner. In Section 3.1, we discuss how web services process personal data through tracking technologies. Section 3.2 presents the applicable rules for consent – the ePrivacy Directive and the GDPR.

### 3.1 Web services process personal data

The digital economy is increasingly dominated by service providers that collect and process vast amounts of personal data. Web services are a central part of the interface of any organization for the dissemination of information, collection of input and more complex transactions. We assume that web services process personal data.[18] Therefore, these web services must be operated in compliance with the privacy and data protection principles, so that the fundamental rights to privacy and to the protection of personal data are guaranteed. Examples of personal data abound: data that enables users to log in into the web service for authentication and customization purposes, IP addresses, user identifiers, timestamps, URLs of the visited pages and other parameters that enable the user to be singled-out. Usage of cookies for storing identifiers are explicitly mentioned in Recital 30 of the GDPR:

> "Natural persons may be associated with online identifiers provided by their devices, applications, tools and protocols, such as internet protocol addresses, cookie identifiers or other identifiers. (…) This may leave traces which, in particular when combined with unique identifiers and other information received by the servers, may be used to create profiles of the natural persons and identify them".

[17] Article 29 Working Party, "Working Document 02/2013 providing guidance on obtaining consent for cookies" (WP 208, 2 October 2013) 4 (henceforth named 29WP 208).
[18] Personal data means any information relating to an identified or (directly or indirectly) identifiable natural person. In determining whether the information relates to an identifiable individual, website publishers need to consider any means that could reasonably be used by them or any third party to enable the identification of an individual, according to Art. 4(1) and Recital 26 of the GDPR. For a deeper analysis of this concept, see Article 29 Working Party, "Opinion 4/2007 on the concept of personal data" (WP 136, 20 June 2007).

It is noteworthy that personal data do not consist only in the data originally collected via the web service, but also in any other information that the controller collected through other means and that can be linked to personal data collected through the web service. It also means any other information inferred that relates to an individual. The European Data Protection Supervisor (hereinafter named EDPS) declares that the use of device fingerprinting can lead to a certain percentage of assurance that two different sets of data collected belong to the same individual.[19] Thus, the GDPR applies to data that can identify users (i.e. when identification of users is likely), whether they are meant or used to track the online activity of such users.

In general, any use of tracking technologies[20] which involves the processing of personal data, whether to identify directly (e.g. an email address) or more often to identify indirectly (e.g. unique cookie identifier, IP address, device identifier or component of the device, device fingerprinting, identifier generated by a software program or operating system) must comply with the GDPR. While many cookies indeed contain unique identifiers, it does not hold to all types of data; for example, some of them carry information which is too coarse to identify users, while several of them can be combined to uniquely identify users. As such, website operators need to consider cookies as storage mechanisms that may potentially contain personal data and therefore protect it accordingly. Cookies used for tracking users' online activities are unique identifiers used to single them out and recognize returning website visitors. As a result, such tracking cookies are personal data as defined in the GDPR, even if the traditional identity parameters (name, address, etc.) of the tracked user are unknown or have been deleted by the tracker after collection.

### 3.2 Applicable rules for consent: the ePrivacy Directive and the GDPR

The ePD prescribes that websites obtain users' informed consent before using any kind of tracking technology. Article 2(f)[21] and Recital 17[22] of the 2002 ePD define consent in reference to the one set forth in Directive 95/46/EC[23], the GDPR predecessor. The GDPR points out the conditions for obtaining valid consent in Articles 4(11) and 7. Article 4(11) of the GDPR provides for the elements composing a valid consent: "any freely given, specific, informed and unambiguous indication of the data subject's wishes by which he or she, by a statement or by a clear affirmative action, signifies agreement to the processing of personal data relating to him or her". The GDPR provides additional guidance in Article 7 and in Recitals 32, 33, 42, and 43 as to how the controller must act to comply with the main elements of the consent requirement.

On websites, consent for cookies is usually presented in a form of cookie banners. A cookie banner is a means for getting user's consent on the usage of cookies and potentially other web application technologies that can store data or use browser attributes to recognize the user's browser, such as browser fingerprinting.[24]

[19] European Data Protection Supervisor, "Opinion 6/2017 on the Proposal for a Regulation on Privacy and Electronic Communications (ePrivacy Regulation)", Abril 2017, 14 (henceforth named EDPS Opinion).
[20] Irene Kamara, Eleni Kosta, "Do Not Track initiatives: regaining the lost user control" (2016) *International Data Privacy Law*, Volume 6, 276–290.
[21] Art. 2(f) reads that "consent by a user or subscriber corresponds to the data subject's consent in Directive 95/46/EC*"*.
[22] Recital 17 provides that "for the purposes of this Directive, consent of a user or subscriber, regardless of whether the latter is a natural or a legal person, should have the same meaning as the data subject's consent as defined and further specified in Directive 95/46/EC".
[23] Directive 95/46/EC of the European Parliament and of the Council of 24 October 1995 on the protection of individuals with regard to the processing of personal data and on the free movement of such data, OJ 1995 L 281/31.
[24] Pierre Laperdrix, Nataliia Bielova, Benoit Baudry, Gildas Avoine, "Browser Fingerprinting: A survey" (2019) <https://arxiv.org/abs/1905.01051> accessed 7 May 2020.

## 4 Scoping browser-based tracking technologies

This section presents the important elements of tracking technologies: user and/or subscriber, terminal equipment, browser-based tracking technology and provider of an information society service.

This paper focuses on legal requirements relating to the processing of personal data from/onto users' devices through cookies and similar technologies. In particular, within the scope of this work, we refer to the use of cookies, and any similar technologies (browser-based tracking technology) to be stored, executed and read on the user´s terminal device, and thus falling within the scope of Article 5(3) of the ePrivacy Directive, which is worded as follows:

> Member States shall ensure that the storing of information, or the gaining of access to information already stored, in the terminal equipment of a subscriber or user is only allowed on condition that the subscriber or user concerned has given his or her consent, having been provided with clear and comprehensive information, in accordance with Directive 95/46/EC, inter alia, about the purposes of the processing. This shall not prevent any technical storage or access for the sole purpose of carrying out the transmission of a communication over an electronic communications network, or as strictly necessary in order for the provider of an information society service explicitly requested by the subscriber or user to provide the service.

Article 5(3) of the ePD applies to *providers* that store or gain access to information in the *terminal equipment* of the *subscriber* or *user*. Account must be taken to these four framing elements below:

- *Subscriber and/or user,*
- *Terminal equipment,*
- *Browser-based tracking technology,*
- *Provider of an information society service.*

**Subscriber and/or user.** The *subscriber*[25] means the person who pays the bill for the use of the online service. The *user* is the person using either the computer or any other device to access the online service. In many cases, the subscriber and the user can coincide, e.g. when an individual uses the broadband connection to access a website on his computer or mobile device – this person would be both the "user", as well as the "subscriber", if he or she pays for the connection. However, this is not always the case, since end-users might include employees, tenants, hotel guests, family members, visitors, and any other individuals who are using the service, for private or business purposes, without necessarily having subscribed to it. Following the example given by the UK DPA, if a family member or a visitor visits this subscriber's home and uses his internet connection to access that service from their own device, he would be the user.[26]

The ePD does not specify from whom the consent is required. The legislator did not preview which consent takes precedence (the user's or the subscriber's), nor if that choice should be at the discretion of the entity that stores or gains access to the information.[27] Whilst the web publisher, in principle, is not meant to distinguish between a consent provided by the subscriber or the user, what is relevant is that one of the parties must deliver a valid consent against BTT-related information in the landing page. Surmounting this qualification, the EDPS[28] recommends including a stand-alone definition of end-user in the forthcoming ePrivacy Regulation, for purposes of providing consent, to ensure that it is the individuals effectively using the service, rather than those subscribing to it. In this paper, we use *user* and *data subject* interchangeably.

**Terminal equipment**. *Terminal equipment* refers to a device where information is accessed or stored, e.g. desk computers, laptop, pads, smartphones, but also other equipment such as wearable technologies, smart TVs, game consoles, connected vehicles, voice assistants, as well as any other object that is connected to an electronic communication network open to the public. Our understanding of the term *web service* refers to any type of information service made accessible over the internet with which users interact usually through web browsers, mobile apps or other client software. IoT web services, accessed by IoT devices, are included.

[25] This paper uses "subscriber" and "user" interchangeably.
[26] UK DPA (also known as ICO), "Guidance on the rules on use of cookies and similar technologies", Privacy and Electronic Communications Regulations, (2019) 9 <https://ico.org.uk/media/for-organisations/guide-to-pecr/guidance-on-the-use-of-cookies-and-similar-technologies-1-0.pdf> accessed 7 May 2020 (henceforth named ICO Guidance).
[27] Eleni Kosta, "Peeking into the cookie jar: the European approach towards the regulation of cookies" (2013) *International Journal of Law and Information Technology*, Volume 21, Issue 4, 380–406 <https://doi.org/10.1093/ijlit/eat011> accessed 7 May 2020.
[28] cf. EDPS Opinion (n 19) 14.

**Browser-based Tracking Technology.** A *Browser-based Tracking Technology* (henceforth named BTT), the third element of this quadrant, is commonly acknowledged as any technology which enables tracking of the user while she visits a website using a Web browser. From a legal perspective, BTT is defined as the reading or storing of information from/onto the users' devices for tracking purposes, in line with the text of Article 5(3) of the ePD. From a computer science perspective, BTT is a technology that enables tracking of the user by either depositing identifiers on their computer or using fingerprinting methods to identify them. For a technology to successfully track a browser user, trackers need to have two key capabilities:

1. The ability to store a unique identifier (or to re-create it) on a user's machine,
2. The ability to communicate that identifier, as well as visited sites, back to the domain, controlled by the tracker.

The most common type of BTT is "*stateful*" tracking. A typical example of it are *browser cookies*. They are used to store a unique identifier and communicate it to their owner's domain when they are automatically sent by the browser, or via JavaScript programs running on a visited website. Alternative tracking technologies that rely on other browser storages are also actively used by trackers today – they include HTML5 local storage, browser cache and many others. Most of technologies rely on JavaScript programs, since this is the most convenient and portable way to both store and send the unique identifier to the tracker's domain.

Alternatively, instead of storing an identifier, a tracker can re-create it based on the browser's and machine's properties, accessible via the HTTP protocol and also via JavaScript. Such tracking is called "*stateless*", and for Web browsers is represented by browser fingerprinting.[29]

In this paper, we unify all such technologies under the common terminology of *BTT*. In the scope of current BTT, the risk to data protection comes from the purpose(s) of processing (29WP (WP194)).[30]

**Provider of an information society service**. The *provider of an information society service* (i.e. a publisher) provides a website content service, at the request of a user, either paid or unpaid, remotely and electronically.

### 4.1 Browser-based tracking technologies requiring or exempted from consent

In this paper, we only refer to the use of BTT requiring consent. According to Article 5(3) of the ePD, consent is not required when the purpose of trackers is:

- ***Communication**:* used for the sole purpose of enabling the communication on the web; and
- ***Strict necessity***: cookies strictly necessary to enable the service requested by the user: if BTT is disabled, the service will not work.

The above mentioned 29WP (WP194)[31] analyzed these two exceptions accordingly (considering browser cookies, but it is extended to all BTTs):

- The *communication exemption* applies when the transmission of the communication is impossible without the use of the BTT (e.g. load-balancing cookie). Hence, using BTT to merely "assist" or "facilitate" the communication is insufficient.
- The *strict necessity exemption* involves a narrow interpretation. It means that the use of BTT must be restricted to what is strictly necessary (and hence essential) to provide a service explicitly requested by a user. Thus, using BTT that is *reasonably necessary* or *important* – this implies that the service provided by the website operator would not function without the BTT. In this regard, the choice of a certain functionality that relies on BTT is not enough to justify the *strict necessity* if the web publisher has a different implementation choice that would work without a BTT.

Both the 29WP and DPAs provide explicit examples of BTT's purposes that require the user's consent. They assert that the following purposes are usually not strictly necessary to the user visiting a website, since they are usually related to a functionality that is distinct from the service that has been explicitly requested:

[29] cf. Laperdrix et al. (n 24).
[30] Article 29 Working Party, "Opinion 04/2012 on Cookie Consent Exemption" (WP194), June 2012 (henceforth named "29WP (WP194)").
[31] cf. 29WP (WP194) (n 30).

"advertising, and use of the data for marketing, research and audience measurement" are not strictly necessary to deliver a service that is requested by a user (29WP (WP240)).[32]

We will further analyze which BTTs are exempted from consent based solely on their *purpose*, and not on their technical abilities. Ultimately, as the 29WP (WP194) exposes, "it is thus the purpose and the specific implementation or processing being achieved that must be used to determine whether or not a cookie can be exempted from consent". The 29WP (WP194) clarifies further that when applying the exemptions for obtaining consent, it is important to examine what is strictly necessary *from the point of view of the user*, not of the service provider. Regarding *multipurpose* BTT, whenever a BTT covers different purposes, some of which require consent (e.g. can be used for the purpose of remembering user preferences and for the purpose of tracking), the website still needs to seek user consent for such multipurpose BTT. The 29WP recalls that in practice, this should encourage website owners to use a different BTT for each purpose.

For this classification of the purposes of BTT, we relied on the guidance from the 29WP.[33]-[34] For a comparative analysis, we also consulted the recent guidelines from DPAs (ICO, CNIL, German and Dutch DPA). This classification is shown in Table 5.

**Table 5** Examples of purposes of BTT exempted and non-exempted of consent.

| Purposes exempted of consent | Purposes needing consent |
|---|---|
| ***Local Analytics*** – These are statistical audience measuring tools for providing information on the number of unique visits to a website, how long users stay in the site, what parts and pages of the website they browse, detecting main search keywords, track website navigation issues. The 29WP and the EDPS[35] exempt these from consent insofar they are limited to first party (website owner) anonymized and aggregated statistical purposes, as these are not likely to create a privacy risk. The CNIL[36] points out that certain analytic cookies can be exempted if they meet a list of cumulative requirements. The Irish[37] and Dutch DPAs[38] state that these may have little privacy effects on users. | ***Non-local Analytics*** – Even if a website owner relies on self-claims of "strictly necessary" first-party analytics, the 29WP[39] says that they are not strictly necessary to provide a functionality explicitly requested by the user, because the user can access all the functionalities provided by the website when such cookies are disabled. As a consequence, these cookies do not fall under the exemption of consent if they are not limited to the website owner. Moreover, both the ICO[40] and the German DPA[41] held that third-party analytics cookies are not *strictly necessary.* The Greek DPA[42] says that third-party web analytics trackers, such as the Google Analytics service as not *strictly necessary*, hence requiring consent. |

[32] Article 29 Working Party, "Opinion 03/2016 on the evaluation and review of the e-Privacy Directive (2002/58/EC)", (WP240, 19 July 2016).
[33] Article 29 Working Party, "Opinion 2/2010 on online behavioural advertising", (WP 171, 22 June 2010).
[34] cf. 29WP (WP194) (n 30).
[35] European Data Protection Supervisor, "Guidelines on the protection of personal data processed through web services provided by EU institutions" (2016) 10-13 (henceforth named "EDPS Guidelines").
[36] The French DPA (also known as CNIL), prescribes that these cookies may be exempt from consent if the conditions of Article 5 are met, such as: the user is informed thereof and has the possibility to refuse them; such cookies exclude any form of unique targeting of individuals; the collected data must not be combined or merged with other types of data, nor disclosed to third parties; the use of trackers must be strictly limited to producing anonymous statistics; the trackers may only be used by one publisher and must not enable tracking a user over different websites or mobile apps; an IP address cannot be used to geolocate the user more precisely than the city, otherwise must be deleted or anonymized once the user has been located to avoid this data from being used or combined with other data. See CNIL Guidelines on cookies and other trackers, (2019) <www.legifrance.gouv.fr/affichTexte.do?cidTexte=JORFTEXT000038783337> accessed 7 May 2020.
[37] Irish DPA, Guidance note on the use of cookies and other tracking technologies (2020) <https://www.dataprotection.ie/sites/default/files/uploads/2020-04/Guidance%20note%20on%20cookies%20and%20other%20tracking%20technologies.pdf> (henceforth named "Irish DPA Guidance")
[38] An explanation of the legal requirements for cookies (besides tracking cookies) is available on the website of the Netherlands Authority for Consumers and Markets (ACM), "Cookies" (2019) <www.acm.nl/nl/onderwerpen/telecommunicatie/internet/cookies> accessed 7 May 2020.
[39] Regarding first-party analytics, the 29WP (WP194) (n 30) considers that these are not likely to create a privacy risk when they are strictly limited to aggregated statistical purposes and when users are informed thereof and can opt out therefrom.
[40] The ICO declares that it is "unlikely that priority for any formal action would be given to uses of cookies where there is a low level of intrusiveness and low risk of harm to individuals" and first party analytics cookies are given as an example of cookies that are potentially low risk, cf. ICO Guidance (n 26).
[41] cf. German DPA Guidelines (n 10).
[42] Greek Data Protection Authority, "Guidelines on Cookies and Trackers" (2020) http://www.dpa.gr/APDPXPortlets/htdocs/documentSDisplay.jsp?docid=84,221,176,170,98,24,72,223.

| | |
|---|---|
| ***Session User input*** – The 29WP states these are used to keep track of the user's input (session-id) when filling online forms over several pages, or as a shopping cart, to keep track of the items the user has selected by clicking on a button. These BTT are clearly needed to provide a service explicitly requested by the user, for the duration of a session. Additionally, they are tied to a user action (such as clicking on a button or filling a form). | ***Advertising*** – The 29WP affirms that third-party advertising BTTs require consent, as well as operational purposes related to third-party advertising, such as frequency capping, financial logging, ad affiliation, click fraud detection, research and market analysis, product improvement and debugging.[43] Even though the 29WP only distinguishes third-party advertising, we believe that the category of purposes should only be called "Advertising". We insist on it because it has been observed that first-party cookies are also often synchronized with third-party cookies[44], and moreover, publishers started hiding advertising content under the first-party content (typical case is with DNS redirection). The ICO[45] posits that while advertising cookies may be crucial in the eyes of a website or mobile app operator as they bring in revenue to fund the service, they are not "*strictly necessary*" from the point of view of the website user and hence, the law. The Dutch DPA[46] names these as *tracking cookies* and advises companies to request consent to place tracking cookies. The same reasoning holds for the German[47] and Irish[48] DPA. |
| ***User-security for a service explicitly requested by the user*** – The 29WP names these due to their function on providing security functionalities for a service the user has requested (e.g. online banking services) and for a limited duration, e.g. to detect repeated failed login attempts on a website, or other similar mechanisms designed to protect the login system from abuses. | ***User-security for a service not explicitly requested by the user*** – The 29WP[49] refers to cookies providing security for content not explicitly requested by the user. For example, if a website uses advertising content that contains user-security cookies, such as those of Cloudflare, then the user consent is required.[50] |
| ***Social media plugin for a functionality explicitly requested by the user*** – The 29WP refers that many social networks propose "social plug-in modules" that website owners integrate in their platform, to provide some services than can be considered as "explicitly requested" by their members, e.g. to allow them to share content they like with their "friends" (and propose other related functionalities such as publishing comments). These plugins store and access cookies in the user's terminal equipment in order to allow the social network to identify its members when they interact with them. | ***Social media plugin for a functionality not requested by the user*** – The 29WP refers that these "social plug-in modules" can also be used to track users: logged-in, "non-logged-in" users, and also non-members. We conclude however that even logged-in members can be tracked and therefore name this category as "functionality not requested by the user". The German DPA has the same position.[51] |
| ***Session Authentication*** – The 29WP describes these as the ones used to identify the user once he has logged in into websites, for the duration of a session. They allow users to authenticate themselves on successive loads of the website and gain access to authorized content or functionality, such as viewing their account balance, transactions in an online banking website, online shopping. This authentication functionality is an essential part of the service a user explicitly requests. | ***Persistent Authentication*** – The 29WP says also that persistent login cookies which store an authentication token across browser sessions are not exempted of consent. This is an important distinction because the user may not be immediately aware of the fact that closing the browser will not clear their authentication settings. They may return to the website under the assumption that they are anonymous whilst in fact they are still logged in to the service. |

[43] cf. 29WP (WP194) (n 30) 9-10.
[44] Imane Fouad, Nataliia Bielova, Arnaud Legout, and Natasa Sarafijanovic-Djukic. Missed by Filter Lists: Detecting UnknownThird-Party Trackers with Invisible Pixels. In proceedings on Privacy Enhancing Technologies ; 2020 (2):499–518, https://petsymposium.org/2020/files/papers/issue2/popets-2020-0038.pdf
[45] cf. ICO Guidance (n 26) 39.
[46]Autoriteit Persoonsgegevens, "Cookies" (2019) <https://autoriteitpersoonsgegevens.nl/nl/onderwerpen/internet-telefoon-tv-en-post/cookies#mag-ik-als-organisatie-een-cookiewall-gebruiken-7111> accessed 7 May 2020 (henceforth named "Dutch DPA").
[47] cf. German DPA Guidelines (n 10).
[48] cf. Irish DPA Guidance (n 37).
[49] According to the 29WP the consent exemption does not cover the use of cookies that relate to the security of websites or third-party services that have not been explicitly requested by the user, (WP194) (n 30) 7.
[50]The purpose of the cookie "__cfduid" is used by Cloudflare for detection of malicious visitors. Such cookie requires consent when it is used by Cloudflare in advertising content or other content not explicitly requested by the user, Cloudflare, "Understanding the Cloudflare Cookies" (2019) <https://support.cloudflare.com/hc/en-us/articles/200170156-What-does-the-Cloudflare-cfduid-cookie-do-> accessed 2 December 2019.
[51] cf. German DPA Guidelines (n 10).

| | |
|---|---|
| ***Short-term User Interface Customization (personalization, preferences)*** – According to the 29WP, these are used to store a user's preference regarding a service across web pages and not linked to other persistent identifiers such as usernames. These are explicitly enabled by the user, e.g. by clicking on a button or ticking a box to keep a language, display format, fonts, etc. Only session (or short term) cookies storing such information are exempted. | ***Long-term User Interface Customization*** – The 29WP says that the addition of information to remember the user's preference for a longer duration will not be exempted of consent. |
| ***Load Balancing*** – The 29WP says that load balancing is a technique that allows distributing the processing of web server requests over a pool of machines instead of just one. Among several techniques, a cookie may be used to identify the server in the pool in order for the load balancer to redirect the requests appropriately. These are session cookies. | |
| ***Session Multimedia Content Player*** – The 29WP clarifies that these apply to BTT used to keep track of the state of audio/video. When the user visits a website containing related text/video content, this content is equally part of a service explicitly requested by the user and is exempted of consent. As there is no long-term need for this information, they should expire once the session ends. | |

## 5 Requirements for valid consent for consent banners

This section presents our interdisciplinary legal and technical analysis of the requirements applied to consent banner design. We present the seven high-level requirements, followed by the definition of 22 low-level requirements.

We convey the respective *legal sources* upon which each requirement is based. The sources are either: binding (GDPR, ePD and CJEU case-law), non-binding (EDPB and DPA guidelines) and grounded in our own subjective interpretation of legal sources (L) or from a technical computer science perspective (CS). We present the *procedure for compliance verification*. For each requirement, we describe the procedure that needs to be put in place in order to detect violations. Such procedure can be assessed in three ways:

- Manual, relying only on a human operator (M);
- Technical, an expert using computer tools able to detect a violation (T);
- Performing user studies to evaluate perceptions of end users (U).

Table 6 describes all the high- and low-level requirements, their provenance, position in the paper that describes them in detail and how they can be assessed.

**Table 6** Requirements for a valid consent on consent banner design, assessment and source

| Requirements | | Assessment | Sources at low-level requirement | | | Location in the paper (page) |
|---|---|---|---|---|---|---|
| **High-Level Requirements** | **Low-Level Requirements** | **Manual (M), Technical (T) or User study (U)** | **Binding** | **Non-binding** | **Interpretation: Legal (L) or Computer Science (CS)** | |
| **Prior** | R1 Prior to storing an identifier | M (partially) or T (partially) | ✔ | ✔ | - | 17 |
| | R2 Prior to sending an identifier | T (partially) | - | - | CS | 19 |
| **Free** | R3 No merging into a contract | M (fully) or T (partially) | ✔ | ✔ | - | 22 |
| | R4 No tracking walls | M (fully) | - | ✔ | - | 24 |
| **Specific** | R5 Separate consent per purpose | M (fully) | ✔ | ✔ | - | 28 |
| **Informed** | R6 Accessibility of information page | M (fully) or T (partially) together with U | - | ✔ | - | 34 |
| | R7 Necessary information on BTT | M (fully) or T (partially) | ✔ | ✔ | - | 35 |
| | R8 Information on consent banner configuration | M (fully) or T (partially) | - | ✔ | - | 37 |
| | R9 Information on the data controller | M (fully) or T (partially) | ✔ | ✔ | - | 38 |
| | R10 Information on rights | M (fully) or T (partially) | ✔ | ✔ | - | 39 |
| **Unambiguous** | R11 Affirmative action design | Combination of M and T (partially) | ✔ | ✔ | - | 40 |
| | R12 Configurable banner | M or T (partially) | - | ✔ | L | 43 |
| | R13 Balanced choice | M (fully) | - | ✔ | L | 45 |
| | R14 Post-consent registration | T (partially) | - | ✔ | CS | 47 |
| | R15 Correct consent registration | Combination of M and T (partially) | - | ✔ | CS | 49 |
| **Readable and accessible** | R16 Distinguishable | M (fully) or T (partially) | ✔ | ✔ | - | 52 |
| | R17 Intelligible | U | ✔ | ✔ | - | 52 |
| | R18 Accessible | U | ✔ | ✔ | | 52 |
| | R19 Clear and plain language | U | ✔ | ✔ | - | 53 |
| | R20 No consent wall | M (fully) or T (partially) | - | ✔ | L | 53 |
| **Revocable** | R21 Possible to change in the future | M (fully) | ✔ | ✔ | - | 57 |
| | R22 Delete "consent cookie" and communicate to third parties | Not possible | - | - | CS | 59 |

Table 7 depicts the positioning of DPAs (French, UK, Irish, German, Spanish, Greek, Danish, Belgium) in relation to the 22 low-level requirements proposed in this paper.

**Table 7** DPAs positioning in relation to the low-level requirements

| **Requirements** | | **DPAs positioning** | | | | | | | |
|---|---|---|---|---|---|---|---|---|---|
| **High-Level Requirements** | **Low-level Requirements** | **French (CNIL)** | **UK (ICO)** | **Irish** | **German** | **Spanish** | **Greek** | **Danish** | **Belgium** |
| Prior | **R1 Prior to storing an identifier** | ✔ | ✔ | ✔ | ✔ | - | ✔ | ✔ | ✔ |
| | **R2 Prior to sending an identifier** | - | - | - | - | - | - | - | - |
| Free | **R3 No merging into a contract** | ✔ | ✔ | ✔ | ✔ | - | | ✔ | ✔ |
| | **R4 No tracking walls** | ✔ | **?** | ✔ | ✔ | **?** | ✔ | ✔ | ✔ |
| Specific | **R5 Separate consent per purpose** | ✔ | ✔ | ✔ | ✔ | ✔ | ✔ | ✔ | ✔ |
| Informed | **R6 Accessibility of information page** | ✔ | ✔ | ✔ | ✔ | ✔ | ✔ | ✔ | ✔ |
| | **R7 Information on BTT** | ✔ | ✔ | ✔ | ✔ | ✔ | ✔ | ✔ | ✔ |
| | **R8 Information on consent banner configuration** | ✔ | ✔ | ✔ | - | ✔ decision | ✔ | ✔ | ✔ |
| | **R9 Information on the data controller** | ✔ | ✔ | ✔ | ✔ | ✔ | ✔ | ✔ | ✔ |
| | **R10 Information on rights** | ✔ | ✔ | ✔ | ✔ | ✔ | ✔ | ✔ | ✔ |
| Unambiguous | **R11 Affirmative action design** | ✔ | ✔ | ✔ | ✔ | X | ✔ | ✔ | ✔ |
| | **R12 Configurable banner** | ✔ | ✔ | ✔ | ✔ | ✔ decision | ✔ | ✔ | - |
| | **R13 Balanced choice** | ✔ | ✔ | ✔ | - | | ✔ | ✔ | - |
| | **R14 Post- consent registration** | ✔ | ✔ | ✔ | - | ✔ | - | ✔ | ✔ |
| | **R15 Correct consent registration** | ✔ | - | - | - | ✔ | ✔ | - | - |
| Readable and accessible | **R 16 Clearly distinguishable** | ✔ | ✔ | ✔ | ✔ | ✔ | - | - | ✔ |
| | **R17 Intelligible** | ✔ | ✔ | - | - | ✔ | - | ✔ | - |
| | **R18 Easily accessible** | ✔ | ✔ | ✔ | - | ✔ | - | - | ✔ |
| | **R19 Clear and plain language** | ✔ | ✔ | - | - | ✔ | - | ✔ | - |
| | **R20 No consent wall** | ? | ✔ | - | - | - | - | - | - |
| Revocable | **R21 Possible to change in the future** | ✔ | ✔ | ✔ | ✔ | ✔ | ✔ | ✔ | ✔ |
| | **R22 Delete "consent cookie", communicate to third parties** | - | - | - | - | - | - | - | - |

Every subsequent subsection is structured as follows:

**Analyzing legal sources for high-level requirement.** We present each high-level requirement, followed by derived low-level requirements presented in a table with the respective sources: binding (GDPR, ePD and CJEU case-law), non-binding (EDPB and DPA guidelines) and grounded in our own subjective interpretation of legal sources (L) or from a technical computer science perspective (CS).

**One subsection for each low-level requirement.** For every low-level requirement, we present its description and

1. Explain the correspondent violation in a "requirement box" (in a consolidated form, for ease of reading).

2. Provide two example websites: one demonstrating compliance, and one presenting a violation. We illustrate (where possible) screenshots or technical content from each website to explain the technical details of each requirement. Each example is extracted from real-world websites, illustrated in figures duly dated.
3. Describe the procedure (manual, technical, or with user studies) to detect violations.

### 5.1 Prior consent

Before storing information or gaining access to information on a user's terminal, website publishers need to request prior consent to data subjects in order to guarantee that the user has some control over the processing of their information.[52] Even if no explicit provision was made manifest both in the GDPR and the ePD, the "prior" timing is confirmed through the combined analysis of both legislative instruments.

Under the GDPR aegis, the 29WP (WP259 rev.01)[53] claims that "prior consent" can be derived from Article 6 by the wording "has given",

> Although the GDPR does not literally prescribe in Article 4(11) that consent must be given prior to the processing activity, this is clearly implied. The heading of Article 6(1) and the wording "has given" in Article 6(1)(a) supports this interpretation. It follows logically from Article 6 and Recital 40 that a valid lawful basis must be present before starting a data processing.

From the ePD stance, such understanding of a "prior consent" is derived from Article 5(3) of the ePD, according to the 29WP guidance,[54]-[55]

> Article 5(3) contains a specific rule regarding the storing of information or gaining of access to information on a user's terminal, including for the purpose of tracking the user's on-line activities. While Article 5(3) does not use the word prior, this is a clear and obvious conclusion from the wording of the provision. (…) It makes good sense for consent to be obtained prior to the starting of the data processing.

In the light of the above, a consent request needs to be presented before BTT are deployed. Seconding this rule, the 29WP (WP208)[56] asserts that "consent should be sought before cookies are set or read. As a result, a website should deliver a consent solution in which no cookies are set to user's device (other than those that may not require user's consent) before that user has signaled their wishes regarding such cookies".

Moreover, processing is unlawful if carried out before the request for consent due to the lack of legal ground, as denoted by the 29WP (WP147)[57]:

> Otherwise, the processing carried out during the period of time from the moment the processing had started until the moment that consent had been obtained would be unlawful because of lack of legal ground. Furthermore, in such cases, if the individual decided against consenting, any data processing that had already taken place would be unlawful for that reason as well.

Notice that instead of specifying a certain type of BTT, such as cookies, we resort to describe low-level requirements in terms of the usage of user identifiers because BTTs are simply mechanisms that store and/or transfer identifiers, thus allowing tracking the users across the Web. We have therefore subdivided the requirement of "prior consent" into two low-level requirements, as shown in Table 8.

- first, consent must be obtained before user identifier is set or stored (those requiring consent);
- second, consent must be obtained before user identifier is sent, i.e. before the content of the webpage that is associated to such identifier is loaded.

[52] The CNIL recalls that many site publishers have reported difficulties in obtaining prior consent from Internet users before depositing and reading cookies for two main reasons: 1. this would prevent the display of certain advertisements, resulting in a significant loss of income; 2. cookies do not come from their own servers, being linked to the activity of third-party partners, over which they have no control. As a result, publishers alone cannot bear full responsibility for enforcing tracer rules as "third-party cookies" because they originate from third-party companies, "Cookies: CNIL extends its controls beyond site publishers" (2016) <www.cnil.fr/fr/cookies-la-cnil-etend-ses-controles-au-dela-des-editeurs-de-sites> accessed 11 December 2019.
[53] cf. 29WP (WP259 rev.01) (n 4) 17.
[54] Article 29 Working Party, "Opinion 15/2011 on the definition of consent" (WP187, 13 July 2011).
[55] Article 29 Working Party, "Working Document 02/2013 providing guidance on obtaining consent for cookies" (WP 208, 2 October 2013) 4.
[56] cf. 29WP (WP208) (n 55) 4.
[57] cf. 29WP (WP187) (n 54) 31.

**Table 8** Derived low-level requirements and their sources

| Requirements | | Sources at low-level requirement | | |
|---|---|---|---|---|
| **High-Level Requirements** | **Low-level Requirements** | **Binding** | **Non-binding** | **Interpretation: Legal (L) or Computer Science (CS)** |
| **Prior** | **R1 Prior to storing an identifier** | 4(11), 6(1)(a) GDPR | 29WP (almost all DPAs) | - |
| | **R2 Prior to sending an identifier** | - | - | CS |

### R1 Prior to storing an identifier

It follows from the foregoing subsection that consent must be collected before an identifier is stored in the user's device (other than those that may not require user's consent). This requirement has often been considered in legal sources as "prior to setting cookies", however cookies is just one example of a stateful BTT. Therefore, we rename this requirement as "prior to storing an identifier" because what technically happens is that a user identifier is stored on her device.

| | |
|---|---|
| **Requirement** | **Prior to storing an identifier**<br>Consent must be obtained before a user identifier is stored |
| Violation | A user identifier is stored before consent is given |
| | |

**Examples.** Figures 1 and 2 depict the case of violation of this requirement. While accessing the eBay webpage, a banner appears affirming that by using the website, the user accepts the use of cookies to enhance their services. This overlay includes a link to "learn more". This consent mechanism does not allow a user to make a choice before an advertising cookie that requires consent, named "IDE", stores a user identifier in the user's browser.

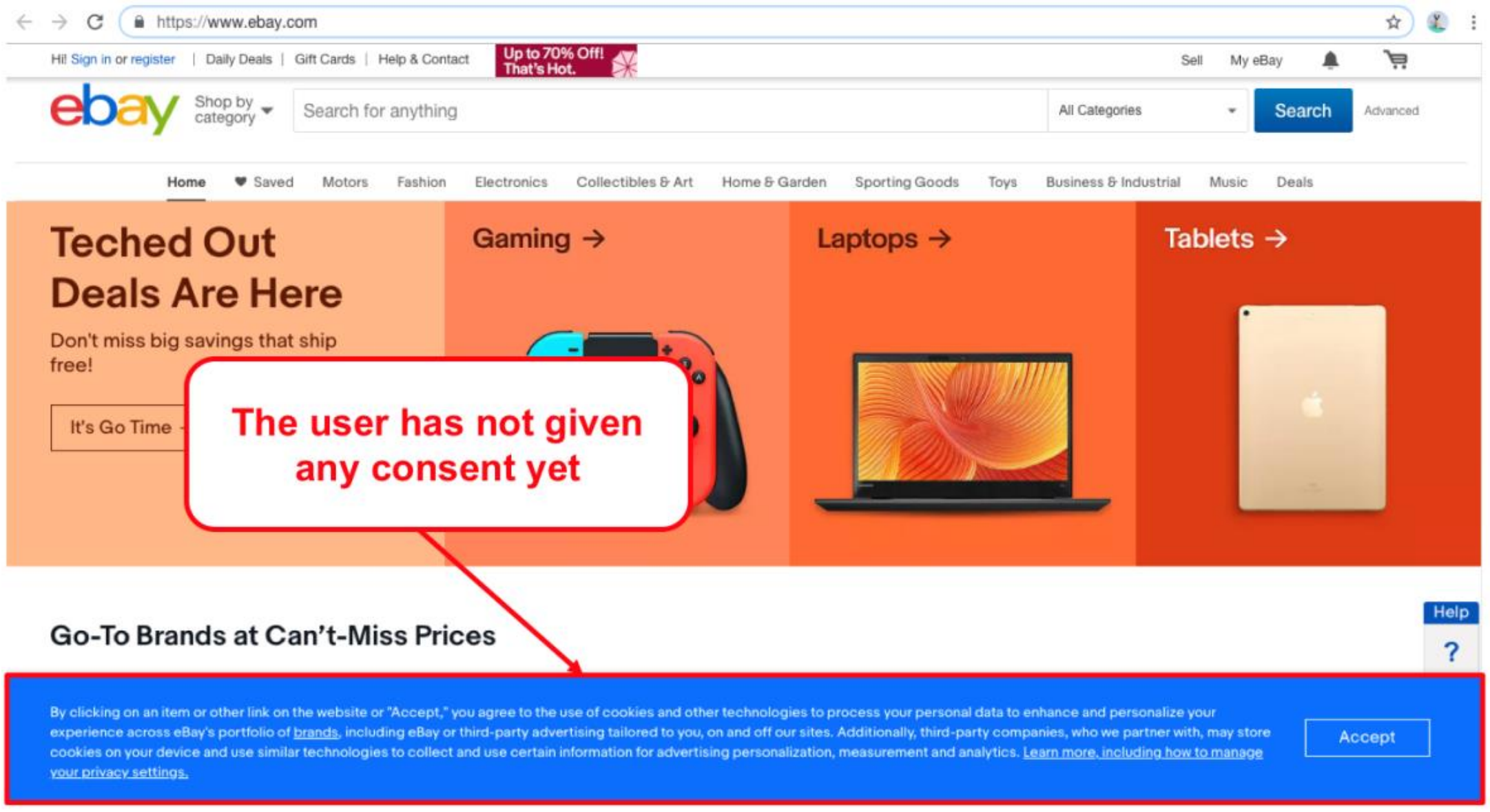

**Figure 1** Access to the eBay website (<www.ebay.com/> accessed 27 July 2019)

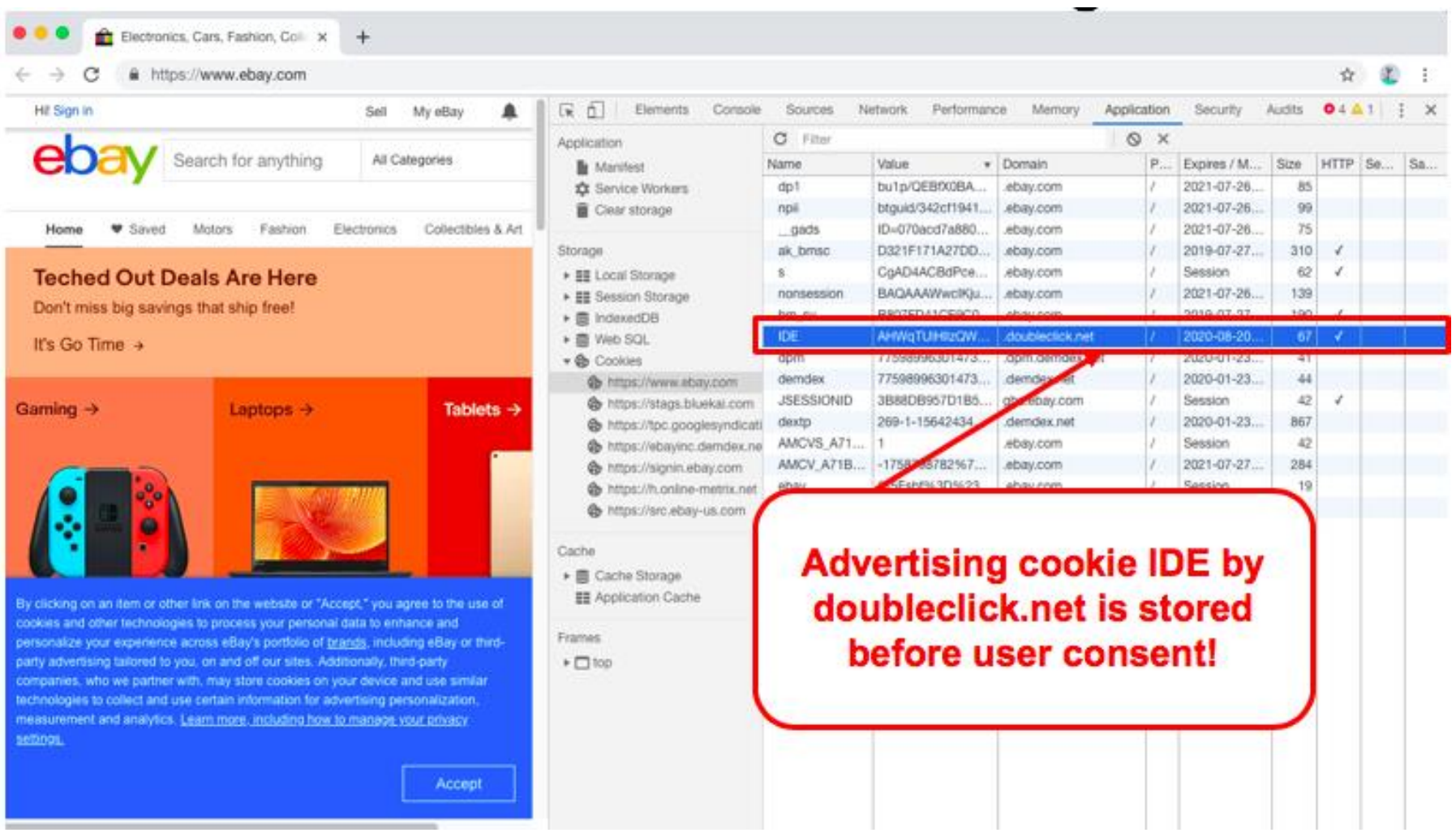

**Figure** 2 Violation of the requirement "Prior to setting cookies" by eBay website (<www.ebay.com> accessed 27 July 2019)

**How to detect violations?** One could detect a violation of the "Prior to storing an identifier" requirement by visiting a website with an empty browser storage (no cookies, empty cache, all other storages are empty) and analyzing all the elements stored in the browser (in all storages) that are set upon visiting the website (as shown in Figure 2). Such verification, however, contains two complex tasks:

1. **Detecting whether a stored element is a user identifier.** It is a very complex question of what constitutes an identifier and whether a specific element stored in a browser storage is indeed an identifier. Computer science researchers resorted to heuristics[58]-[59]-[60]-[61] that included length and expiration date for each stored element in the browser or relied on entropy measures. Nevertheless, it is not possible to know with certainty whether a given string stored in a browser storage is indeed a user identifier.
2. **Analyzing all possible browser storages.** Very specific browser tools would be required to detect such violations when various browser storage mechanisms are used, like Web caching mechanisms. It is possible to detect the setting of cookies by a technical expert with the corresponding browser tools or even fully automatically, but the complexity grows as trackers use other storages or even combinations of them (e.g. storing a piece of an identifier in one storage, and another piece in another storage). Computer science researchers have mostly analyzed cookies, and other basic storages and raised concerns about the usage of more advances techniques, such as HTTP Strict Transport Security (HSTS).[62]
3. **Identifying the purpose of an identifier.** Finally, the *purpose* of each stored identifier needs to be declared and known in order to determine whether consent is required. In general, it is rarely possible to detect a cookie's purpose automatically or with technical tools. Even manually, it is complex to estimate whether a cookie requires consent or not by reading its purpose in the cookie policy.[63] Also, the purposes of cookies described in cookie policies are often not clear, too vague or incomplete.[64]

[58] Steven Englehardt and Arvind Narayanan, "Online Tracking: A 1-million-site Measurement and Analysis", ACM CCS 2016.< https://senglehardt.com/papers/ccs16_online_tracking.pdf> accessed June 19, 2020.

[59] Steven Englehardt, Dillon Reisman, Christian Eubank, Peter Zimmerman, Jonathan Mayer, Arvind Narayanan, Edward Felten, "Cookies that Give You Away: Evaluating the surveillance implications of web tracking" , WWW 2015. <https://senglehardt.com/papers/www15_cookie_surveil.pdf> accessed June 19, 2020.

[60] Gunes Acar, Christian Eubank, Steven Englehardt, Marc Juarez, Arvind Narayanan, Claudia Diaz, "The Web Never Forgets: Persistent tracking mechanisms in the wild", ACM CCS 2014. <https://securehomes.esat.kuleuven.be/~gacar/persistent/the_web_never_forgets.pdf> accessed June 19, 2020.

[61] Tobias Urban, Dennis Tatang, Martin Degeling, Thorsten Holz, Norbert Pohlmann, "The Unwanted Sharing Economy: An Analysis of Cookie Syncing and User Transparency under GDPR". *ArXiv e-prints*, November 2018. arXiv:1811.08660.

[62] P. Syverson and M. Traudt. HSTS supports targeted surveillance. In USENIX Workshop on Free and Open Communications on the Internet (FOCI), 2018.

[63] Imane Fouad, Cristiana Santos, Feras Al Kassar, Nataliia Bielova, Stefano Calzavara. On Compliance of Cookie Purposes with the Purpose Specification Principle. IWPE, Jul 2020, Genova, Italy. ⟨hal-02567022⟩ accessed June 19, 2020.

[64] For instance, the privacy policy on the pubmatic.com website indicates that the "*repi*" cookie is "a short-lived cookie that is used to determine if repixeling is in progress". This description is obscure and makes it difficult to qualify the

For automatic verification, one would need a self-declaration of the purpose of each cookie in a standard format.

**Conclusion:** Detection of an identifier storage is a very complex task, for which technical tools do not exist today due to impossibility to detect an identifier, to technical analysis of all browser storages and to the difficulty of identifying the purpose of an identifier. Manual analysis is neither possible for the same reasons. Therefore, this requirement can only be partially assessed with technical or manual analysis.

### R2 Prior to sending an identifier

Consent must be obtained before identifiers are sent to the third parties. This requirement originates from cookies that are sent automatically when the third-party content is loaded (hence, cookies are "read"), however we generalize it to sending of identifiers that require consent via any means, including JavaScript code, for example via XMLHTTPRequest,[65] or any request that is performed on a visited website.

Such requirement is not based on any legal source but derives from its technical implementation. Law is based on a general "document read/access" view which is not in line with how the Web operates technically. Advertisers do not visit the user's browser to read their cookies, but the opposite happens: users visit websites, and their browsers send user identifiers (for example, cookies are sent automatically). Thus, we need to add this supplementary requirement that consent must be obtained before identifiers are *sent*, and not *read*. We note that respecting such a requirement demands important adaptation of current technical tools. Browsers automatically attach cookies to requests that fetch Web content, and also run JavaScript code that includes identifiers in requests, which makes it complicated for cookie banners implementation to prevent cookie transmission prior to consent.

| **Requirement** | **Prior to sending an identifier**<br>Consent must be obtained before an identifier is sent |
|---|---|
| Violation | Identifiers that require consent are sent before consent is obtained |

**Examples.** Figures 3 and 4 show how google.com sets cookies in the user's browser. Notice that google.com is a default search engine in most browsers, hence such experience is common to many users. Google.com is setting a "NID" cookie prior to the user's consent – this cookie now belongs to google.com (see Figure 4). After visiting google.com, a user goes to a different website that contains some content from google.com. Figure 5 shows an example website https://www.w3schools.com/, commonly consulted by Web developers. While accessing this website (with Firefox 69.0.1 in our experiment), no banner is shown to the user, however requests are sent to cse.google.com in order to fetch Google Customized Search Engine that helps the user to search inside this website. Figure 6 shows a violation of the 'Prior to sending an identifier' requirement because the NID cookie (see (1)) that contains a user identifier is now sent to cse.google.com (2) without user's consent while fetching some (supposedly functional) content from cse.google.com (3).

purpose of this cookie. Another example is "*centerVisitorId*" on the learnworlds.com website, whose only description states: "used by site's popups and download forms".

[65] XMLHttpRequest Living Standard, https://xhr.spec.whatwg.org/.

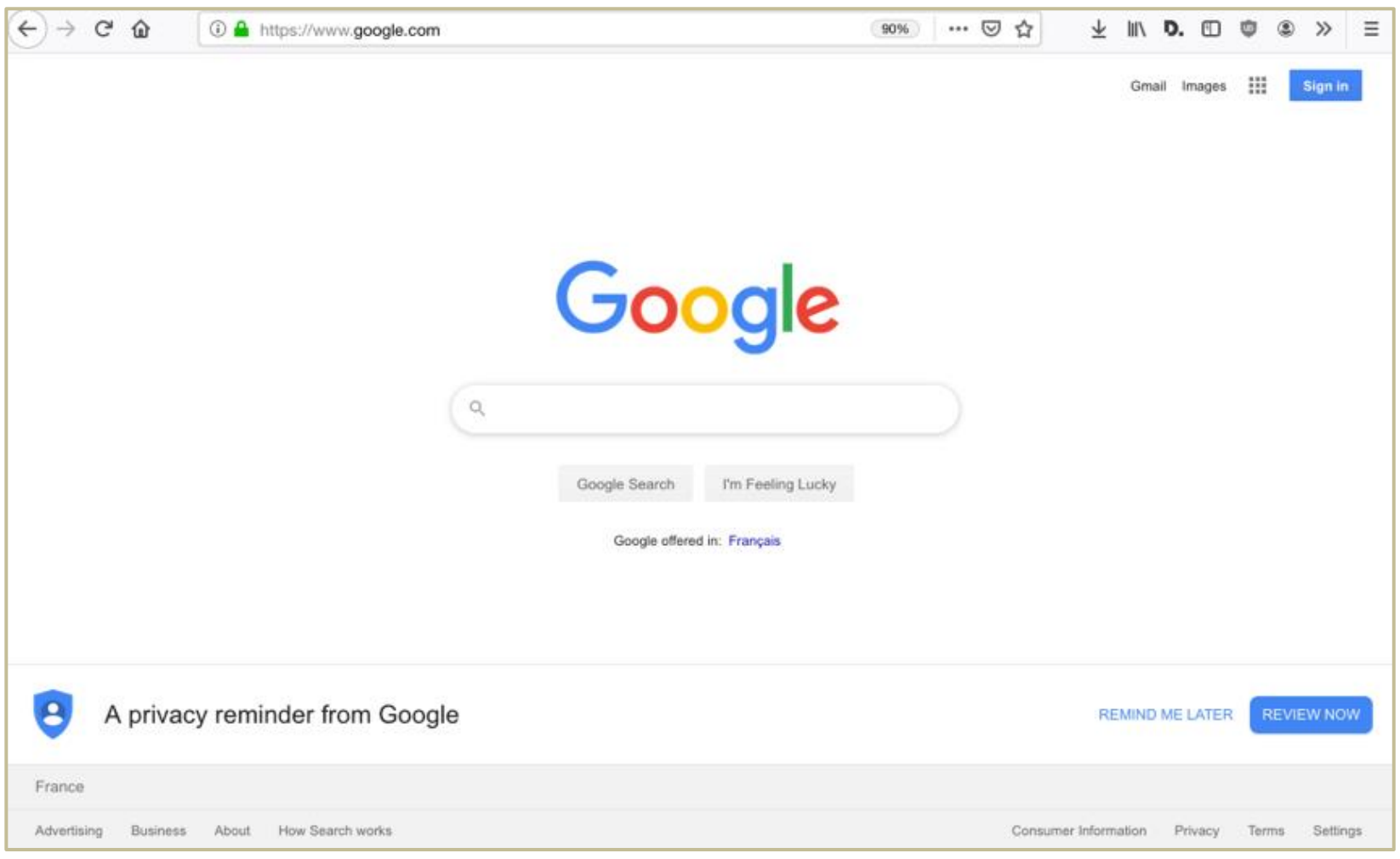

**Figure 3** Access to the google.com website (<https://google.com> accessed 24 September 2019)

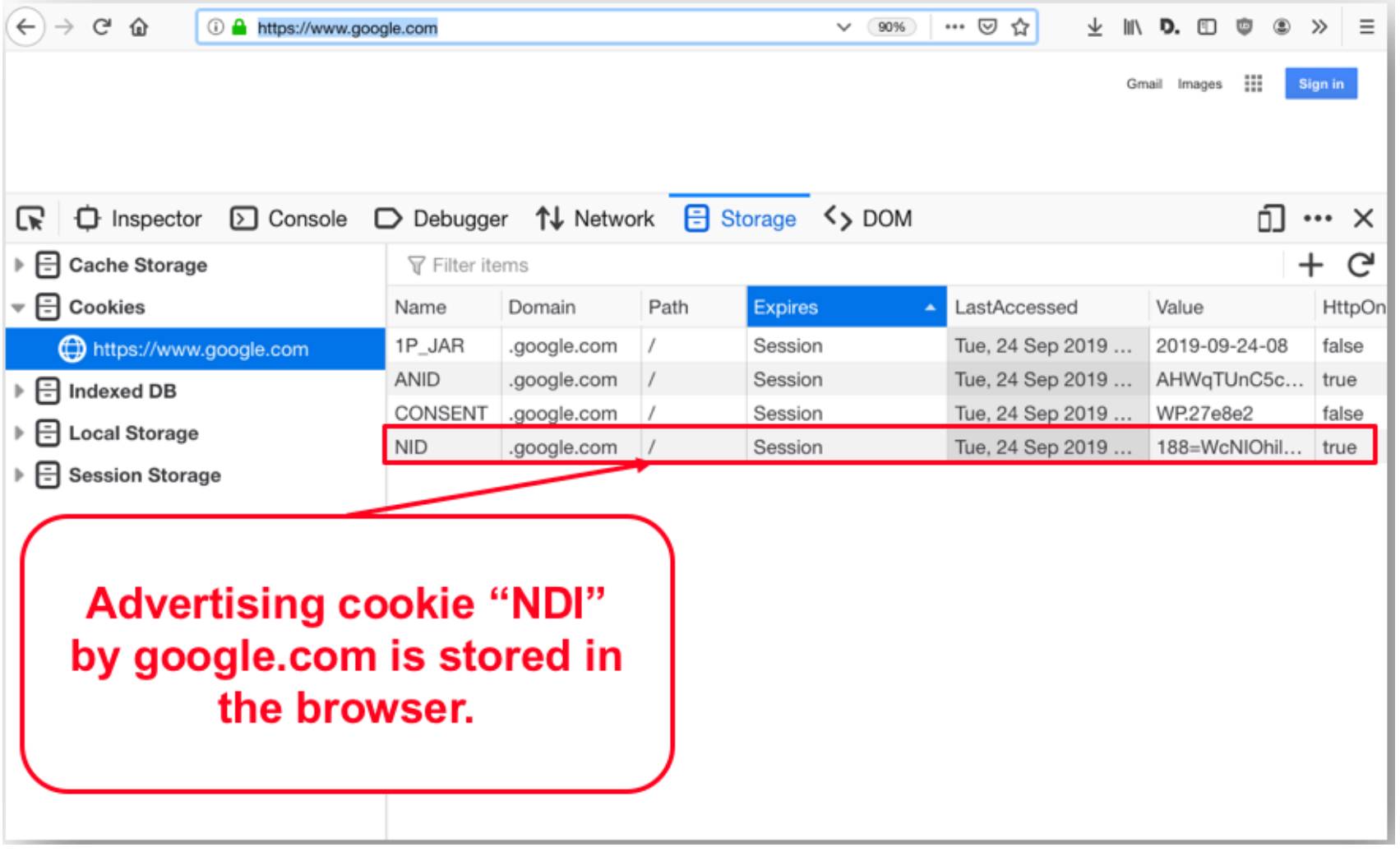

**Figure 4** Access to the google.com website: advertising cookie NID is stored in the browser (<https://google.com> accessed 24 September 2019)

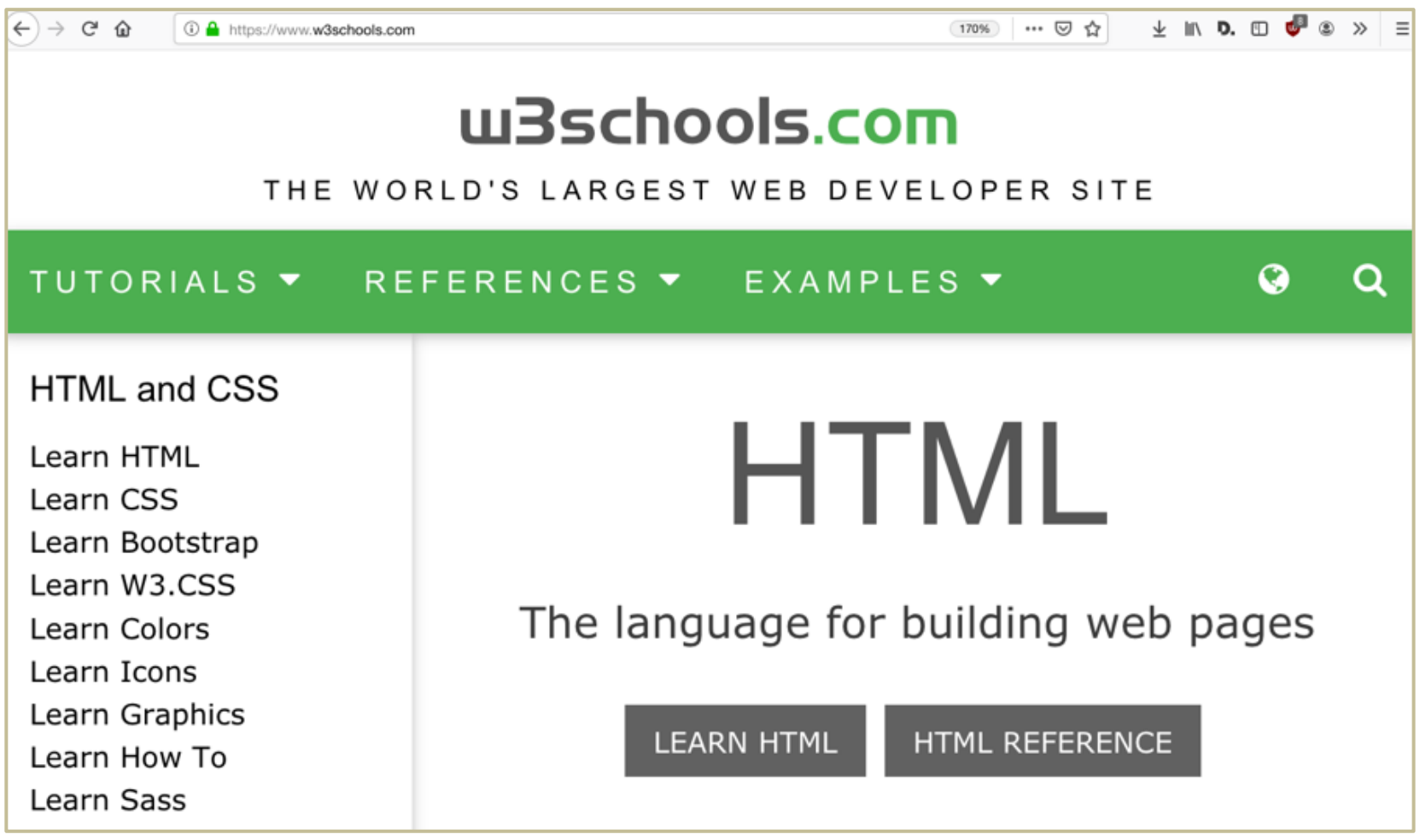

**Figure 5** Access to the W3Schools.com website (<https://www.w3schools.com/ > accessed 24 September 2019)

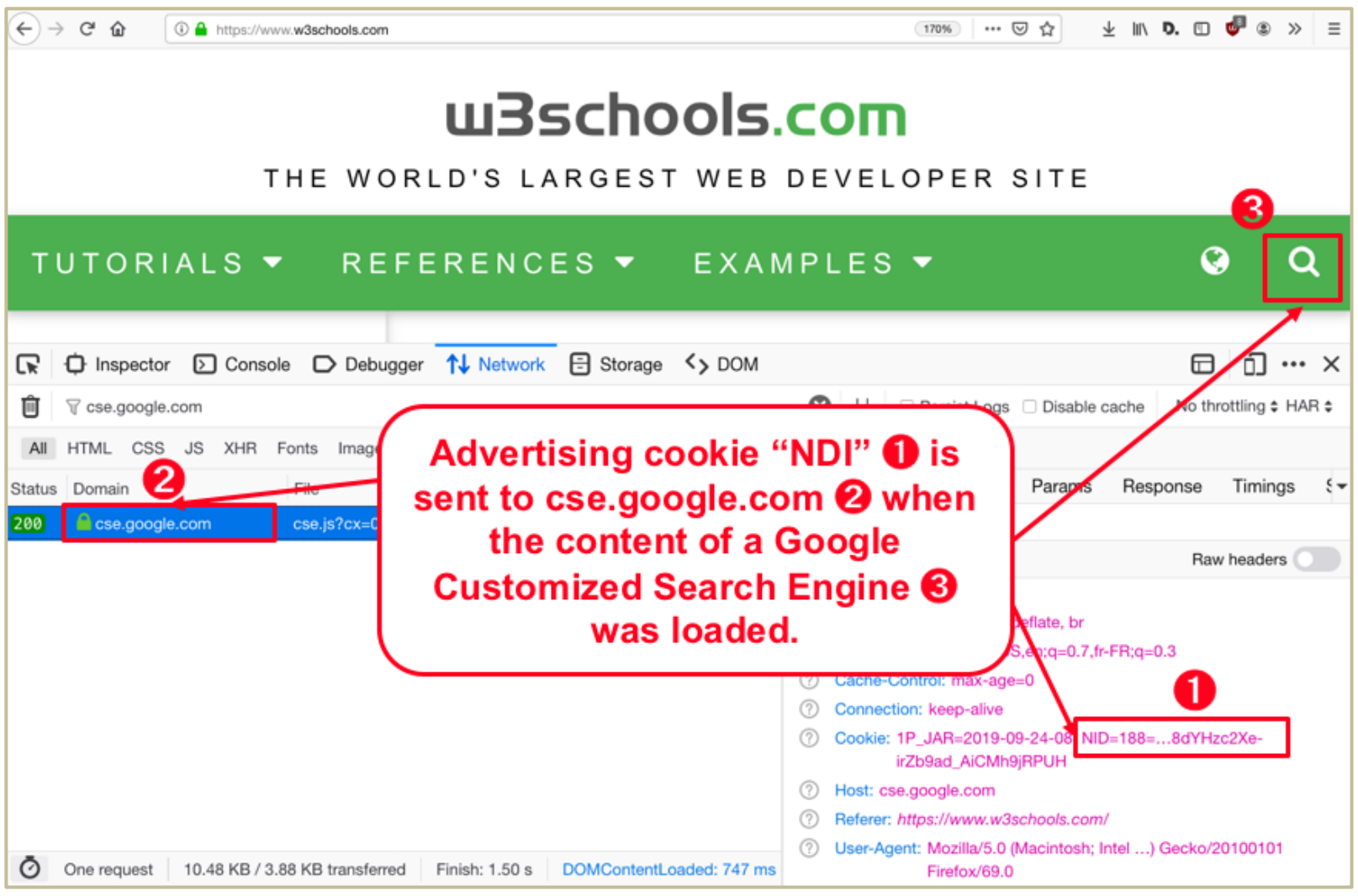

**Figure 6** Violation of the requirement “Prior to sending an identifier” by W3Schools.com website (<https://www.w3schools.com/ >, accessed 24 September 2019)

**How to detect violations?** Detecting violations of this requirement is a very complex task, and no technical solutions exist today that are able to assess this requirement. Apart from the difficulties we have raised in Section R1, the assessment of this requirement requires further technical investigations:

1. **Extensive testing of all browser storages with all possible identifiers set by various domains is needed**. Even for HTTP cookies, one would need to test the website with the corresponding cookies already set in the browser and analyze all the loaded content in order to detect what content is sending such cookies. This procedure might sound easy when cookies are simply attached by the browser when the content is loading. However, computer science researchers[66] have shown that when cookies are sent via JavaScript requests, they are often encrypted or obfuscated. To the best of our knowledge, as of June 2020, no one has measured whether companies encrypt and send identifiers from other storages to third parties.
2. **Language-based security analysis[67] for JavaScript[68], such as taint-tracking[69] and information flow monitors[70] can be used** to monitor when identifiers are read and further sent to other third parties. However, these technologies can be used only if it is known how to detect identifiers and what is their purpose (and hence it is clear whether such identifiers require consent).
3. **Browser fingerprinting also falls into this requirement**: no information is explicitly stored in the user’s browser; however, a unique identifier built from a browser fingerprint can be constructed and sent. It is well known in the computer science research community that detection of fingerprinting is a complex challenge and as of today, there is no technique to detect browser fingerprinting accurately, as summarized in a recent extensive survey by computer scientists.[71] Similar to browser storages, when browser fingerprinting is used, the purpose must be clearly defined and it must be clear when an identifier is created from a browser fingerprint. As a result, it must be clear whether browser fingerprinting is used for a purpose that requires consent, or is exempted of consent (for example, when fingerprinting is used for a security purpose, such as enhanced authentication).

[66] Panagiotis Papadopoulos, Nicolas Kourtellis, and Evangelos Markatos, “Cookie Synchronization: Everything You Always Wanted to Know But Were Afraid to Ask”. In The World Wide Web Conference 2019, pages 1432-1442.
[67] Andrei Sabelfeld, Andrew C. Myers. Language-based information-flow security. IEEE J. Sel. Areas Commun. 21(1): 5-19 (2003)
[68] Daniel Hedin, Luciano Bello, Andrei Sabelfeld. Information-flow security for JavaScript and its APIs. J. Comput. Secur. 24(2): 181-234 (2016)
[69] B. Livshits. Dynamic taint tracking in managed runtimes. TechnicalReport MSR-TR-2012-114, Microsoft, November 2012.
[70] Jonas Magazinius, Daniel Hedin, Andrei Sabelfeld: Architectures for Inlining Security Monitors in Web Applications. ESSoS 2014: 141-160, 2012.
[71] cf. Laperdrix (n 24).

## 5.2 Freely given

Consent must be freely given, as prescribed in the GDPR in Article 4(11) and further specified in Article 7(4). The request for consent should imply a voluntary choice to accept or decline the processing of personal data, taken in the absence of any kind of pressure or compulsion[72] on the user in persuading to give his consent.

The same holds for processing personal data through BTT. The 29WP (WP208)[73] refers to this "freedom of choice" of the users in choosing cookie settings; it asserts that "the user should have an opportunity to freely choose between the option to accept some or all cookies or to decline all or some cookies and to retain the possibility to change the cookie settings in the future." The Finnish DPA[74] adds that consent is not freely given when there is "any undue pressure on or influence on the user's free will to consent".

As a consequence of not having a freely given consent, the request becomes invalid, as cautioned by the WP29 (WP187): "any pressure or inappropriate influence exerted on the person (in different ways) preventing them from exercising their will shall invalidate consent", and "cannot be claimed to be a legitimate ground to justify the processing".

Forced consent is decomposed in the 29WP guidelines considering three elements: imbalance of power[75], unconditional and non-detrimental. In this paper, we analyze both the unconditional and the non-detrimental elements, as shown in Table 9. Imbalance of power is a subjective requirement that can be only evaluated in a case-per-case manner and is dependent on a specific context when consent is given. Hence, we excluded this analysis, as explained in Section 2.3.

**Table 9** Derived low-level requirements and their sources

| Requirements | | Sources at low-level requirement | | |
|---|---|---|---|---|
| **High-Level Requirements** | **Low-level Requirements** | **Binding** | **Non-binding** | **Interpretation: Legal (L) or Computer Science (CS)** |
| **Free** | **R3 No merging into a contract** | 7 (2) (4), Recital 43 | 29WP; DPAs: Danish, French, UK, Irish, Belgium | - |
| | **R4 No tracking wall** | - | Recital 42 GDPR, Recital 25 ePD; EDPB, EDPS; BEUC; EU Parliament; DPAs: Dutch, French, German, Danish, Greek, Irish, Belgian | - |

### R3 Unconditionality related to a contract

Article 7(4) and Recital 43 of the GDPR confer a *presumption* of a not freely given consent in the presence of a contract or service. Article 7(4) reads as: "When assessing whether consent is freely given, utmost account shall be taken of whether, inter alia, the performance of a contract, including the provision of a service, is conditional on consent". Recital 43 recites as "consent is presumed not to be freely given (…) if the performance of a contract, including the provision of a service, is dependent on the consent despite such consent not being necessary for such performance".

[72]The 29WP opinions provide examples of a non-freely given consent can reveal different conducts: compulsion, pressure or inability to exercise free will; being put under pressure, be it social, financial, psychological or other; deception; intimidation; inappropriate influence; coercion; significant negative consequences if he does not consent (e.g. substantial extra costs), 29WP (WP187 (n 54), and WP259 rev.01 (n 4)).
[73] cf. 29WP (WP208) (n 55) 5.
[74]Finish DPA (n 13).
[75] Recital 43 of the GDPR clarifies situations in which consent cannot be seen as freely given "where there is a clear imbalance between the data subject and the controller (…) and it is therefore unlikely that consent was freely given in all the circumstances of that specific situation." The Recital concerns authorities, but also corporations in a dominant market position (e.g. in the area of social networking service of relevance, as in the case of Facebook), and/or in a closed and proprietary network where the data subject is factually forced to join or maintain a profile with the controller, to be able to interact with persons that are not available on other services. A representative related complaint on forced consent was issued by NOYB against Facebook, See NOYB, "Complaint filed against Facebook Ireland Ltd." (2018) <https://noyb.eu/wp-content/uploads/2018/05/complaint-facebook.pdf> accessed 7 May 2020.

The purpose of these provisions is to ensure that services are not offered upon the condition that users give personal information which are not necessary for the offering of these services. Article 7(4) prohibits any form of bundling of a service with a request for consent, when the consent is not necessary for the delivery of that service. For example, if a website makes online transactions (together with marketing purposes) dependent on the user consent for processing personal data that is not necessary for these purposes, it can be reasonably assumed that consent is forced. As a result of the established presumption, any controller has to prove that consent was freely given.

In practice, this requires consent for processing to be clearly distinguishable (untied, unbundled) from contracts or agreements[76] or privacy policies and terms of contract (as posited in Article 7(2) GDPR). The Danish DPA[77] reiterates that "all-purpose acceptance of general terms and conditions cannot be taken as a clear affirmation whereby the user consents to the processing of personal data". Consent would be deprived of any meaning if services are only offered in exchange for mandatory consent to the exploitation of personal data. As the 29WP (WP159 rev.01)[78] reasserts, the "GDPR ensures that the processing of personal data for which consent is sought cannot become directly or indirectly the counter-performance of a contract".

From Article 7(4), we denote the words *conditionality* and *inter alia* (i.e. "among others"). It follows therefrom that the European legislator chose to explicitly list "conditionality" as an instructive example of a non-freely given consent. In addition, the word "*inter alia*" refers to other cases rather than the case of conditionality.

Drawing on this guidance, we deduce the requirement that the consent request should not be merged into a contract or terms of service, as depicted in the requirement box.

| | |
|---|---|
| Requirement | **No merging into a contract**<br>A request for consent cannot be merged into a contract |
| Violation | When both consent and a contract (for which consent is not needed) are merged |

**Example.** Figure 7 represents a case of a bundled consent request where the website offers news service, provided by the Washington Post website, and requests consent of the user.

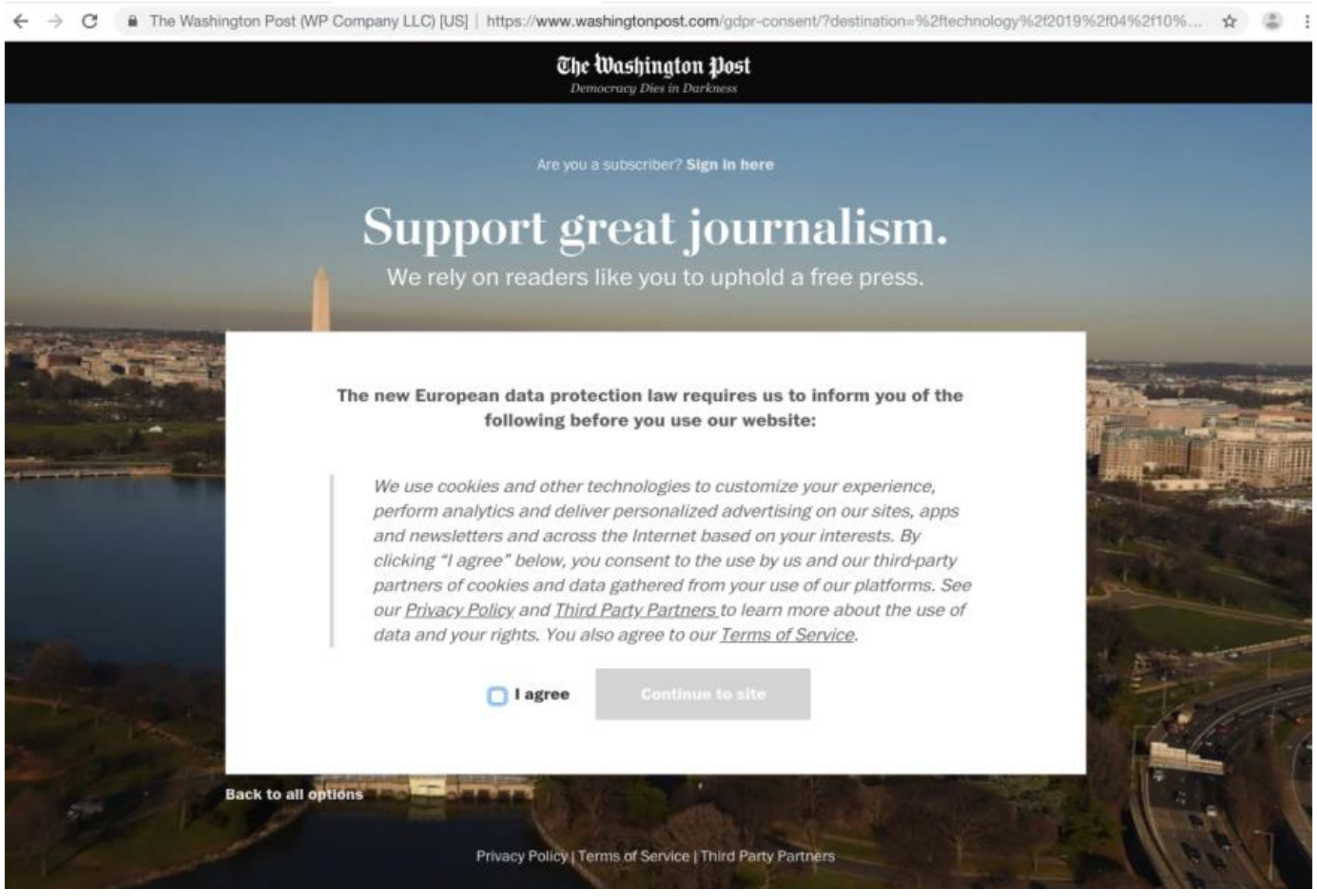

**Figure 7** Violation of the requirement "No merging into a contract" by Washington Post website (homepage <www.washingtonpost.com/gdpr-consent/?noredirect=on&utm_term=.3161abcd072b> accessed 17 May 2019)

[76] An illustrative example is the complaint filed by NOYB against Google that we transcribe for the practical relevance of this requirement "bundling happens when the controller requires the data subject to consent to the privacy policy and to the terms as a whole, which in fact cover all the "services", that the controller offers e.g. YouTube, Chrome Browser, Google Services, Google Maps, Google Search, Google News, Gmail, AdWords, as well as several other services", NOYB, "Complaint filed against Google LL" (2018) <https://noyb.eu/wp-content/uploads/2018/05/complaint-android.pdf> accessed 7 May 2020.
[77] Danish DPA, "Guide on consent" (2019) <www.datatilsynet.dk/media/6562/samtykke.pdf> accessed 7 May 2020 (hereafter named "Danish DPA Guide").
[78] cf. 29WP (WP259 rev.01) (n 4) 8.

**How to detect violations?** Please see section 6, where we describe automatic means that can be used to assess all the language-based requirements, including "No merging into a contract".

### R4 Non detrimental – the case of tracking walls

A freely given consent also implies the consent request to be non-detrimental. "Detrimental consent" refers to the case where the data subject is unable to refuse or withdraw consent without detriment, which means facing significant negative consequences (Recital 42 of the GDPR). For the purposes of this paper, detrimental practices occur is different situations, suchlike:

- When users, even before expressing any choice, face a *tracking wall* blocking access to an online service's content (e.g. stating: "to access our site you must agree to our use cookies");
- When users, after refusing tracking cookies, have denied access to the webpage they want to consult, or the user is redirected to another website, or service is downgraded[79];
- Paid services or extra costs.[80]

The first listed practice refers to the appearance of a barrier page and is known by the designation of *tracking wall* (also known as *cookie wall*, *take-it-or-leave-it-choices* approaches). A tracking wall means that users who do not accept tracking across other sites will be denied access to the websites they seek to access.[81] It occurs when both of the two following conditions hold:

- the consent banner blocks access to the website,
- the banner gives only the option to accept, without any option to refuse.

However, users should have the possibility to refuse cookies and still be able to browse the page.[82] As mentioned in Section 5.2, if certain cookies are not necessary for the services requested and only provide for additional benefits of the website operator, the user should be in a position to refuse them (29WP 208).[83]

Recently, the EDPB[84] (05/2020) established that tracking walls are invalid and complements this claim with an example:

> "In order for consent to be freely given, access to services and functionalities must not be made conditional on the consent of a user to the storing of information, or gaining of access to information already stored, in the terminal equipment of a user (so called cookie walls). Example: A website provider puts into place a script that will block content from being visible except for a request to accept cookies and the information about which cookies are being set and for what purposes data will be processed. There is no possibility to access the content without clicking on the "Accept cookies" button. Since the data subject is not presented with a genuine choice, its consent is not freely given. This does not constitute valid consent, as the provision of the service relies on the data subject clicking the "Accept cookies" button. It is not presented with a genuine choice."

The ePrivacy Directive refers to the "*conditional access to website content*" in Recital 25. It states: "access to specific website content may be made conditional on the well-informed acceptance of a cookie or similar device, if it is used for a legitimate purpose". A literal interpretation of this excerpt apparently legitimizes conditional access to a website and this literal reading is sometimes used to justify the use a cookie wall.[85]-[86] Notably, this interpretation derives from an incorrect analysis of this Recital, for it makes access to a website conditional on the acceptance of cookies[87], and such conditionality renders a non-freely given

[79] cf. 29WP (WP259 rev.01) (n 4) 11.
[80] Regarding extra costs, such an obligation could foster social/economic discrimination (i.e. the rich, who can pay to protect their privacy, and the poor, who cannot) which would run against the universal nature of the fundamental rights to privacy and data protection. Forcing websites to offer a paid subscription service could also interfere with the development of new innovative business models that might be advantageous to consumers.
[81] cf. EDPS Opinion (n 19) 17.
[82] Ronald Leenes, "The Cookiewars: From regulatory failure to user empowerment?" (M. van Lieshout, & J-H. Hoepman (Eds.), *The Privacy & Identity Lab: 4 years later*, 3, The Privacy & Identity Lab, Nijmegen (2015) 31-49.
[83] cf. 29WP (WP208) (n 46) 6.
[84] EDPB, "Guidelines 05/2020 on consent under Regulation 2016/679" (2020) <https://edpb.europa.eu/sites/edpb/files/files/file1/edpb_guidelines_202005_consent_en.pdf> (henceforth called "EDPB 05/2020").
[85] cf. Kosta (n 20) 1.
[86] Frederik Borgesius, Sanne Kruikemeier, Sophie Boerman and Natali Helberge, "Tracking Walls, Take-It-Or-Leave-It Choices, the GDPR, and the EPrivacy Regulation" (2017) *European Data Protection Law Review*, Volume 3, Issue 3, 353-368.
[87] cf. Leenes (n 82).

consent. In this regard, the 29WP (WP126)[88] recommends clarification or review of this Recital. In the 29WP (WP 240) understanding, these *take it or leave it* approaches rarely[89] meet the requirements for freely given consent. It specifically stated that "if the consequences of consenting undermine individuals" freedom of choice, consent would not be free. The Working Party invites the EC to develop a specific prohibition on such "take it or leave it" choices with regard to electronic communications, where such choices would undermine the principle of freely given consent."

The resulting analysis, also consolidated by the positioning of the majority of the stakeholders shown in the next Section 5.2.1, sustains that websites need to give access to content when a user does not consent to BTT beyond strictly necessary to provide the service, and hence, consent requests should not present a tracking wall.

| **Requirement** | **No tracking walls**<br>Blocking access to a website unless the user gives a positive consent is not a valid consent. |
|---|---|
| Violation | Existence of a tracking wall for BTT that require consent. |

**Examples.** Figure 8 shows an example of a cookie wall on the MedicalNewsToday website. When the page first loads, the website prevents a visitor from viewing any other page unless the user clicks the displayed *Accept and continue to site* button. Figure 9 shows the resulting page after the user clicked on the *Deny permission* link: the website only provides access to 10 articles, preselected by the website (and not the article requested by the user). The banner above reminds the user that he has a limited access to the website because he disallowed cookies and proposes to update the privacy settings.

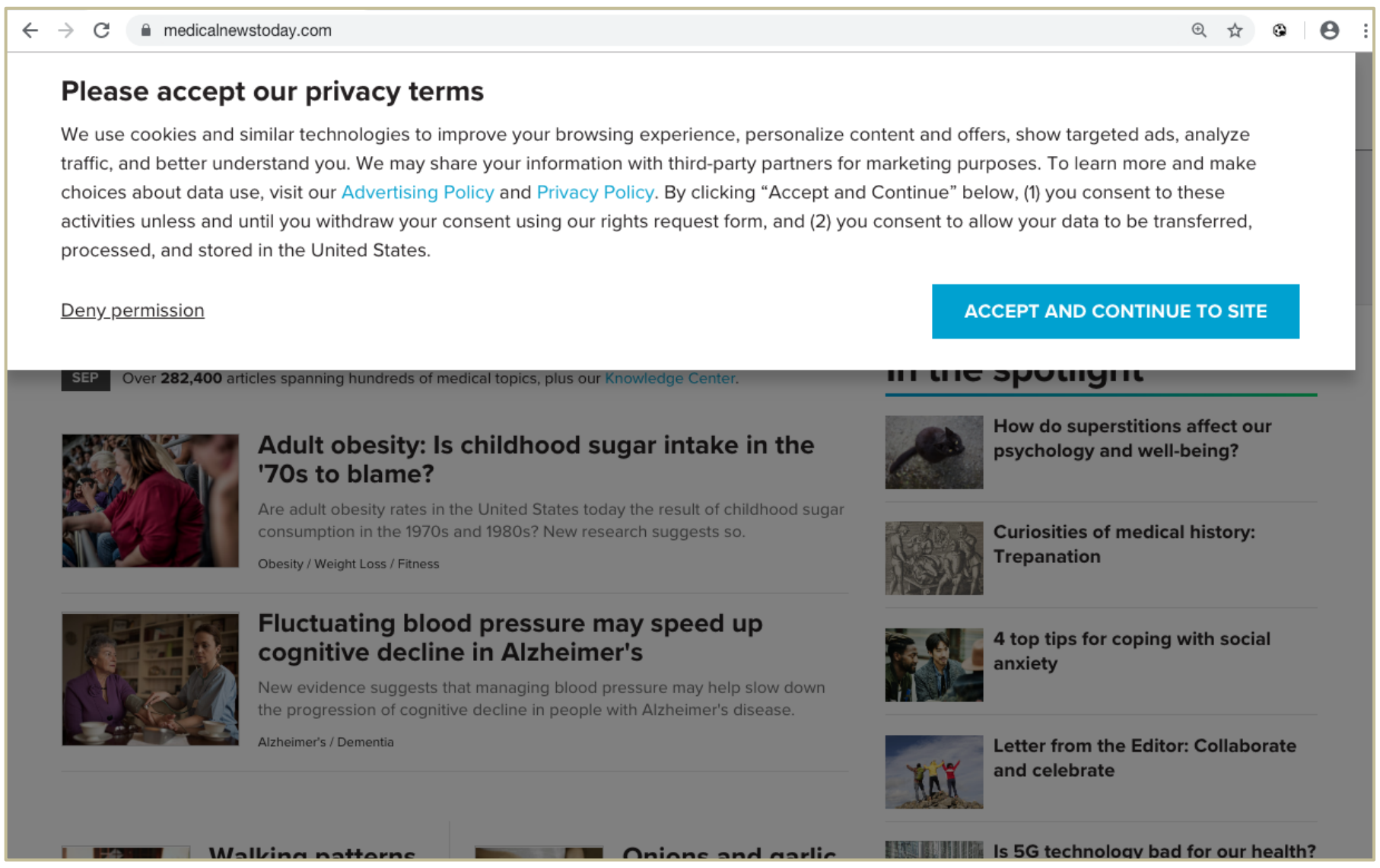

**Figure 8** Violation of the requirement "No tracking wall" by the MedicalNewsToday website (<www.medicalnewstoday.com/> accessed on 25 September 2019)

[88] The 29WP states that "the last paragraph of Recital 25, stipulating that access to specific website content may be made conditional on the acceptance of a cookie, might be contradictory with the position that the users should have the possibility to refuse the storage of a cookie on their personal computers and therefore may need clarification or revision", Article 29 Working Party, "Opinion 8/2006 on the review of the regulatory Framework for Electronic Communications and Services, with focus on the ePrivacy Directive" (WP 126, 26 September 2006) 3.

[89] The 29WP identifies five circumstances in which forced consent should be specifically prohibited, namely: 1. Tracking on websites, apps and or locations that reveal information about special categories of data. 2. Tracking by unidentified third parties for unspecified purposes. 3. All government funded services; 4. All circumstances identified in the GDPR that lead to invalid consent; 5. Bundled consent for processing for multiple purposes. Finally, the 29WP alerts to the position of news media, since they seem to be the heaviest users of tracking cookies and cookie walls, see 29WP (WP240) (n 32) 17.

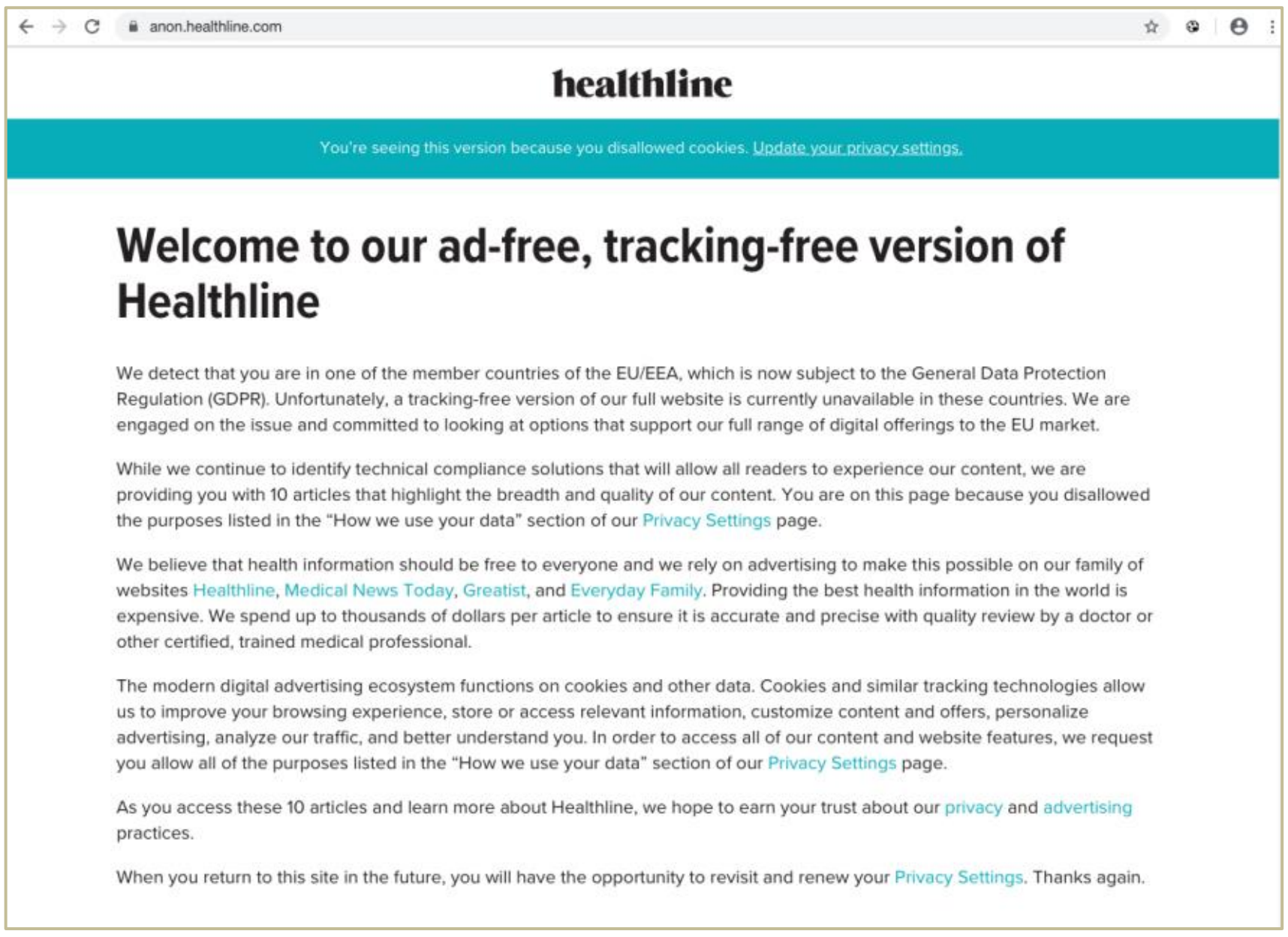

**Figure 9** Result of denying consent on MedicalNewsToday website (<www.medicalnewstoday.com/> accessed 25 September 2019)

**How to detect violations?** Detection of such a violation is possible manually. The user needs to understand whether a website allows the user to access it without expressing consent and whether there is an option to refuse consent. We also consider a violation of this requirement when refusing consent leads to a restrictive access to the service, as in the example of MedicalNewsToday website.

### 5.2.1 Stakeholders positioning on tracking walls

There is some inconsistency in the positions taken by EU DPAs and other stakeholders on whether a tracking/cookie wall consists in a violation of a valid consent. The European Data Protection Supervisor,[90] the European Parliament[91], and the Bureau Européen des Unions de Consommateurs[92] (BEUC) are of the opinion that tracking walls and any other type of detrimental rendering of consent should be forbidden, as the GDPR mandates. Noyb.eu[93] filed four complaints over forced consent against Google, Instagram, WhatsApp and Facebook.

The CNIL in its current guidelines[94] for cookies (article 3) considers that consent can only be valid if the concerned person is able to validly exercise his choice and does not suffer major inconvenience in the absence or withdrawal of consent. In its draft recommendation on the use of cookies[95], this authority proposes that users should not be exposed to any prejudice should they decide not to accept cookies. It states that "the ability to express refusal as easily is indeed the counterpart of the ability to express free consent".

The Greek DPA[96] confirms that the website should not provide for "no option for declining/rejecting cookies and trackers".

[90] cf. EDPS Opinion (n 19) 17.
[91] In the Proposal for the ePrivacy Regulation of the European Parliament, it is proposed that "the Regulation should prevent the use of so- called "cookie walls" and "cookie banners" that do not help users to maintain control over their personal information and privacy or become informed about their rights", Draft European Parliament Legislative Resolution <www.europarl.europa.eu/doceo/document/A-8-2017-0324_EN.html?redirect> accessed 7 May 2020.
[92]BEUC Position Paper, "Proposal For A Regulation On Privacy And Electronic Communications (E-Privacy)" (2017)<www.beuc.eu/publications/beuc-x-2017-059_proposal_for_a_regulation_on_privacy_and_electronic_communications_e-privacy.pdf> accessed 7 May 2020.
[93] NOYB, "GDPR: noyb.eu filed four complaints over "forced consent" against Google, Instagram, WhatsApp and Facebook" (2018) <https://noyb.eu/wp-content/uploads/2018/05/pa_forcedconsent_en.pdf> accessed 7 May 2020.
[94] CNIL, Guidelines on cookies and other tracers (n 36).
[95] Commission Nationale Informatique et Libertés (CNIL). On the practical procedures for collecting the consent provided for in article 82 of the French data protection act, concerning operations of storing or gaining access to information in the terminal equipment of a user (recommendation "cookies and other trackers") https://www.cnil.fr/sites/default/files/atoms/files/draft_recommendation_cookies_and_other_trackers_en.pdf (January 2020), (hereafter named "CNIL draft recommendation 2020").
[96] cf. Greek DPA (n 42).

The Irish DPA[97] confirms that a banner merely giving "the user the option to click "accept" to say yes to cookies and which provides no other option is not compliant. This means banners with buttons that read "ok, got it!" or "I understand", and which do not provide any option to reject cookies or to click for further, more detailed information, are not acceptable and they do not meet the standard of consent required".

The ICO[98] in its recent guidance states that consent which is forced via a cookie wall is "unlikely to be valid". However, it also notes that the GDPR must be balanced against other rights, including freedom of expression and freedom to conduct a business. The ICO seems to adopt a wait and see approach, as it argues that:

> In some circumstances, this approach is inappropriate; for example, where the user or subscriber has no genuine choice but to sign up. (…) If your use of a cookie wall is intended to require, or influence, users to agree to their personal data being used by you or any third parties as a condition of accessing your service, then it is unlikely that user consent is considered valid.

The Dutch DPA[99] published on its website in December 2019 its viewpoint that websites must remain accessible when refusing tracking cookies and that cookie walls are not permitted under the GDPR. It adds that with a cookie wall, websites, apps or other services cannot receive valid permission from their visitors or users. The regulator explains that the inspected websites are involved in an ongoing investigation into cookie walls. Alongside, the Minister for Legal Protection[100] of the Netherlands adverts that when a website is visited, the visitor cannot be denied access to the content of the website if he does not agree with the placement of the cookies (cookie wall). Only functional cookies and non-privacy sensitive cookies do not need permission. It states further that the government is arguing in the European Council for a ban on cookie walls in the new ePrivacy Regulation.

In the same light, the Belgian DPA[101] states that blocking a user's access to a website, on the basis that the user did not consented to cookies, is not a compliant solution.

The German DPA[102] contends that a visit to a website should still be possible if data subjects decide against the setting of cookies. The same reasoning is upheld by the Danish DPA.[103]

Conversely, the Austrian DPA[104] issued a decision on 30 November 2018, pronouncing that consent was freely given via a cookie wall in the case of an Austrian newspaper, "*Der Standard*", that gave users the option to either: i) accept cookies and receive full access to the website; ii) refuse cookies and receive a limited access to the website; or iii) pay a fee for a monthly subscription without accepting cookies. The authority indicated that cookie walls are not prohibited because the newspaper's own settings provide a degree of choice. First, *Der Standard* only places cookies after the user makes an informed decision to allow the placement of cookies. Second, the individual can withhold consent by either entering into a paid subscription or leaving *Der Standard*'s website. Third, the DPA considered *Der Standard*'s prices to be "not unreasonably high." In fact, giving consent to cookies results in a positive outcome for the individual, because they gain unlimited access to the newspaper's articles. The Austrian DPA did not, however, discuss what would happen if an individual withdrew their consent to the usage of cookies.

The Spanish DPA[105] recognizes as a valid practice the blocking access to the website if a user rejects consent, as depicted in the excerpt below (when information duties were duly complied with). We consider

[97] cf. Irish DPA Guidance (n 37).
[98] cf. ICO Guidance (n 26) 31.
[99] cf. Dutch DPA "Cookies" (n 46); and "Many websites incorrectly request permission to place tracking cookies" (2019) <https://autoriteitpersoonsgegevens.nl/nl/nieuws/ap-veel-websites-vragen-op-onjuiste-wijze-toestemming-voor-plaatsen-tracking-cookies> accessed 7 May 2020.
[100] House of Representatives of the Netherlands, "Answer to questions from members Middendorp and Van Gent about a possible cookie wall ban" (2019) <www.tweedekamer.nl/kamerstukken/kamervragen/detail?id=2019D49667&did=2019D49667> accessed 7 May 2020.
[101] Belgian DPA, (2020) "Guidance Materials and FAQs on Cookies and Other Tracking Technologies" <https://www.autoriteprotectiondonnees.be/recueillir-valablement-le-consentement-des-personnes-concernees>, accessed 7 May 2020 (henceafter named "Belgian DPA Guidance").
[102] cf. German DPA Guidelines (n 10).
[103] cf. Danish DPA Guide (n 77).
[104] Austrian DPA decision on the validity of consent (2018) <www.ris.bka.gv.at/Dokumente/Dsk/DSBT_20181130_DSB_D122_931_0003_DSB_2018_00/DSBT_20181130_DSB_D122_931_0003_DSB_2018_00.pdf > accessed 7 May 2020.
[105] Spanish DPA "Guide on the use of cookies" (2019) <www.aepd.es/media/guias/guia-cookies.pdf> accessed 7 May 2020 (author's translation of the Spanish version) (henceforth named "Spanish DPA Guide").

this scenario being equivalent to a tracking wall because the user is not able to access the service unless she gives her consent:

> "In certain cases, not accepting cookies shall entail being entirely or partially prevented from using the service; users must be appropriately informed of this situation. However, access to services may not be denied due to cookie refusal in those cases in which such refusal prevents the user to exercise a legally recognised right, since such website is the only means provided to users to exercise such rights".

Borgesius et al., in their commissioned study[106] on the Proposal for the ePrivacy Regulation, mentioned a circumstance catalogue composed of a non-exhaustive *blacklist* of circumstances in which tracking walls are banned (list of illegal practices), supplemented with a *grey list* (practices presumed to be illegal). The study refers that if a situation is on the grey list, there is a legal presumption that a tracking wall makes consent involuntary, and therefore invalid. Hence, the legal presumption of the grey list shifts the burden of proof. For example, for situations on the grey list, it is up to the company deploying the cookie wall to prove that users gave a free consent, even though the company installed a tracking wall. We are instead of the opinion of a complete ban to cookie walls.

Further developments need to be consolidated through case law from the European Court of Justice. In addition, businesses using tracking/cookie walls to obtain consent may want to consider preemptively streamlining their method for obtaining consent (e.g. by switching to a cookie banner that allows to refuse consent). Table 10 summarizes the different positionings made public from some stakeholders.

**Table 10** Positioning of stakeholders on cookie walls

| Stakeholders | Positioning on tracking wall |
|---|---|
| EDPB, EDPS, BEUC, EU Parliament, DPAs: Dutch, French, German, Danish, Greek, Irish, Belgian | Violation of a freely given consent |
| UK, Spanish DPAs | Not clear |
| Austrian DPA | Valid consent |

## 5.3 Specific

Specific consent involves granularity of the consent request in order to avoid a catch-all purpose acceptance. In Table 11 and the following subsections we further specify this requirement.

**Table 11** Derived low-level requirements and their sources

| Requirements | | Sources at low-level requirement | | |
|---|---|---|---|---|
| **High-Level Requirements** | **Low-level Requirements** | **Binding** | **Non-binding** | **Interpretation: Legal (L) or Computer Science (CS)** |
| **Specific** | **R5 Separate consent per purpose** | 4(11), 6(1) (a); Planet 49 ruling | 29WP, Recitals 32, 43, all DPAS | - |

### R5 Separate consent per purpose

The request for consent should be granular in the options for consenting to cookies, so that the user is able to give consent for an independent and *specific purpose* (29WP WP208).[107] This reasoning is given by the following recitals of the GDPR. Recital 43 clarifies the need for a separate consent for different processing operations. Recital 32 of the GDPR states that consent should be given per purpose (or set of purposes). The provision is worded as follows: "consent should cover all processing activities carried out for the same purpose or purposes. When the processing has multiple purposes, consent should be given for all of them".

[106] Frederik Borgesius, Joris van Hoboken, Ronan P. Fahy, Kristina Irion, Max Rozendaal, "*An Assessment of the Commission's Proposal on Privacy and Electronic Communications*" (Study for the LIBE Committee. Brussels: European Parliament, Directorate-General for Internal Policies, Policy Department C: Citizens' Rights and Constitutional Affairs, Chapter 3.5.5, 2017) <www.europarl.europa.eu/thinktank/en/document.html?reference=IPOL_STU(2017)583152> accessed 7 May 2020.
[107] cf. 29WP (WP208) (n 54) 3.

This element of a specific consent relates to the *purpose limitation principle* observed in Article 5(1)(b) of the GDPR. Therein rely two elements: i) data must be collected for specified, explicit and legitimate purposes only; and ii) data must not be further processed in a way that is incompatible with those purposes. This Article reads: “personal data shall be collected for specified, explicit and legitimate purposes and not further processed in a manner that is incompatible with those purposes […] (“purpose limitation”)”.

In this same line, the 29WP (WP 203)[108] analyzes this principle of “purpose limitation” and explains that any purpose must be *specified*, i.e. be precisely and fully identified. The 29WP (WP259 rev.01)[109] additionally comments on the needed *consent for each purpose* to comply with the conditions of a valid consent:

> “Data subjects should be free to choose which purpose they accept, rather than having to consent to a bundle of processing purposes. (…) If the controller has conflated several purposes for processing and has not attempted to seek separate consent for each purpose, there is a lack of freedom. This granularity is closely related to the need of consent to be specific. (...) When data processing is done in pursuit of several purposes, the solution to comply with the conditions for valid consent lies in granularity, i.e. the separation of these purposes and obtaining consent for each purpose”.

The 29WP (WP259 rev.01) instructs further that “a controller that seeks consent for various different purposes should provide a separate opt-in for each purpose, to allow users to give specific consent for specific purposes”.

Planet49 Judgment of the Court of Justice of the EU[110] determined that *specific* consent means that “it must relate specifically to the processing of the data in question and cannot be inferred from an indication of the data subject’s wishes for other purposes”. This means that consent should be granular for each purpose of processing.

The resulting analysis sustains that the banner should present each purpose separately (but also, it should allow accepting or rejecting each purpose separately), as depicted in the requirement box.

| **Requirement** | **Separate consent per purpose**<br>Consent should be separately requested for each purpose. |
|---|---|
| Violation | General consent request under conflated or bundled purposes; a user shall not agree or disagree to all at once (Danish DPA[111]) |

**Examples.** Figure 10 shows the wordreference.com website, where a user cannot give consent per purpose, but instead is presented with a “Learn More & Set Preferences” link that only allows to give consent per third party. Figure 11 shows (a part of) the list of vendors, which is several-screen-long and is obviously overwhelming and not usable for an average user. Figure 12 outlines the Dailymail website banner, which conflates together different data processing purposes (e.g. personalization, ad selection, content selection and measurement) under a single acceptance request, therefore violating the requirement that consent should be given per purpose. On the other hand, Figure 13 depicts a compliant design banner from the senscritique.com website.

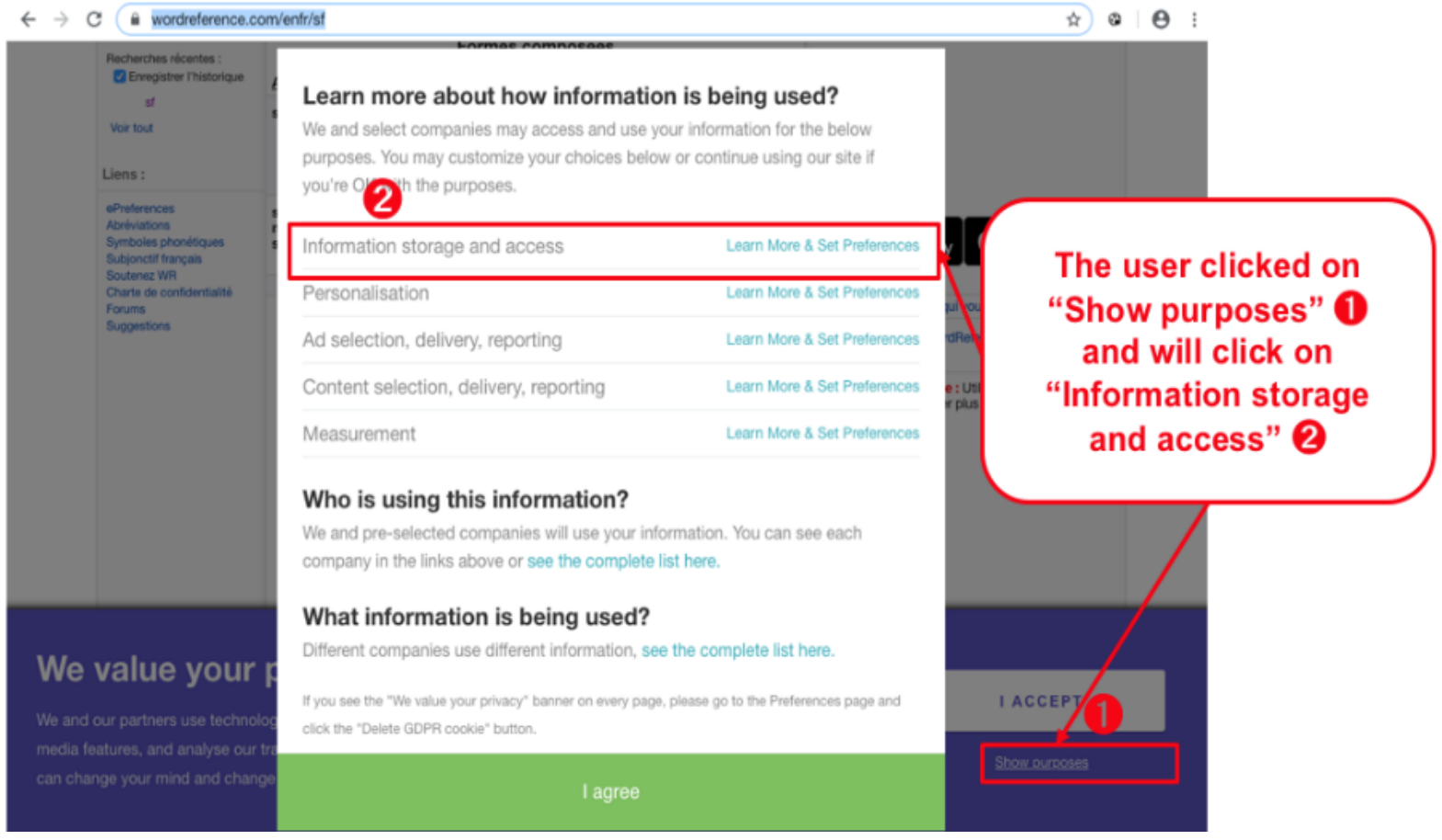

[108] Article 29 Working Party, “Opinion 03/2013 on purpose limitation” (WP 203, 2 April 2013).
[109] cf. 29WP (WP259 rev.01) (n 4) 11.
[110] Planet49 Case (n 11).
[111] cf. Danish DPA Guide (n 77).

**Figure 10** A settings accessible from the cookie banner on website of wordreference.com (<www.wordreference.com/enfr/sf> accessed on 24 September 2019)

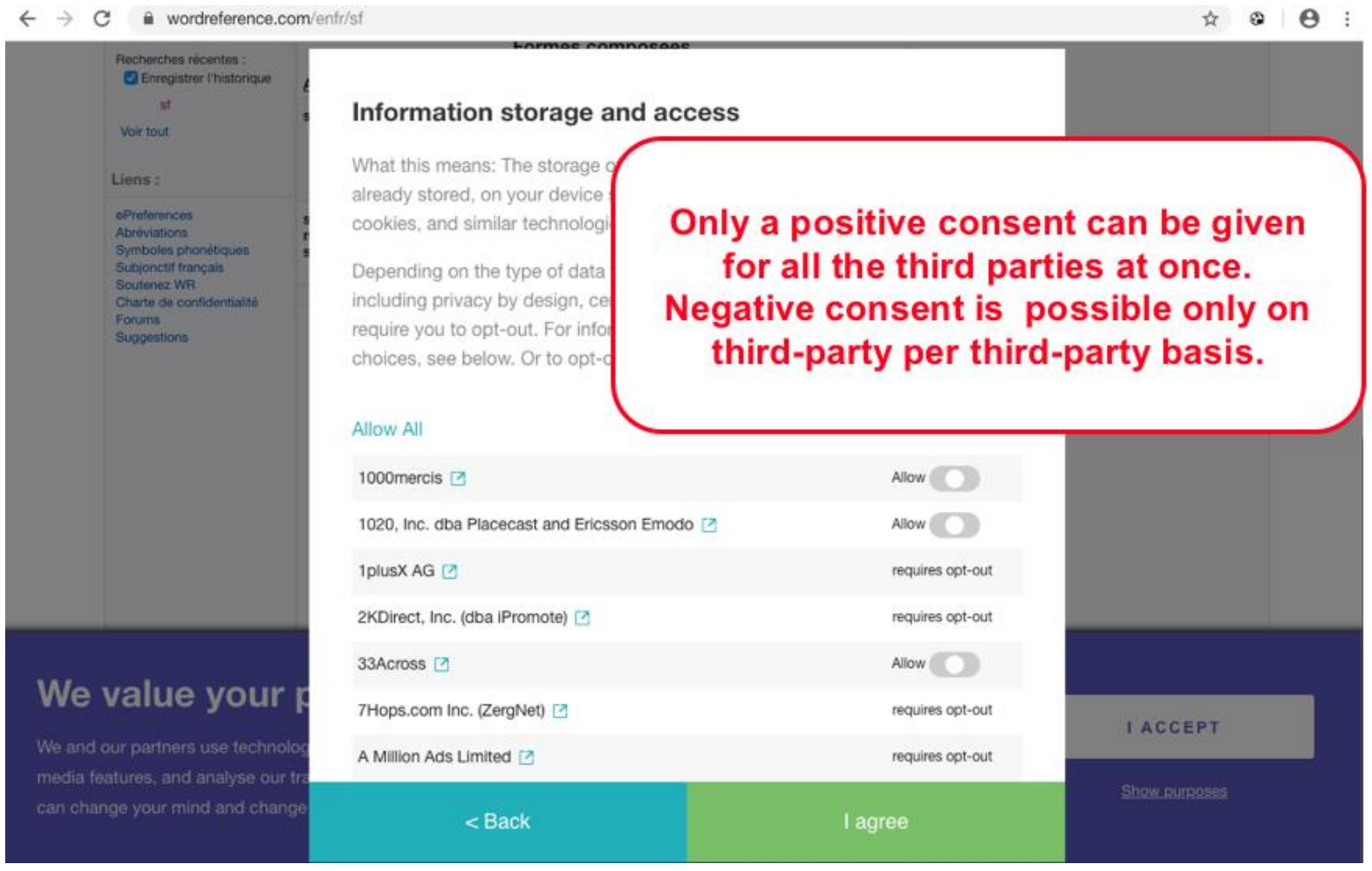

**Figure 11** The cookie banner of wordreference.com does not allow to refuse consent for all third parties at once, only on a "per third party" basis. (<www.wordreference.com/enfr/sf> accessed 24 September 2019)

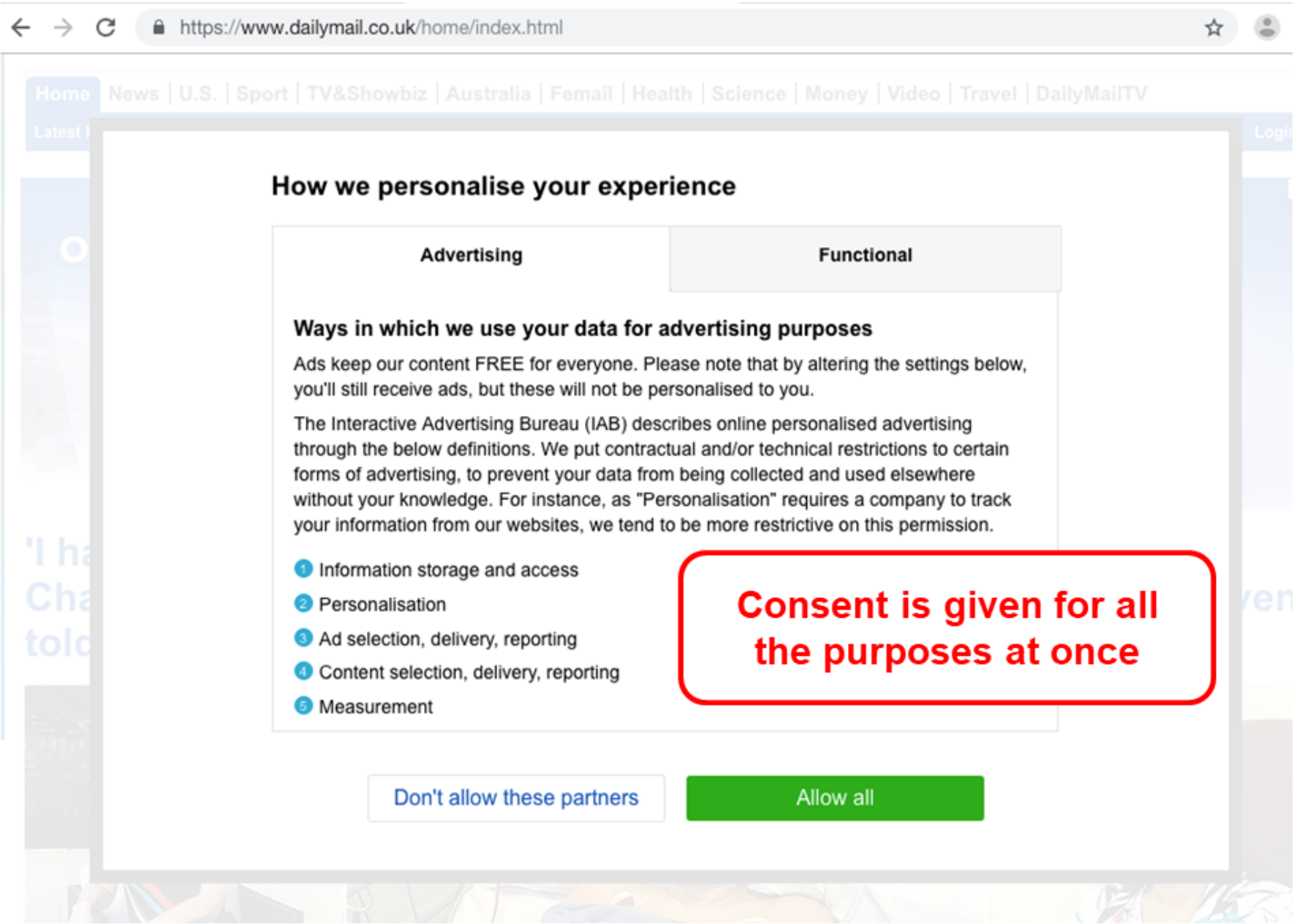

**Figure 12** Non-compliance with the "separate consent per purpose" requirement (<www.dailymail.co.uk/home/index.html> accessed on 17 May 2019)

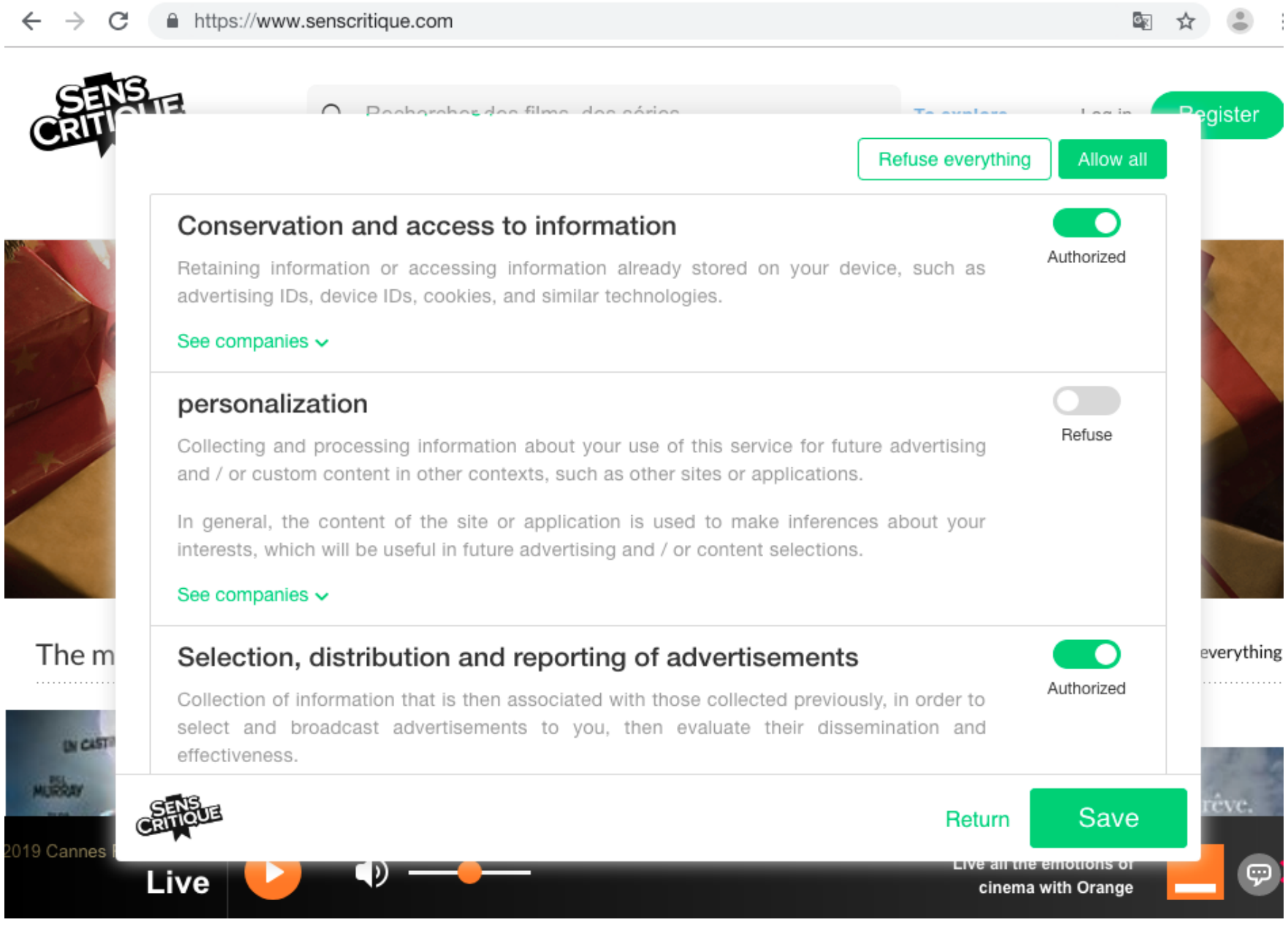

**Figure 13** Compliance with the "separate consent per purpose" requirement (<www.senscritique.com/> accessed on 18 May 2019)

**How to detect violations?** A human operator can observe violations with no technical support. However, it would be possible to detect such violations automatically if the user interface of consent banner had a standardized design, which is not the case nowadays.

### 5.3.1 Consent not required per cookie, per publisher, per third party

Under the following three subheadings we add the observation that a request for consent per purpose *does not* include a request: *per cookie*, *per publisher*, nor *per third-party*, for the reasons explained below.

- ***Not per cookie***. We argue that the requirement of granular purposes does not mandate that the consent request should be provided for each cookie. We claim that the consent request for each cookie is not user-friendly and it might be too overwhelming for users. Moreover, few users are familiar with the concept of cookies and tracking technologies. Therefore, this may lead to certain choices as a consequence of users' lack of knowledge. We derive this conclusion from several bases. The text of the Recital 25 of the ePD states that the cookie consent request covers its further uses, insofar as these uses are compatible with the initial purposes for which the consent is provided. The 29WP (WP208)[112] mentions that each website could prominently display a link to a location where all cookies used by the website are presented through types (and hence, not per cookie). In the same line, the ICO[113] gives the same reasoning when referring to cookie categories:

  > Some sites might use tens or even hundreds of cookies and therefore it may also be helpful to provide a broader explanation of the way cookies operate and the categories of cookies in use. For example, a description of the types of things you use analytics cookies for on the site will be more likely to satisfy the requirements than simply listing all the cookies you use with basic references to their function.

  The Belgian[114], Irish[115] and Danish DPAs[116] accord that consent does not need to be given per cookie, but instead per purpose. The latter refers an example of a specific consent per purpose (and not per cookie):

  > [A] website has a cookie pop-up in which the user can accept or decline cookies by purpose, i.e. the user can freely decide whether he or she wants functional, statistical and/or marketing

[112] cf. 29WP (WP208) (n 54) 3 and 5.
[113] cf. ICO Guidance (n 26) 10.
[114] cf. Belgian DPA Guidance (n 101).
[115] cf. Irish DPA Guidance (n 37).
[116] cf. Danish DPA (n 77).

cookies to be set by the website. The user can easily toggle cookies by purpose on and off. Then the website's cookie consent is specific.

- ***Not per publisher*.** The need of a separate and renewed consent per publisher is also discussable: if one publisher receives consent, it is questionable that it might share the consent with other publishers. In this regard, we refer to the case law of the European Court of Justice and adapt its reasoning to our consent-cookie request context. The Court (in its two decisions of Tele 2 and Deutsche Telekom[117] in the context of electronic public directories), refers to the extension of the initial consent to the subsequent processing of the data by third-party companies, provided that such processing pursues that same purpose, and that the user was informed thereof. The Court holds that where a user consented to the passing of his personal data to a given company, the passing of the same data to another company, with the same purpose and without renewed consent from that user, does not violate the right to protection of personal data. The Court adds that a user will generally not have a selective opinion to object to the sharing of the same data through another, yet similar, provider. From these arguments, we conclude that there is no need for a separate and renewed consent per publisher whenever further processing follows that same purpose, and the user was informed thereof. In these cases, consent could be shared with other publishers.
- ***Not per third party*.** We believe a fine-grained customization per third party is not required. In fact, showing the full advertisers list configures a deceptive design. The 29WP (WP259 rev.01)[118] suggests that the categories of third parties who receive personal data and wish to rely upon the original consent should be listed by category (or be individually named).

  It is possible to conclude that a consent request does not require the user's consent for third-party cookies, but only aims to *inform* users of third-party cookie usage, or third party access to data collected by the cookies on the website: "necessary information would be the purpose(s) of the cookies and, if relevant, an indication of possible cookies from third parties or third party access to data collected by the cookies on the website" 29WP (WP208).[119]

  The Italian DPA[120] adopted the same reasoning and postulated that "publishers may not be required to include, on the home page of their websites, also the notices relating to the cookies installed by third parties via the publishers' websites".

### 5.4 Informed Consent

Whenever Browser-based Tracking Technology (BTT) are accessed or stored on a user's device, the user must be given clear and comprehensive information, and the content information must comprise what is accessed or stored, the purposes and means for expressing their consent, pursuant to Article 5(3) of the ePD.

The need to present information on the processing operations is triggered by the principles of lawfulness, fairness, and transparency depicted in Article 5(1)(a) and the recitals of the GDPR. In particular, Recital 60 explains that "a data controller should provide a data subject with all information necessary to ensure fair and transparent processing, taking into account the specific circumstances (…)".

The rationale behind the requirement to provide information relies in the premise that providing it puts the user in *control* of the data on their own device. As argued by the General Advocate Szpunar,[121] the data subject must be informed of all circumstances surrounding the data processing and its consequences: "crucially, he or she must be informed of the consequences of refusing consent", including a reduced service. He proceeds by asserting that "a customer does not choose in an informed manner if he or she is not aware of the consequences". The 29W (WP131)[122] envisioned that the data subject's consent is "based

[117] C-543/09 *Deutsche Telekom AG v Bundesrepublik Deutschland* [2011] EU:C:2011:279, para 62 to 65; and C-536/15 *Tele2 (Netherlands) BV and Others v Autoriteit Consument en Markt (ACM)* [2017] ECLI:EU:C:2017:214.

[118] cf. 29WP (WP259 rev.01) (n 4) 14.

[119] cf. 29WP (WP208) (n 54) 3 and 5.

[120] Italian DPA, "Simplified Arrangements to Provide Information and Obtain Consent Regarding Cookies" (2014) <www.garanteprivacy.it/web/guest/home/docweb/-/docweb-display/docweb/3167654> accessed 7 May 2020.

[121] Opinion of Advocate General Szpunar. Opinion of the Case C-61/19 Orange România SA v Autoritatea Naţională de Supraveghere a Prelucrării Datelor cu Caracter Personal (ANSPDCP), 2020. ECLI:EU:C:2020:158.

[122] 29WP Working Document on the processing of personal data relating to health in electronic health records (EHR) (WP 131, 15 February 2007) 8.

upon an appreciation and understanding of the facts and implications of an action". The judgment of the Court of Justice of the EU on the Planet49 case[123] elucidated that providing "clear and comprehensive" information means "that a user is in a position to be able to determine easily the consequences of any consent he might give and ensure that the consent given is well informed". It follows therefrom that information must be also "clearly comprehensible and sufficiently detailed so as to enable the user to comprehend the functioning of the cookies employed".

Regarding the *timing* to deliver information, it should be concomitant to the time and place when consent is requested. As posited by the 29WP (WP208), information should be provided "at the time and place where consent is sought, for example, on the webpage where a user begins a browsing session (the entry page). As such, when accessing the website, users must be able to access all necessary information".

From the analysis of the legal provisions, the 29WP guidance and the mentioned case-law, we derive both the approach and the content of the information:

- the *approach* to disclose information, which happens generally with the presence of a privacy or cookie policy (section R6);
- the *content* of the information to be given on BTT. We focus on the information requirements where personal data is collected from the data subject (Article 13 (1) (2) GDPR). The position of the 29WP (260 rev.0.1) is that there is no difference between the status of the information to be provided under sub-article 1 and 2 of Article 13 and therefore all of the information across these sub-articles is of equal importance and must be provided to the data subject. For readability purpose, we decompose the information requirements:

  - necessary information on BTT (section R7),
  - information on consent banner configuration (section R8),
  - information on the data controller (section R9),
  - information on rights (section R10).

Table 12 depicts the information low-level requirements.

**Table 12** Derived low-level requirements and their sources

| Requirements | | Sources at low-level requirement | | |
|---|---|---|---|---|
| **High-Level Requirements** | **Low-level Requirements** | **Binding** | **Non-binding** | **Interpretation: Legal (L) or Computer Science (CS)** |
| **Informed** | **R6 Accessibility of the information page** | - | 29WP, DPAs: Irish, German, Belgium, Finnish | - |
| | **R7 Necessary information of BTT:** | | DPAs: German | |
| | **R7a Identifier name** | - | 29WP; DPAs: Irish Danish | - |
| | **R7b Purposes** | Art.13 (1)(c) | (most DPAs) | - |
| | **R7c Third parties with whom cookies are shared** | Art. 13 (1)(e), Planet 49 | (most DPAs) | - |
| | **R7d Duration of cookies** | Art. 13 (2)(a); Planet 49 | DPAs: Greek, CNIL, Finnish | - |
| | **R8 Information on consent banner configuration** | | 29WP, DPAs: UK, Danish, Irish | - |
| | **R9 Information on the data controller** | Art. 13 (1)(a)(b) | DPAs: Danish, Irish | - |
| | **R10 Information on rights** | Art. 13 (1)(f), (2) (a-f) | almost all DPAs | - |

[123] cf. Planet49 Case (n 11) para 74.

### R6 Accessibility of information page

On the recommended approach, the 29WP (WP208) proposes a visible notice displaying a link to an information page (also known as cookie policy, or entry page) where information on BTT is presented (preferably through a layered approach). The 29WP considers the following built-in possibilities for rendering information:

- the mechanism should provide a visible notice on the use of BTT;
- a link to a designated location where all types of BTT used by the website are presented should be prominently displayed;
- Information should be provided in a layered approach[124], typically through a link (or series of links), where the user can find out more about types of BTT being used.

As for the DPAs' understanding, the Finnish DPA[125] states that no separate pop-up window is required for informing the user, and a cookie policy must also be mentioned on the website so that the user can learn more about it. The Irish DPA[126] advises website publishers to include a link or a means of accessing further information about the use of cookies. We have defined the requirement prescribing that the information page (cookie policy) on the use of BTT should be accessible through a banner, with a clickable link.

| | |
|---|---|
| **Requirement** | **Accessibility of information page**<br>The information page should be accessible through a cookie banner, via a visible link or a button |
| Violation | Inexistence of an information page |

**Examples.** Figure 14 shows a compliant cookie banner example, where the "information page" is accessible through a link.

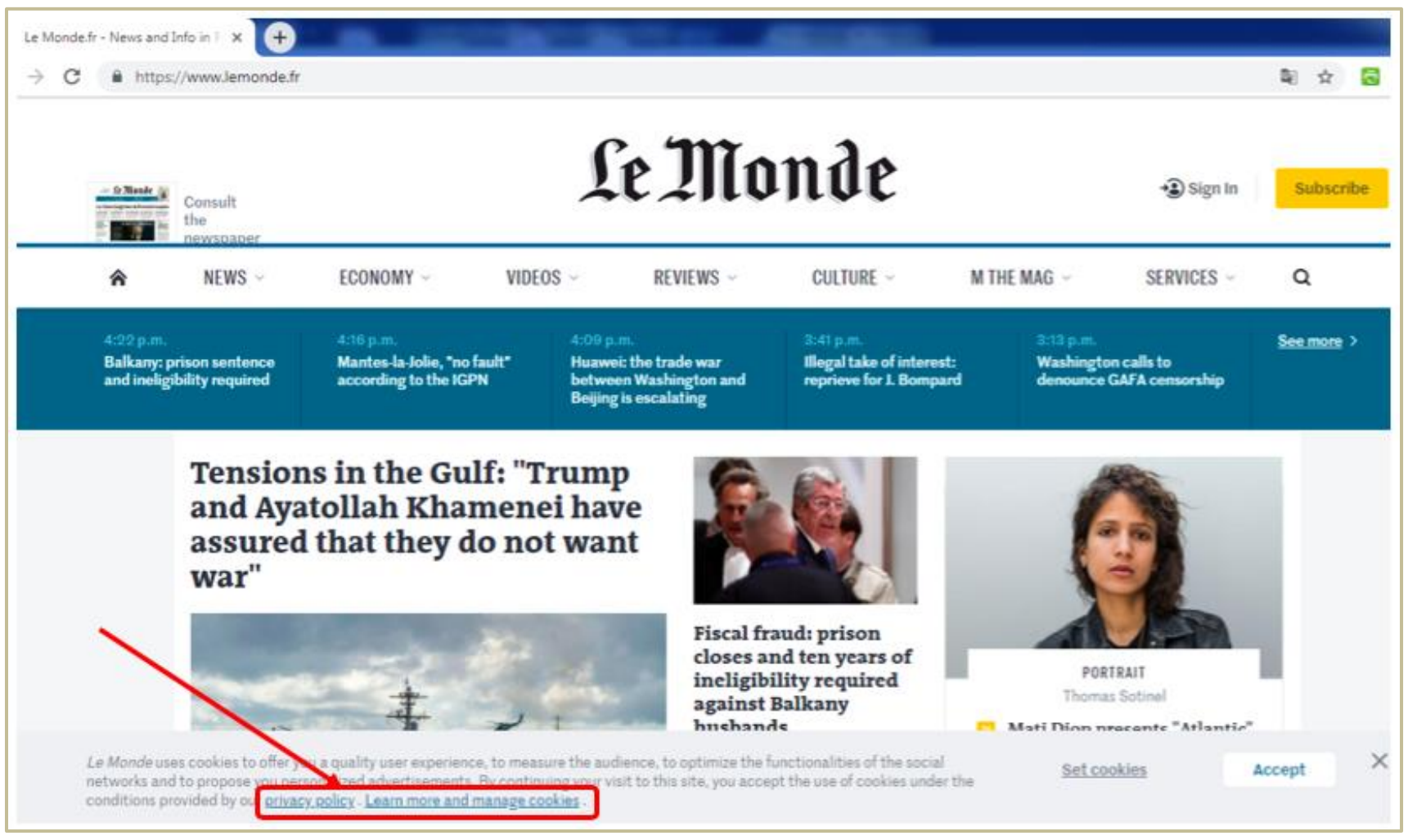

**Figure 14** Compliant example. The "information page" should be accessible through a cookie banner (via a link or a button) (<https://www.lemonde.fr/ > accessed on 18 May 2019)

**How to detect violations?** A manual analysis of a cookie banner content is enough to identify whether a link to the privacy policy is accessible, however such analysis is not scalable when numerous websites must be audited. Moreover, different users may find it more or less difficult to find the same link on a concrete cookie banner. Hence, user studies are needed to evaluate how accessible information for a given target audience is. We provide more details on user studies in Section 6.

From a technical perspective, it is possible to crawl all the links present on a banner interface and detect whether such links lead to a privacy policy with a rather simple detection based on keywords. While this

[124] Interestingly, the EDPS recommends a layered approach, where the information is given at different stages, providing greater detail. The essential information should be present at a sufficient level of detail to already put the user in control at the first layer. A notice providing (the reference to) the first level of information on cookies must be clearly visible to web service users whatever their landing page is. Further, the EDPS strongly recommends that EU institutions provide information on cookies on the web service under their control and not rely on external sources. If, for some reasons, the institution uses external sources, they should set up measures to manage relevant risks, where possible, cf. EDPS Guidelines (n 35) 15.

[125] cf. Finnish DPA (n 13).

[126] cf. Irish DPA Guidance (n 37).

method could produce inaccurate results, it could be usable in a legal procedure after a manual verification. The same analysis holds for all the following requirements in this section.
Computer science researchers have already used keyword-based approaches: Degeling et al.[127] analyzed the availability of privacy policies on the top 500 websites before and after GDPR came in force, while Libert[128] detected and analyzed over 200,000 websites' privacy policies. More sophisticated approaches based on Natural language processing (NLP) can also be applied to analyze whether a page is a privacy policy. We refer to further discussion on NLP in Section 6.

### R7 Necessary information on BTT

Regarding the content of the information to be given on BTT, both the 29WP (WP208) and the recent Planet 49 judgment[129] set the *necessary*[130] information to be disclosed to ensure fair and transparent processing. Planet 49 (in paragraph 76) explicitly quotes Article 13 of the GDPR on the list of information to be provided. The necessary information — specific to BTT — is the following:

- *Purposes* of processing (Article 13 (1)(c)), and their legal basis for each specific processing (under Article 6 GDPR);[131]
- *Recipients,* or categories of recipients, with whom personal data is shared (Articles 4(9) and 13(1)(e)), which can consist of other data controllers, joint controllers, processors and third-party recipients. Planet 49 ruling determines (in paragraph 80) "third parties with whom the cookies are shared with"). The 29WP (WP202)[132] adds that information is needed on whether the data may be reused by other parties, and if so, for what purposes. Regarding the categories of recipients, the 29WP (260 rev.0.1) suggests that if controllers opt to provide the categories of recipients, the information should be as specific as possible by indicating the type of recipient (i.e. by reference to the activities it carries out), the industry, sector and sub-sector and the location of the recipients;
- *Storage period,* interpreted from Article 13 (2)(a). It is explicitly stated in the Planet 49 ruling (paragraph 79) that the "period for which the personal data will be stored, or if that is not possible, to the criteria used to determine that period". The 29WP (260 rev.0.1) declares that this information is linked to the data minimization principle in Article 5(1)(c) and storage limitation requirement in Article 5(1)(e). It adds that the different storage periods should be stipulated for different categories of personal data and/or different processing purposes, including where appropriate, archiving periods;
- *Identifier name* and a responsible party for setting it, as proposed by the 29WP (WP208).[133]

Some DPAs have defined the duration/lifespan for BTT, such as the ones depicted in Table 13:

**Table 13** Positioning of stakeholders on the lifespan of BTT

| DPAs | Lifespan of BTT |
|---|---|
| 29WP (WP194) | "Cookies exempted of consent should have a lifespan that is in direct relation to the purpose it is used for and must be set to expire once it is not needed, taking into account the reasonable expectations of the average user or subscriber". |
| CNIL | "analytic tracers must not have a lifespan exceeding thirteen months and this duration should not be automatically extended during new visits. The information collected through the tracers must be kept for a maximum of twenty-five months". |

[127] M. Degeling, C. Utz, C. Lentzsch, H. Hosseini, F. Schaub, and T. Holz, "We value your privacy ... now take some cookies: Measuring the GDPR's impact on web privacy," in NDSS, 2019.
[128] Timothy Libert, "An Automated Approach to Auditing Disclosure of Third-Party Data Collection in Website Privacy Policies" in Proceedings of the 2018 World Wide Web Conference, p 207-216.
[129] cf. Planet49 Judgment (n 11).
[130] Article 13 sets a distinction between necessary/essential information (1) and possible "further information" (2) which should be provided only to the extent that is necessary to guarantee fair processing having regard to the specific circumstances in which the data are collected. In case of BTT, further informational elements are required as necessary information (cookie name and their duration, third party sharing and their purposes) to ensure transparent processing. Such distinction between necessary and further information is analyzed further in the 29WP Opinion 10/2004 (WP 100) on More Harmonised Information Provisions.
[131] Article 13 (1) (d) posits that when the controller relies on legitimate interests as a legal basis for processing, it should inform the data subject about the interests. The 29WP (29WP 260 rev.01 n 4) adds that at least upon request, provide data subjects with information on the balancing test.
[132] 29WP Opinion 02/2013 (WP202) on apps on smart devices, adopted on 27 February 2013.
[133] cf. 29WP (WP208) (n 54).

| | |
|---|---|
| ICO | Lifespan of cookies must be proportionate in relation to the purpose and limited to what is necessary to achieve it. |
| Belgium | Cookies should be set to expire as soon as they are no longer needed, taking into account the reasonable expectations of the user. Cookies exempt from consent will therefore probably be set to expire when the browsing session ends, or even before. |
| Spain | Lifespan of cookies must be proportionate in relation to the purposes for which they are intended. |
| Irish | Lifespan of a cookie must be proportionate to its function. |

The Greek DPA[134] reiterated that information should include the used tracking categories. The banner (either in the form of a pop-up window or otherwise) should provide specific information for each tracker or tracker category of the same purpose.

The CNIL adds that the categories of data collected through trackers could be specified for each purpose in a way that is easily accessible to the user. This becomes particularly important when the special categories of personal data are used.

Some DPAs advocate that *all* cookies should – as a best practice – declare their purpose. The UK, Greek, Finnish and Belgian DPAs endorse as a good practice disclosure of clear information about the purposes of cookies, including strictly necessary ones. The guidance of the 29WP (WP188)[135] notes that although some cookies may be exempted from consent, they are part of a data processing operation, therefore publishers still have to comply with the obligation to inform users about the usage of cookies prior to their setting.

After listing the necessary information for an informed consent based on the legal sources, we define the low-requirement below:

| **Requirement** | **Information on BTT should contain:**<br>• Purposes<br>• Recipients or categories of recipients with whom personal data is shared with and for what purposes<br>• Storage period<br>• Identifier name |
|---|---|
| Violation | Absence of any of these elements in the information page |

**Examples**. Figure 15 renders a partially compliant example. The "information page" contains some of the required information for each cookie: cookie names, party who dropped the cookie, purposes and retention period. However, it does not show with which parties the cookies are further shared., Besides, for some cookies (e.g. *_ga* or *_gid*), the provided purposes are not explicit enough – i.e. description "to identify the user" does not explain for what concrete purpose this identification will be used. Figure 16 depicts a non-compliant example wherein the "information page" contains only groups of purposes (analytics, social, etc.) of cookies but does not provide detailed information on each cookie.

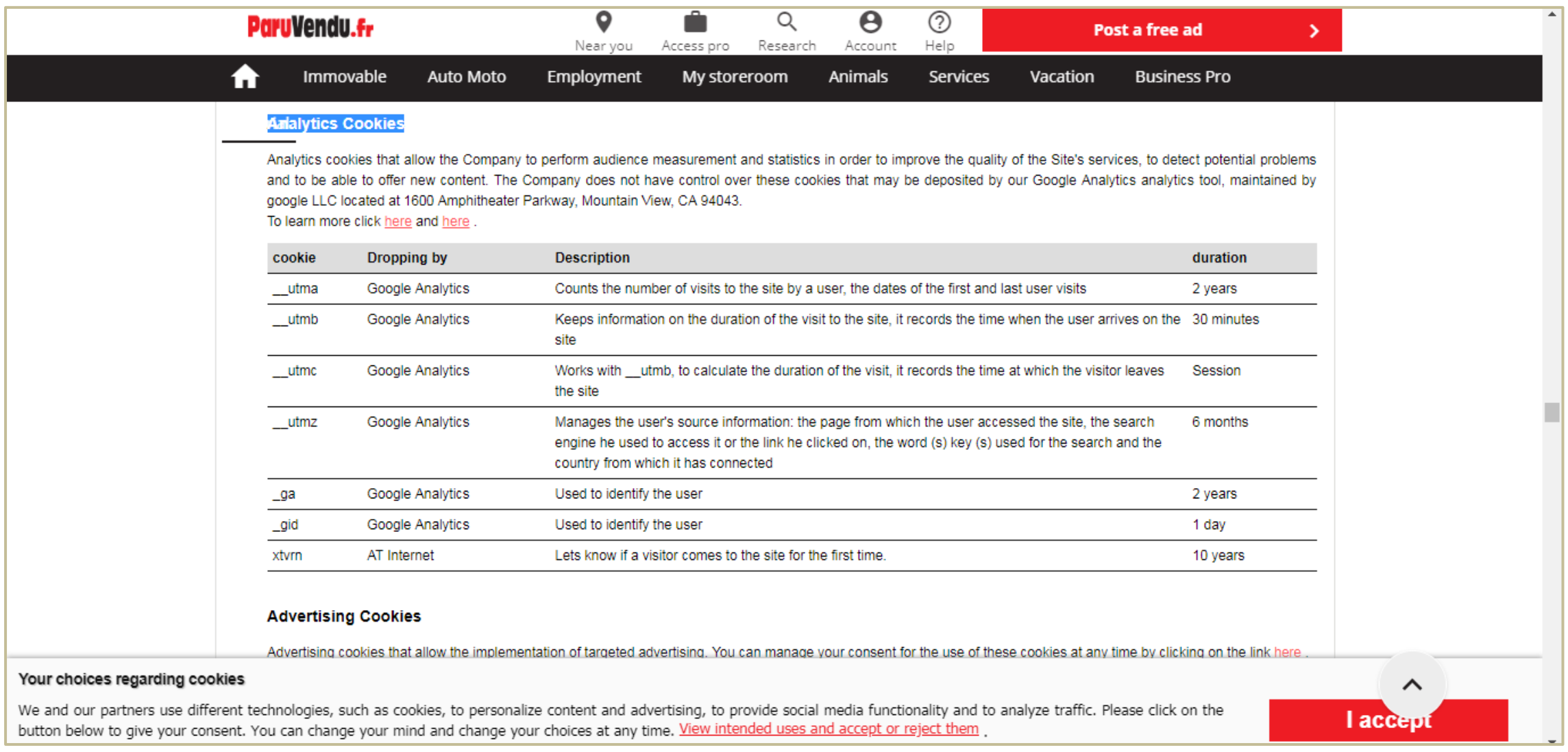

**Figure 15** Partially-compliant example, accessed on 18 May 2019:

[134] cf. Greek DPA (n 42).

[135] Opinion 16/2011 on EASA/IAB Best Practice Recommendation on Online Behavioural Advertising, adopted December, 2011.

(<www.paruvendu.fr/communfo/defaultcommunfo/defaultcommunfo/infosLegales#cookies>)

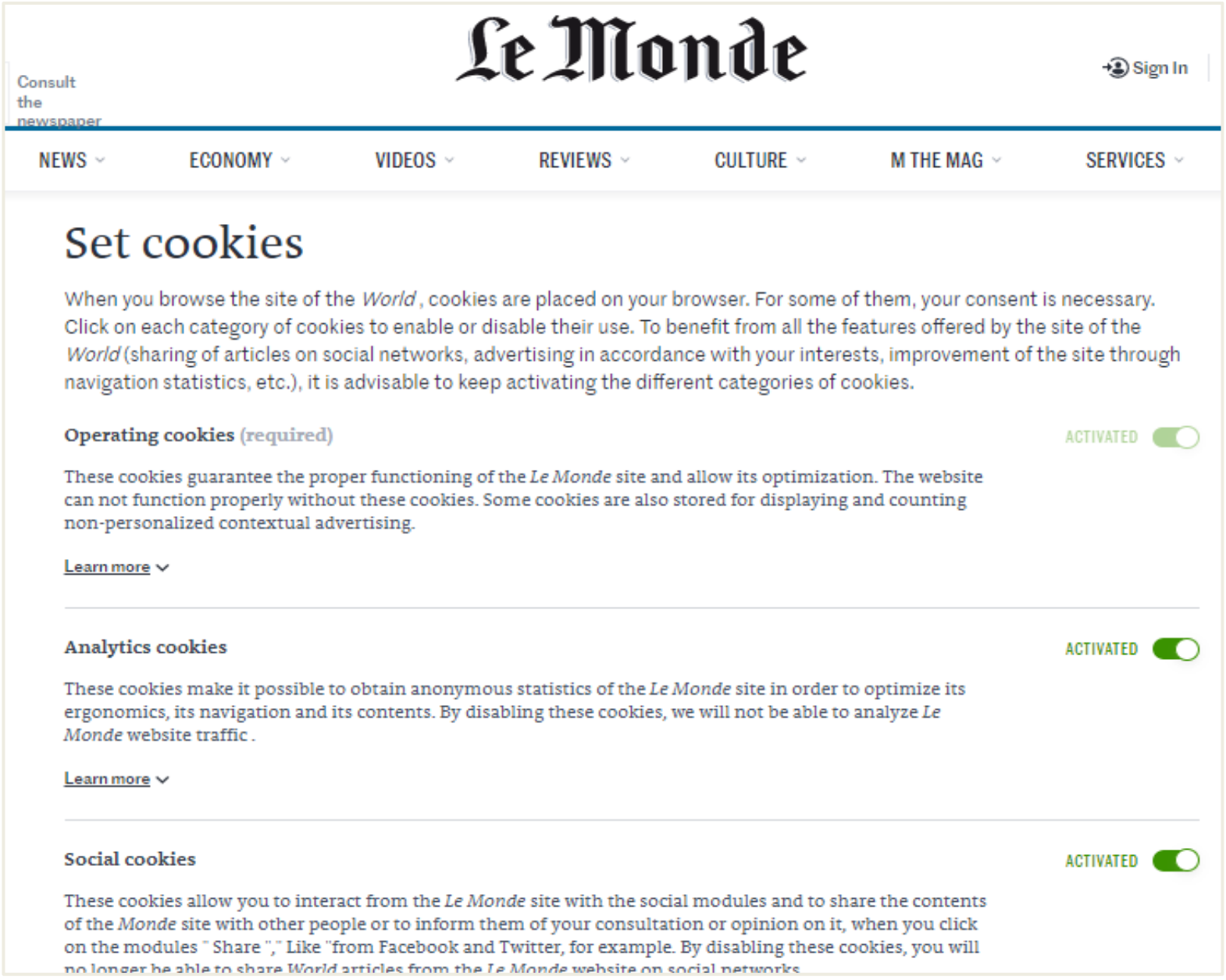

**Figure 16** Non-compliant example: the information page contains only groups of purposes (analytics, social, etc.) of cookies but does not provide detailed information on the purpose for each cookie (<www.lemonde.fr/gestion-des-cookies> accessed on 18 May 2019)

**How to detect violations?** Please see Section 6, where we describe automatic means that can be used to assess language-based requirements.

### R8 Information on consent banner configuration

The 29WP (WP208)[136] instructs that information should refer to *how* the user can express his choice by accepting all-some-or-none BTT and how to change this choice afterwards through the settings. It states that “the ways they can signify their wishes regarding cookies i.e. how they can accept all, some or no cookies and to how change this preference in the future (…) and how to later withdraw their wishes regarding cookies”. Accordingly, information on the possibility of a configuration of the user’s preferences is designed as a low-level requirement presented in the requirement box.

| **Requirement** | **Information on consent banner configurations**<br>The banner or the “information page” should explain how the user can accept all, some or no BTT and how to change this preference in the future. For example, via banner’s buttons or links. |
|---|---|
| Violation | Non-existence of information on configuration possibilities |

**Examples.** Figure 17 depicts a non-compliant banner example. This banner, besides showing general purposes, does not give any information on how the user can accept all, some or no cookies and how to

[136] cf. 29WP (WP208) (n 54).

change this preference in the future. Figure 18 shows a compliant banner that explains how the user can configure his choices.

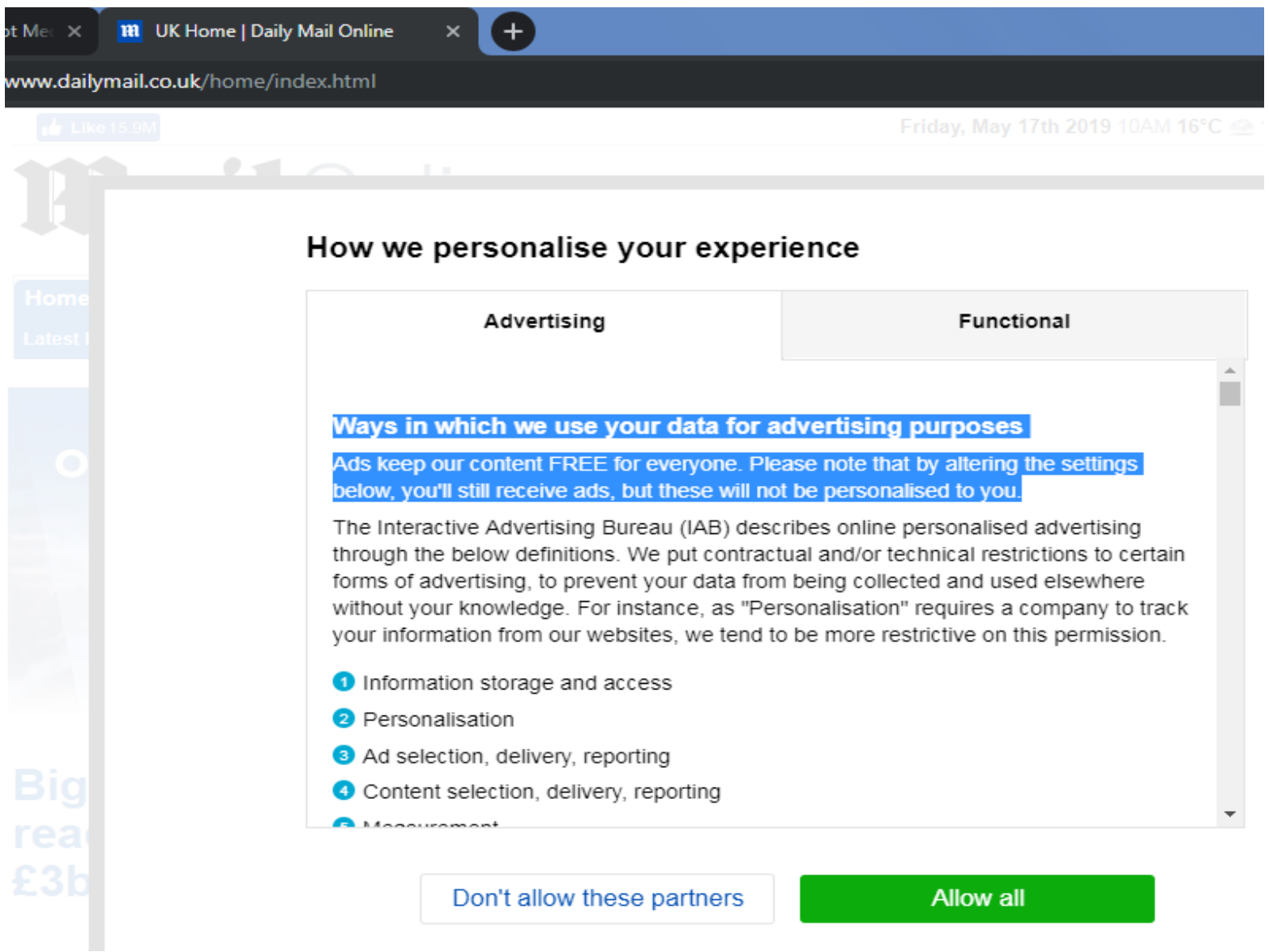

**Figure 17** Violation example: this "information page" only renders general purposes (not requesting consent per purposes) nor any information on how the user can accept all, some or no cookies and how to change this preference in the future (<[www.dailymail.co.uk/home/index.html](www.dailymail.co.uk/home/index.html)> accessed on 18 May 2019)

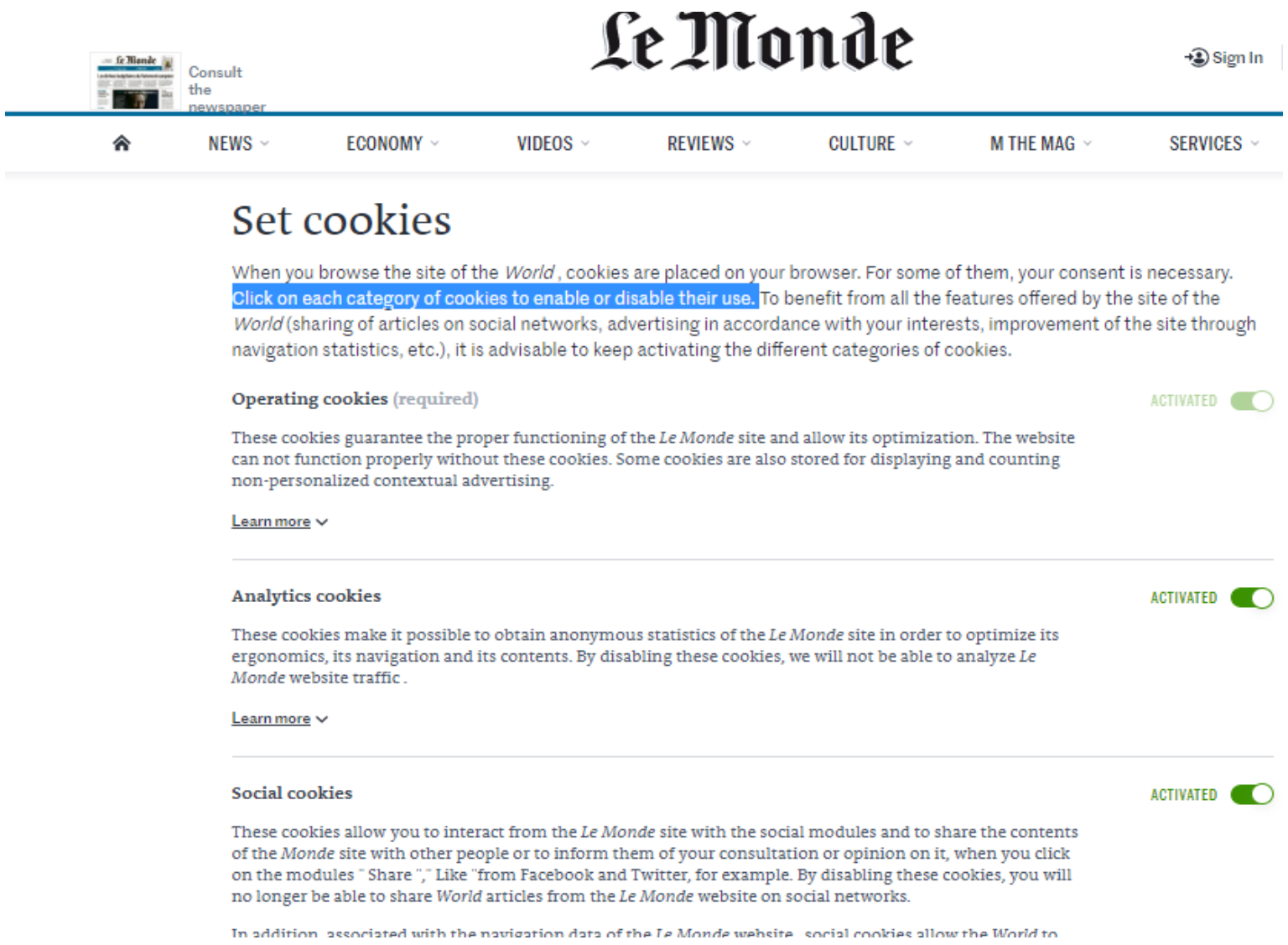

**Figure 18** Compliant example banner: this "information page" explains how the user can configure his choices (<[www.lemonde.fr/gestion-des-cookies](www.lemonde.fr/gestion-des-cookies)> accessed on 18 May 2019)

**How to detect violations?** Please see Section 6, where we describe automatic means that can be used to assess language-based requirements.

### R9 Information on the data controller

Article 13 (1) (a) (b) of the GDPR establishes that the identity and contact details of the controller[137] and the Data Protection Officer (DPO)[138] (when legally obliged to appoint one) are part of the information list to be provided when personal data are collected from the data subject to enable the exercise of the data subject's rights toward the controller (or its representative).

[137] The 29WP (WP 260) lists different forms of contact details of the data controller (e.g. phone number, email, postal address, etc.), and an electronic contact as well, as posited by NOYB, as the service provided is digital, see NOYB, é Report on privacy policies of video conferencing services" (2020) <[https://noyb.eu/sites/default/files/2020-04/noyb_-_report_on_privacy_policies_of_video_conferencing_tools_2020-04-02_v2.pdf](https://noyb.eu/sites/default/files/2020-04/noyb_-_report_on_privacy_policies_of_video_conferencing_tools_2020-04-02_v2.pdf)>

[138] Article 29 Working Party "Guidelines on Data Protection Officers ("DPOs")", WP243 rev.01, adopted in 2017.

We defined as a low-level requirement the need for the information page to incorporate the identity of the controller, contact details and whenever applicable, the representative.

| | |
|---|---|
| Requirement | **Information on the data controller**<br>The "information page" should contain, for each data controller: its identity, contact details, contact of the Data Protection Officer (DPO). |
| Violation | Absence of any reference about the data controller |

**How to detect violations?** Please see section 6, where we describe automatic means that can be used to assess language-based requirements.

### R10 Information on rights

Article 13(2) (a-f) of the GDPR stipulates the need to provide information on the rights of the users: access, rectification, erasure, restriction, object, portability, withdraw consent, lodge a complaint with a DPA, right to be informed about the use of data for automated decision-making and data transfers to a third country or an international organization (and the corresponding safeguards).

We have reproduced these rules into another low-level requirement as shown in the requirement box.

| | |
|---|---|
| **Requirement** | **Information about the data subject rights**<br>The "information page" should contain the user"s rights:<br>1. access<br>2. rectification<br>3. erasure<br>4. restriction on processing<br>5. objection to processing<br>6. portability<br>7. withdraw consent<br>8. lodge a complaint with a DPA<br>9. informed on the use of data for automated decision-making<br>10. informed of data transfers to a third country or an international organization |
| Violation | Absence to any reference on the rights |

**Examples.** Figure 19 shows an information page in which the rights of the subjects are illustrated. However, as shown on Figure 20, it does not provide for all the informative elements, such as the risks of transfers of data.

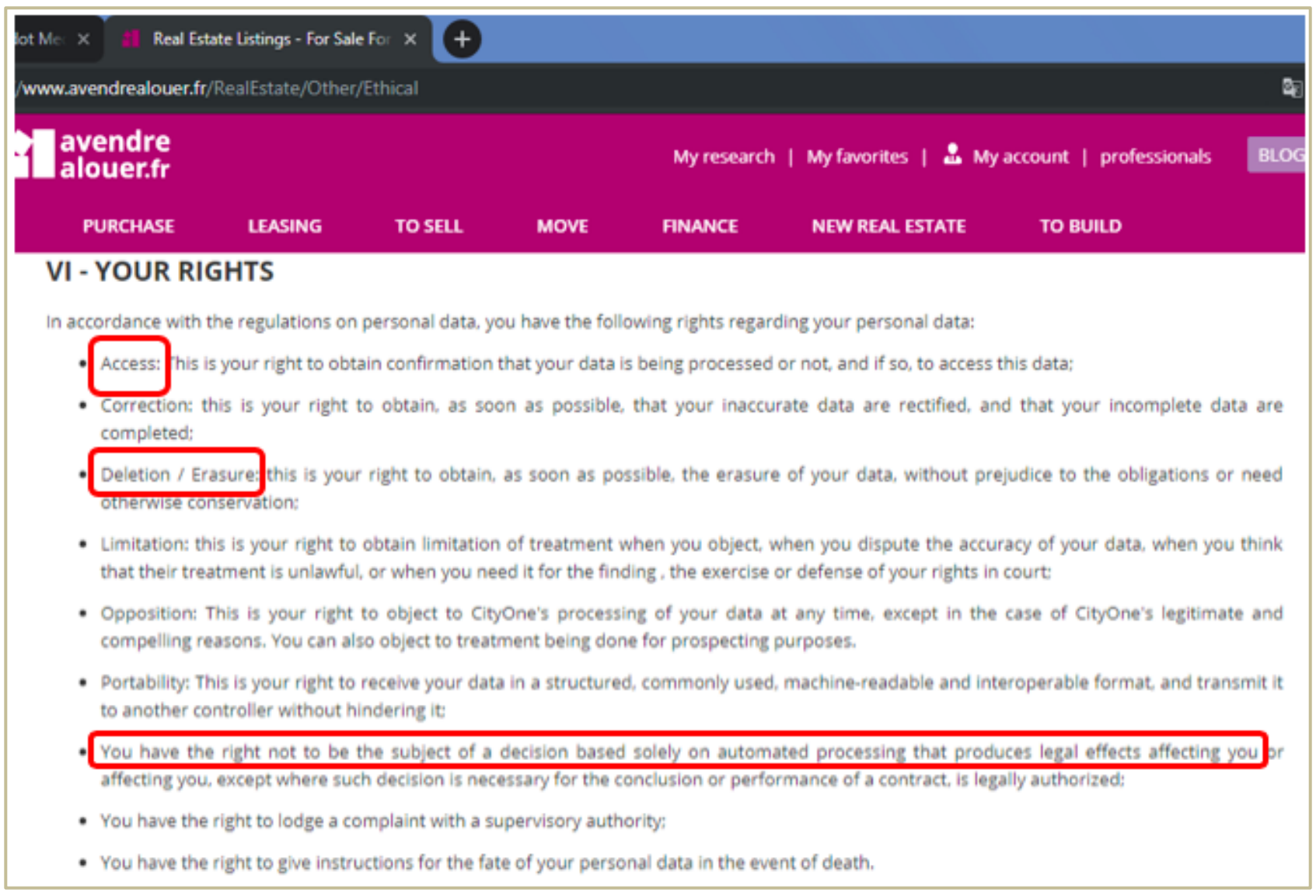

**Figure 19** Compliant example on informed consent: the page refers to the rights of the data subjects, such as the right of access or deletion (<www.avendrealouer.fr/RealEstate/Other/InfosCookies> accessed on 18 May 2019).

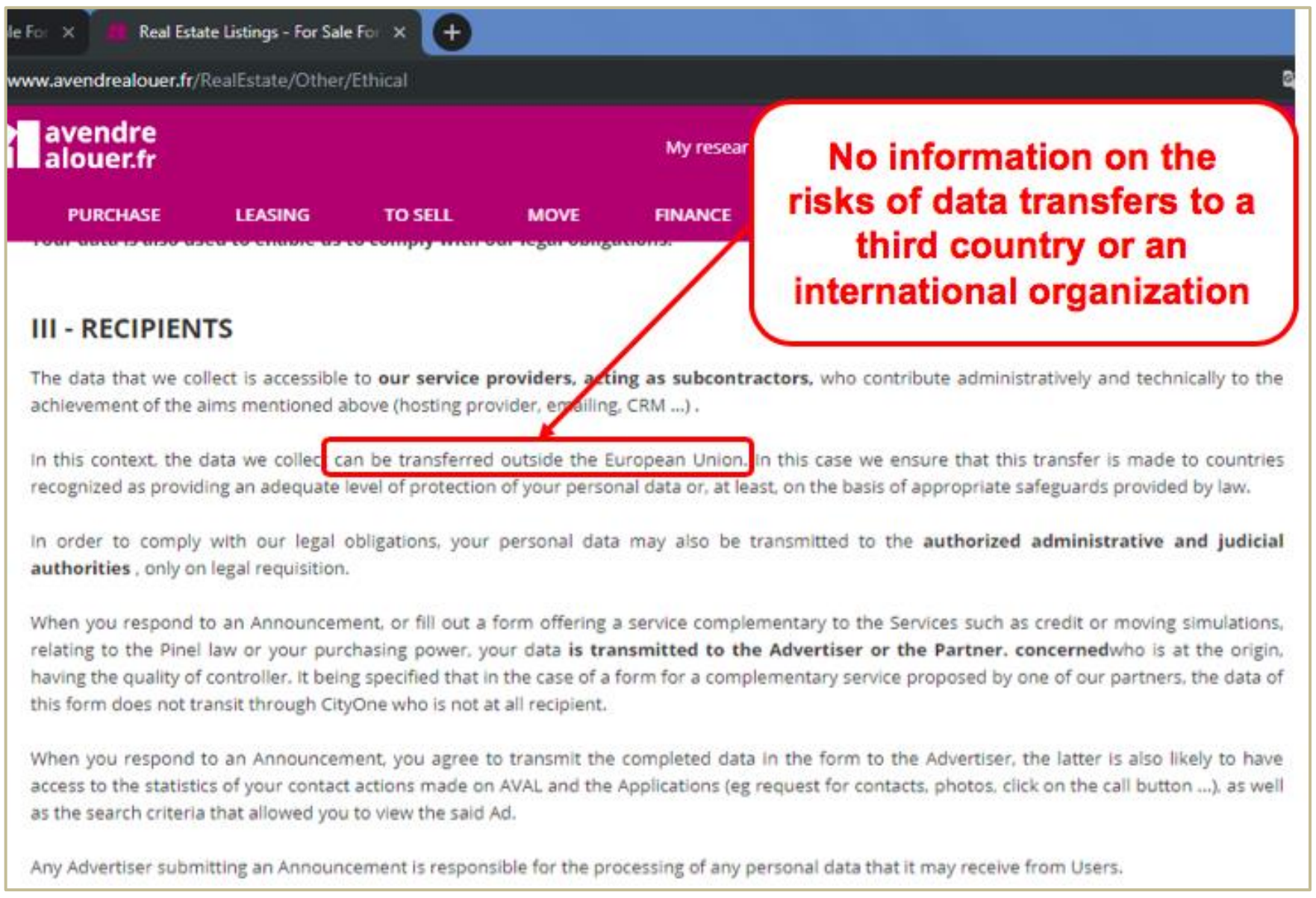

**Figure 20** Non-compliant example of informed consent: the information page does not show the risks of data transfers to a third country (<www.avendrealouer.fr/RealEstate/Other/InfosCookies> accessed on 18 May 2019).

**How to detect violations?** Please see Section 6, where we describe automatic means that can be used to assess language-based requirements.

## 5.5 Unambiguous consent

For the consent to be valid, the user must give an "unambiguous indication" through a "clear and affirmative action" (Article 4(11) of the GDPR). In the following subsections we further decompose this requirement of an unambiguous consent into five low-level requirements: affirmative action design, configurable banner, balanced choice, post-consent registration and correct consent registration, as shown in Table 14.

**Table 14** Derived low-level requirements and their sources

| **Requirements** | | **Sources at low-level requirement** | | |
|---|---|---|---|---|
| **High-Level Requirements** | **Low-level Requirements** | **Binding** | **Non-binding** | **Interpretation: Legal (L) or Computer Science (CS)** |
| **Unambiguous** | **R11 Affirmative action design** | Planet 49 | 29WP; Recital 32<br>DPA: Danish, German | - |
| | **R12 Configurable banner** | - | 29WP; DPAs: French, Spanish, Irish, ICO, Greek, Danish, German | L based on 7(4) |
| | **R13 Balanced choice** | - | DPAs: French, Danish, UK, Greek, Irish | L based on 7(4) |
| | **R14 Post-consent registration** | - | DPAs: French, Greek | CS 7(1) |
| | **R15 Correct consent registration** | - | DPAS: French, Greek, Spanish | CS 7(1) |

### R11 Affirmative Action Design

Unambiguous consent refers to an active behavior of the user through which he indicates acceptance or refusal of BTT (Article 5(3) and Recital 66 of the ePD).

The 29WP (WP208) explains this active behavior:

> "Active behaviour means an action the user may take, typically one that is based on a traceable user-client request towards the website. (…) The process by which users could signify their consent for cookies would be through a positive action or other active behaviour […] The consent mechanism should present the user with a real and meaningful choice regarding cookies on the entry page".[139]

The Advocate General[140] points (in paragraph 44), that the requirement of an "indication" of the data subject's wishes necessitates that the data subject enjoys *a high degree of autonomy* when choosing whether or not to give consent.

Planet49 Judgment[141] made even more precise this requirement. The ruling asserts that "only active behavior on the part of the data subject with a view to giving his consent may fulfil that requirement", and this wording ("with a view to") denotes the element of volition and willfulness towards giving an affirmative consent.

An active behavior leaves *no scope for interpretation of the user"s choice,* which must be distinguishable from other actions. As such, behaviors presenting a margin of doubt do not deliver a choice and therefore are void[142]. To netter ascertain in practice how to distinguish an unambiguous from ambiguous practices, we document examples of both herein. Both Recital 32 GDPR and the 29WP (WP208)[143] provide for concrete and clear examples of an active behavior:

> "clicking on a link, or a button, image or other content on the entry webpage, ticking a box in or close to the space where information is presented (…) or by any other active behavior from which a website operator can unambiguously conclude it means specific and informed consent".

Stakeholders also pinpoint instances of ambiguous behaviors, such as:

- presumed or implied consent from inactivity (or silence) of the data subject[144] (as signed in Recital 32 of the GDPR), e.g.:
  - i) Actions such as browsing, scrolling on a website, swiping a bar on a screen, waiving in front of a smart camera, turning a smartphone around clockwise, or in a figure eight motion (29WP WP 259 rev1.; EDPB 05/2020);
  - ii) Actions like clicking on a "more information" link:[145] "By continuing to use the site we assume you consent to this", or "By accessing the website, you give your consent to our use of cookies". This practice is still acceptable by the Spanish DPA.
  - iii) Disappearance of the cookie banner without an affirmative action of the user, and a positive consent is registered by the fact that the user scrolled the website, visited other pages, clicked on links or other actions on a website;[146]
- when the user's browser is configured to receive cookies;[147]
- pre-ticked boxes[148]. The Advocate General[149] refers that unticking a pre-ticked checkbox on a website is considered too much of a burden for a customer. The ICO[150] stated that pre-selecting any cookie needed of consent, without the user taking a positive action before it is set on their device, does not represent valid consent.

[139] cf. 29WP (WP208) (n 55) 4 and 5.
[140] cf. Opinion of Advocate General Szpunar (n 121).
[141] cf. Planet49 Judgment (n 11) para 54.
[142] cf. 29WP (WP187) (n 54) 35.
[143]The 29WP refers that opt-in consent is the mechanism most aligned to Art. 5(3) of the ePD: "in general users lack the basic understanding of the collection of any data, its uses, how the technology works and more importantly how and where to opt-out. As a result, in practice very few people exercise the opt-out option", cf. 29WP (WP208) (n 55) 4 and 5.
[144] On implied consent, the ICO observes that statements such as "by continuing to use this website you are agreeing to cookies" should not be used as they do not meet the requirements for valid consent required by the GDPR. Pre-ticked boxes or any equivalents, such as sliders defaulted to "on", cannot be used for non-essential cookies. Users must have control over any non-essential cookies, and they must not be set on landing pages before consent is obtained". cf. ICO Guidance (n 26).
[145] Controllers must avoid ambiguity (…). Merely continuing the ordinary use of a website is not conduct from which one can infer an indication of wishes by the data subject to signify his or her agreement to a proposed processing operation, cf. 29WP (WP187) (n 54) 17.
[146] cf. 29WP (29WP208) (n 55) 5.
[147] cf. Greek DPA (n 42).
[148] cf. Planet49 Case (n 11).
[149] Opinion of Advocate General Szpunar (n 121).
[150] cf. ICO Guidance (n 26).

The EDPD[151] (05/2020) establishes that such actions or similar user activities may be difficult to distinguish from other activities or interactions by a user and therefore determining that an unambiguous consent has been obtained will also not be possible.

In the light of the above, we define the "Affirmative Action Design" low-level requirement to make prominent this positive action. The consent must be registered by the controller only after an affirmative action of a user, like clicking on a button, checking a box, or actively selecting settings.

| **Requirement** | **Affirmative action design**<br>Consent must be registered only after an affirmative action of a user, like clicking on a button or checking a box. |
|---|---|
| Violation | The action of closing a cookie banner considered as consent. Allowing only closing the banner and forcing agreement to consent. Pre-ticked boxes. Disappearance of the cookie banner without an affirmative action of the user with a positive consent registered. |

**Examples.** Figure 21 gives a non-compliant example of the requirement of "Active Action Design". It shows the Twitter account of the European Data Protection Board and a cookie banner provided by twitter.com, wherein it is not possible to exercise an active consent since the only possible action is to close the cookie banner, while agreeing to the use of cookies.

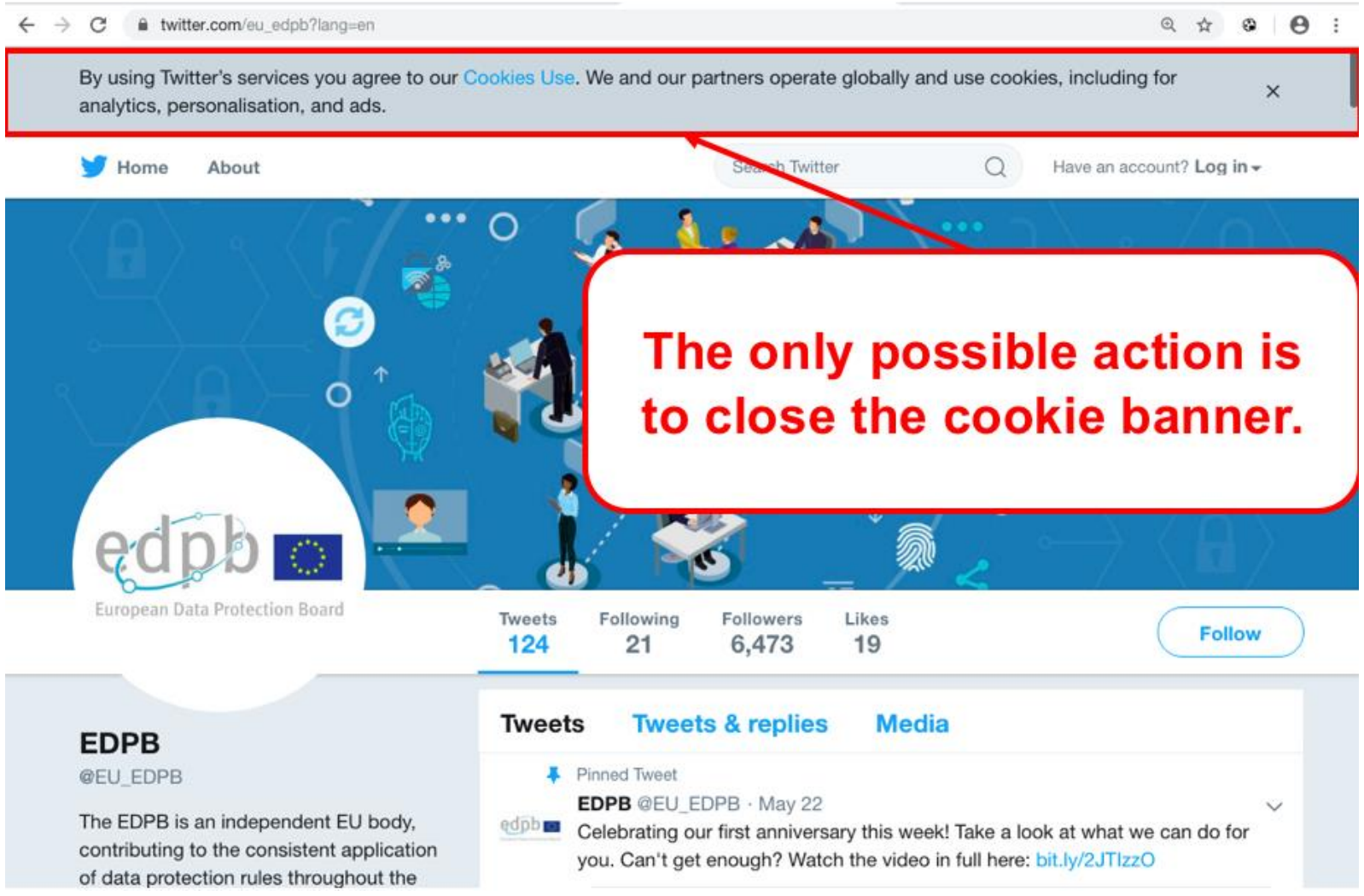

**Figure 21** Violation of the "Affirmative Action Design" requirement (<https://twitter.com/eu_edpb?lang=en> accessed on 24 September 2019)

**How to detect violations?** To detect a violation of this requirement, one needs to perform an action on the website, like closing the banner or scrolling the website and verify whether a positive consent has been registered. While an action on a website must be done by a human operator (because there is no standard design of closing banners that can be automated), verification of a registered consent can be done only with technical means.

However, verification of a registered consent is only possible if it is known a priori which standard or specification is used to store the consent in the user's browser. For instance, this is the case if the publisher is using IAB Europe's Transparency and Consent Framework, as demonstrated by Matte et al.[152] By using the "Cookie Glasses" browser extension[153] (developed by the same authors), a human operator can observe whether the consent is stored in the browser before making an affirmative action in the banner, or upon scrolling the website.

[151] cf. EDPB 05/2020 (n 84).

[152] Celestin Matte, Nataliia Bielova, Cristiana Santos, "Do Cookie Banners Respect my Choice? Measuring Legal Compliance of Banners from IAB Europe's Transparency and Consent Framework" (IEEE Symposium on Security and Privacy (IEEE S&P 2020). To appear. http://www-sop.inria.fr/members/Nataliia.Bielova/papers/Matt-etal-20-SP.pdf

[153] "Cookie Glasses" extension available for Firefox: https://addons.mozilla.org/fr/firefox/addon/cookie-glasses/ and Chrome / Chromium : https://chrome.google.com/webstore/detail/cookie-glasses/gncnjghkclkhpkfhghcbobednpchjifk

### R12 Configurable banner

Several customization implementations are possible, suchlike the ones proposed by different stakeholders in Table 15:

**Table 15** Positioning of the 29WP and DPAs on the configuration of consent dialogs

| Stakeholders | Configurations of consent dialogs (configuration and web design) |
|---|---|
| 29WP (WP208) | "The user should have an opportunity to freely choose between the option to accept some or all cookies or to decline all or some cookies and to retain the possibility to change the cookie settings in the future". |
| French DPA[154] | "By its presentation, the mechanism for obtaining consent must enable the data subject to be aware of the goal and scope of the act enabling him or her to signify his or her agreement or disagreement." (Parag. 51) |
| Spanish DPA[155] | "a management system (or cookie configuration panel) that allows the user to choose in a granular way to manage his preferences, by: enabling a mechanism or buttons to reject all cookies, another to enable all cookies, or to be able to do it in a granular way." |
| Danish DPA[156] | "users are given the option to accept or decline cookies either by an "accept" or "reject" button, or by toggles to accept or reject specific cookie purposes. The option to decline cookies is as easy as it is to accept cookies". |
| Irish DPA[157] | - "if you use a button on the banner with an "accept" option, you must give equal prominence to an option which allows the user to "reject" cookies, or to one which allows them to manage cookies and brings them to another layer of information in order to allow them to do that, by cookie type and purpose".<br>- "manage cookies and brings them to another layer of information in order to allow them to do that, by cookie type and purpose". |
| UK DPA[158] | "A consent mechanism that does not allow a user to make a choice would also be non-compliant, even where the controls are located in a "more information" section". |
| Greek DPA[159] | "the option to decline the use of trackers is only given at a second level, i.e. following the selection of a hyperlink to "more information" or "settings". |

We believe that a sufficient level of granularity of choice is demanded in the consent banner design. We interpret that a consent banner must give the user an option such as: i) one "configure" button; or ii) "accept", "reject" and "customize/configure".

| Requirement | Configurable banner<br>A banner must give the user an option to customize his consent. Several implementations are possible:<br>- One "Configure" button<br>- "Accept" and "Configure" buttons<br>- "Accept", "Reject" and "Configure" buttons |
|---|---|
| Violation | - A banner does not provide a choice in the settings/configuration,<br>- the customization options are not emphasized as a link in the first layer,<br>- the only existing box is "Accept and close",<br>- invisible "Parametrize" button. |

**Examples.** Figure 22 illustrates a non-compliant banner design, curiously from one of the flagship security and privacy conferences (IEEE Symposium on Security and Privacy). In this banner, the only available option is to accept and close the banner, not offering a sufficient level of granularity of choice demanded by the GDPR. This example, however, presents a violation only if cookies that require consent are used on this website, while the banner does not provide an explicit enough purpose (i.e. "to give you the best user

[154] cf. CNIL draft recommendation 2020 (n 95).
[155] The Spanish DPA decision reads accordingly,

> "III. It does not provide a management system or cookie configuration panel that allows the user to eliminate them in a granular way. To facilitate this selection the panel may enable a mechanism or button to reject all cookies, another to enable all cookies or do so in a granular way to manage preferences. In this regard, it is considered that the information offered on the tools provided by several browsers to configure cookies would be complementary to the previous one, but insufficient for the intended purpose of allowing to set preferences in a granular or selective way",

See Spanish DPA decision, "Procedimiento PS/00300/2019" (2019) <www.aepd.es/resoluciones/PS-00300-2019_ORI.pdf?utm_source=POLITICO.EU&utm_campaign=fc1f5e664f-EMAIL_CAMPAIGN_2019_10_17_04_52&utm_medium=email&utm_term=0_10959edeb5-fc1f5e664f-190359285> accessed 7 May 2020 (our translation).
[156] cf. Danish DPA Guide (n 77).
[157] cf. Irish DPA Guidance (n 37).
[158] cf. ICO Guidance (n 26).
[159] cf. Greek DPA (n 42).

experience”) to determine whether consent is needed in this case. Figure 23 shows a banner design which is closer to be compliant with the “Configurable” consent requirement. Through the indication of the “yes” and “configure” buttons, it is possible to accept and customize consent. Configuring the choice at any time in the privacy center is also possible. As the customization of the preferences is easy and user-friendly, the banner seems to comply with the above-mentioned requirement.

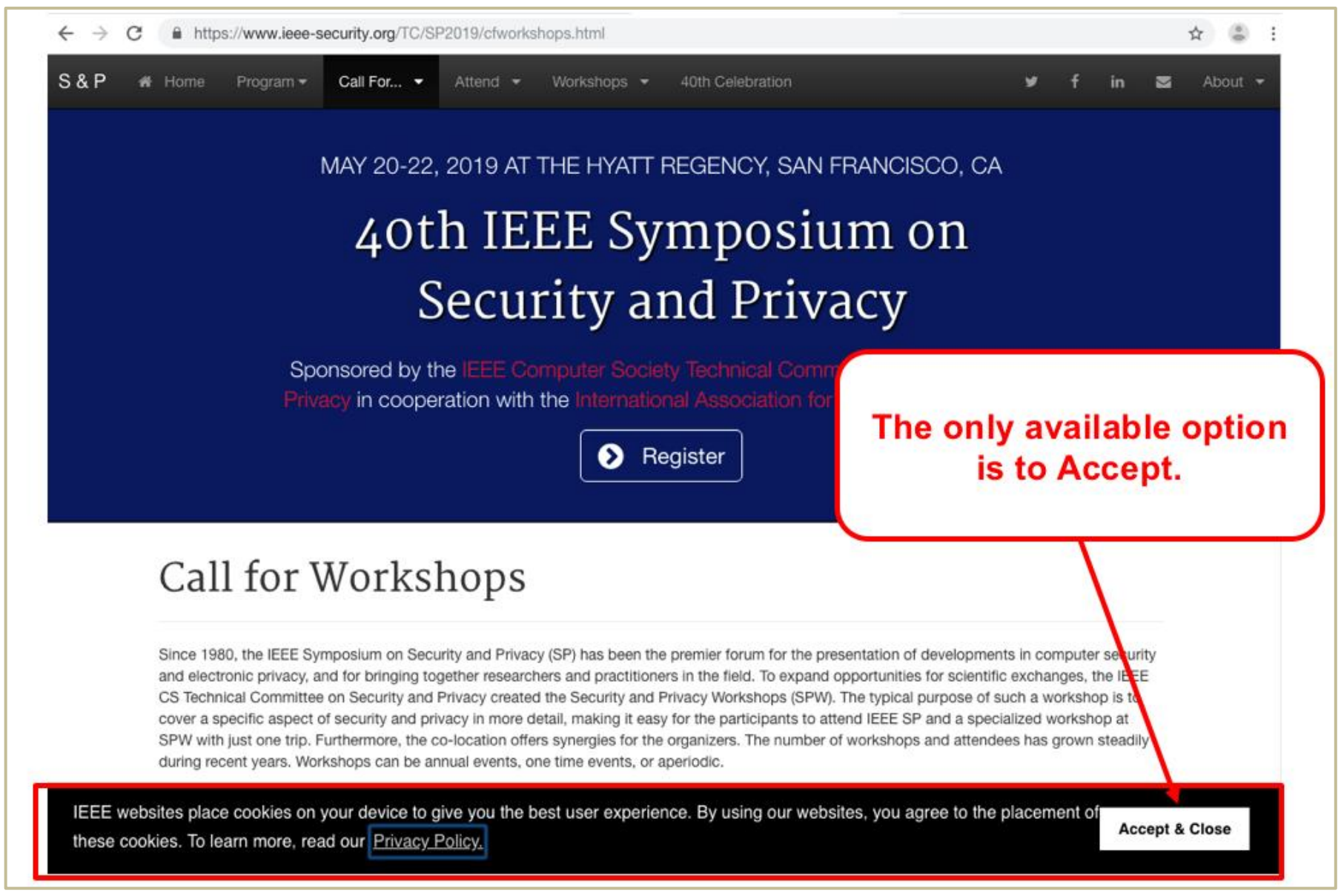

**Figure 22** Violation of the “Configurable consent” requirement (<www.ieee-security.org/TC/SP2019/venue.html> accessed on 17 May 2019)

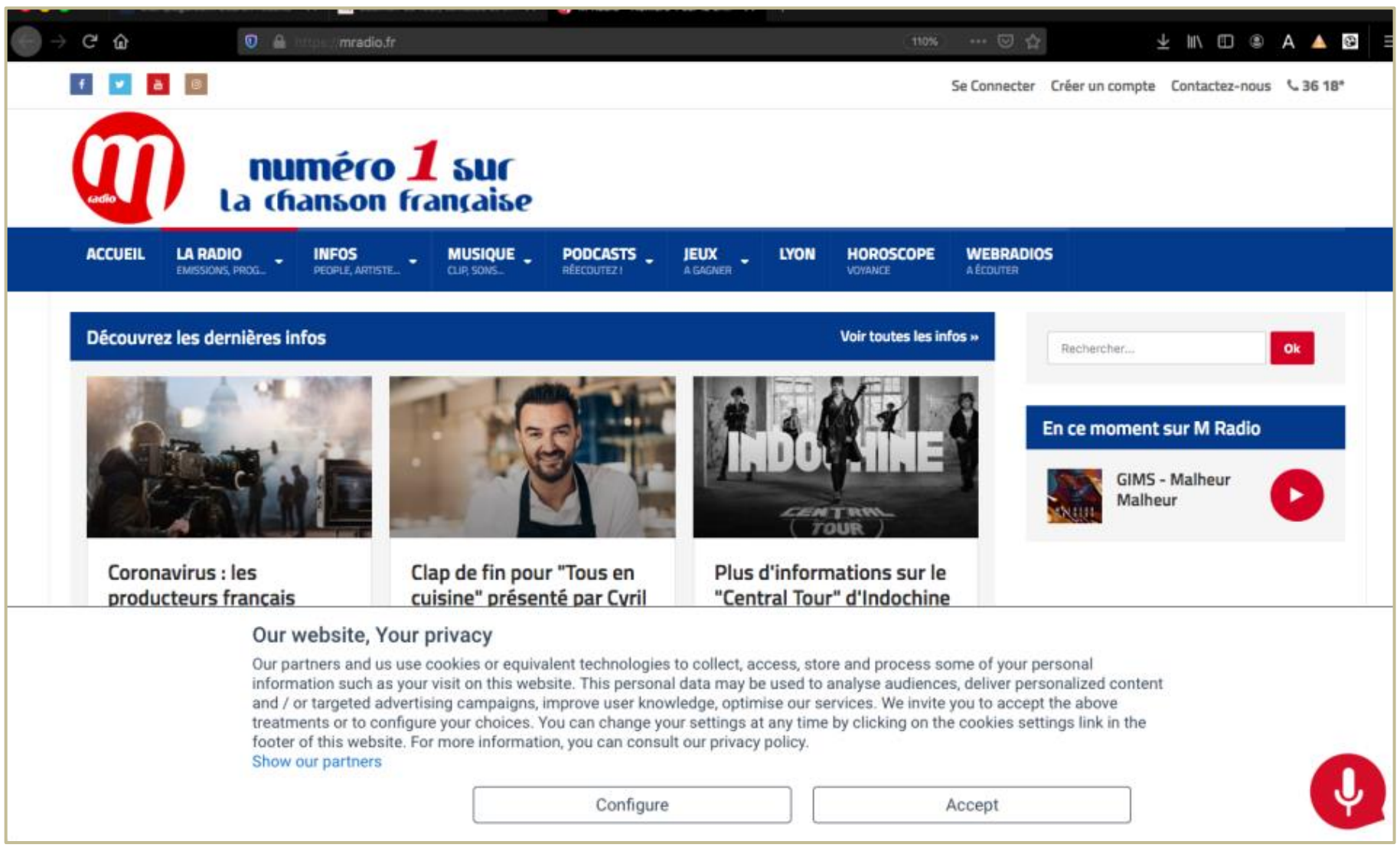

**Figure 23** Example of compliance with the “Configurable consent” requirement (<https://mradio.fr/> accessed on 27 May 2020). It is possible to accept or configure, however this banner does not provide a balanced choice.

**How to detect violations?** To detect violations of this requirement, a human operator needs to evaluate whether a banner gives a set of options to the user. As of today, verifying this requirement fully automatically is not possible because of a lack of standards in cookie banner design. Nevertheless, some technical developments have been made to automatize the process of either fully accepting or fully refusing consent when such option exists in the configuration settings. The “Consent-o-Matic” Web browser extension[160] (and previously the Cliqz browser[161]) implements such functionality to automatically interact with the HTML content of a cookie banner to either refuse or accept consent.

[160] https://github.com/cavi-au/Consent-O-Matic
[161] https://github.com/cliqz-oss/autoconsent

### R13 Balanced choice

From Article 7(4) of the GDPR which states that withdrawing consent should be as easy as giving it, we additionally interpret that the choice between "accept" and "reject" BTT must be consequently *balanced* (or equitable). Our interpretation is also sustained by the recent opinion of the CJEU's Advocate General and by some DPAs, as listed in Table 16.

**Table 16 Positioning of stakeholders regarding the requirement of a balanced banner**

| Stakeholders | Positioning regarding the "balanced banner" requirement |
|---|---|
| CJEU's Advocate General[162] | emphasized the need that actions, "*optically in particular, be presented on an equal footing*" |
| French DPA[163] | Parag. 39. 'interfaces should not use potentially misleading design practices, such as the use of visual grammar that might lead the user to think that consent is required to continue browsing or that visually emphasizes the possibility of accepting rather than refusing. Parag. 40. The user may also have the choice between two buttons presented at the *same level and in the same format*, with "accept" and "refuse", "allow" and "forbid", or "consent" and "do not consent", or any other equivalent wording that is sufficiently clear to the user. Parag. 51. Thus, this mechanism should not involve potentially misleading design practices, such as the use of visual grammar that impedes the user's understanding of the nature of his or her choice. |
| Danish DPA[164] | "users are given the option to accept or decline cookies either by an "accept" or "reject" button, or by toggles to accept or reject specific cookie purposes. The option to decline cookies is as *easy as* it is to accept cookies". |
| UK DPA[165] | -refusal of trackers should be at the *same level* as the "accept" button. "[A] consent mechanism that emphasizes "agree" or "allow" over "reject" or "block" represents a non-compliant approach, as the online service is influencing users towards the "accept" option". |
| Greek DPA[166] | -Parag. 4: The user must be able, with the same number of actions ("click") and from the *same level*, to either accept the use of trackers (those for which consent is required) or to reject it, either all or each category separately. Parag. 7: "To ensure that the user is not affected by website designs favoring the option to consent vis-à-vis the option to decline, buttons of the *same size, tone and color* ought to be used, so as to provide the same ease of reading to the attention of the user". Parag. 6: "The size and colour of the "accept" or "consent" button strongly urges the user to choose, e.g. is very large and / or in bold and / or is pre-ticked." |
| Irish DPA[167] | "no use of an interface that "nudges" a user into accepting cookies over rejecting them. Therefore, if you use a button on the banner with an "accept" option, you must give *equal prominence* to an option which allows the user to "reject" cookies, or to one which allows them to manage cookies and brings them to another layer of information in order to allow them do that, by cookie type and purpose. |

*Potential violations* of a balanced consent banner requirement normally happen when there is unbalance in the design choices given. Design choices related to an unbalanced choice in a consent banner can consist, for example, of "*False Hierarchy*" and "*Aesthetic manipulation*".

According to Gray et al.,[168] *False Hierarchy*

> "gives one or more options visual or interactive precedence over others, particularly where items should be in parallel rather than hierarchical. This convinces the user to make a selection that they feel is either the only option, or the best option".

Some examples of false hierarchy in consent banners are illustrated below, both at the first and second layers of the banner:

- far-away approach: when the sliders or the menu settings are far, e.g. website forwards the user to click on a link for opt-out tools offered by DAA, NAI, and Google;
- click-away approach: requires more clicks and diligence to reach to the parametrization and refuse consent (more than two sub-menus)

[162] Opinion of Advocate General Szpunar on the case of Planet 49, delivered on 21 March 2019, <http://curia.europa.eu/juris/document/document.jsf?docid=212023&doclang=en>, accessed 18th June 2020.
[163] cf. CNIL draft recommendation 2020 (n 95).
[164] cf. Danish DPA Guide (n 77).
[165] cf. ICO Guidance (n 26).
[166] cf. Greek DPA (n 42).
[167] cf. Irish DPA Guidance (n 37).
[168] cf. Gray et al. (n 7).

- box with a bigger “OK” button and a small, less visible “Configure” button, which gives a higher hierarchy to “OK”;
- banner presenting 2 options: “accept all” and “reject all” whereas the “accept all” option comes first, has green color, while “reject all” is in white, indicating some desirability in choosing this one;
- box with “I consent” emphasized in a black box, and “More Options” link on the corner of the banner;
- banner in which the option to refuse is only given at a second level, i.e. following the link to “more information”;
- button for “Refuse” or “Preferences” is a text, while “Accept” is a button;
- use of the same color with dark/light differences, e.g. light-blue and dark-blue sliders to signify accept and refuse;
- the legend in the banner always labels “Activate”, whether the slider is activated or not;
- the same button corresponds to “Activate all” and “Deactivate all” (and the meaning of the slider is unclear), it is not possible to see what is was chosen;
- the only button to disable purposes in barely visible;

*Aesthetic manipulation* consists of an interface design that nudges users into clicking the “accept” button rather than the "refuse" button by using design, colors, hovering properties. The CNIL[169] names it “Attention Diversion” referring to these design choices that draw attention to a point of the site or screen to distract or divert the user from other points that could be useful. The CNIL states that visual saliency is effective and abundant. It gives a concrete example, working on the color of a “continue” green button while leaving the “find out more” or “configure” button smaller or grey, users may perceive green as the preferable choice. This holds particularly if they are conditioned by the traffic light metaphor bias used by designers that assign colors according to the flow of information (“green” = free flowing”; “red”=stop”), which bring ambiguity to the choice. Some examples of false manipulation in consent banners are illustrated below, both at the first and second layers of the banner:

- hovering over a button: “accept” has a hover background while “reject” does not
- bright and attractive “accept” button and grey/white “reject” button
- all information is written in a very small window difficult to navigate size and colour of the “accept” or “consent” button strongly urges the user to choose it, e.g. is very large and / or in bold and / or is pre-ticked.

| **Requirement** | **Balanced choice**<br>A banner must present a fair or balanced design choice |
|---|---|
| Violation | A banner does not provide a fair choice. |

**Examples.** Figure 24 presents an example of non-compliant banner. Even though this banner provides an option to configure privacy settings, the provided choice is not balanced – it contains “Accept” and “Configure” buttons – and hence guides the user towards acceptance. Figure 25 provides an example of a balanced choice since both options, “Yes” and “No” are present in the banner interface.

[169] cf. CNIL report (n 8).

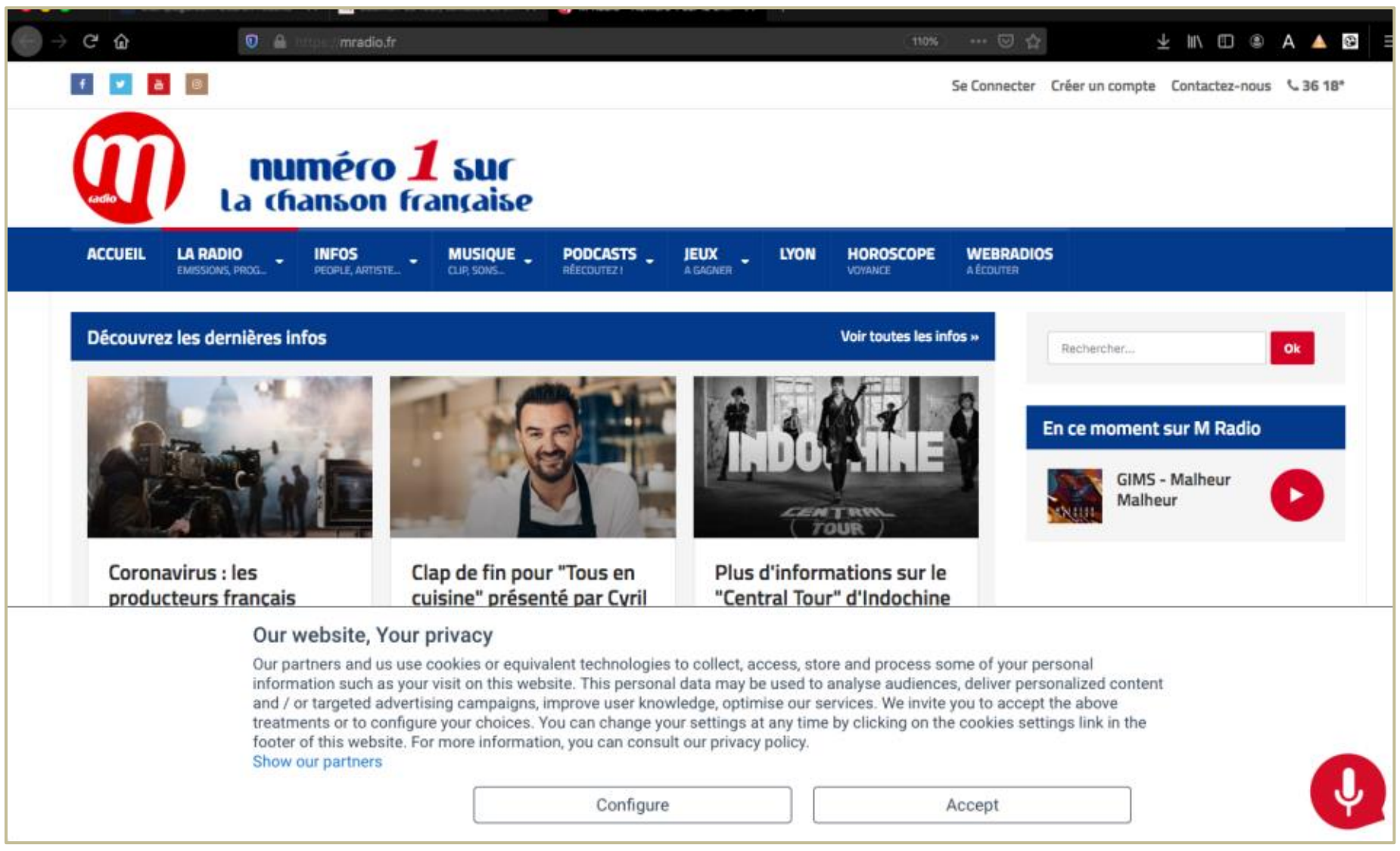

**Figure 24** Violation of the requirement "Balanced choice" (<https://mradio.fr/> accessed 27 May 2020).

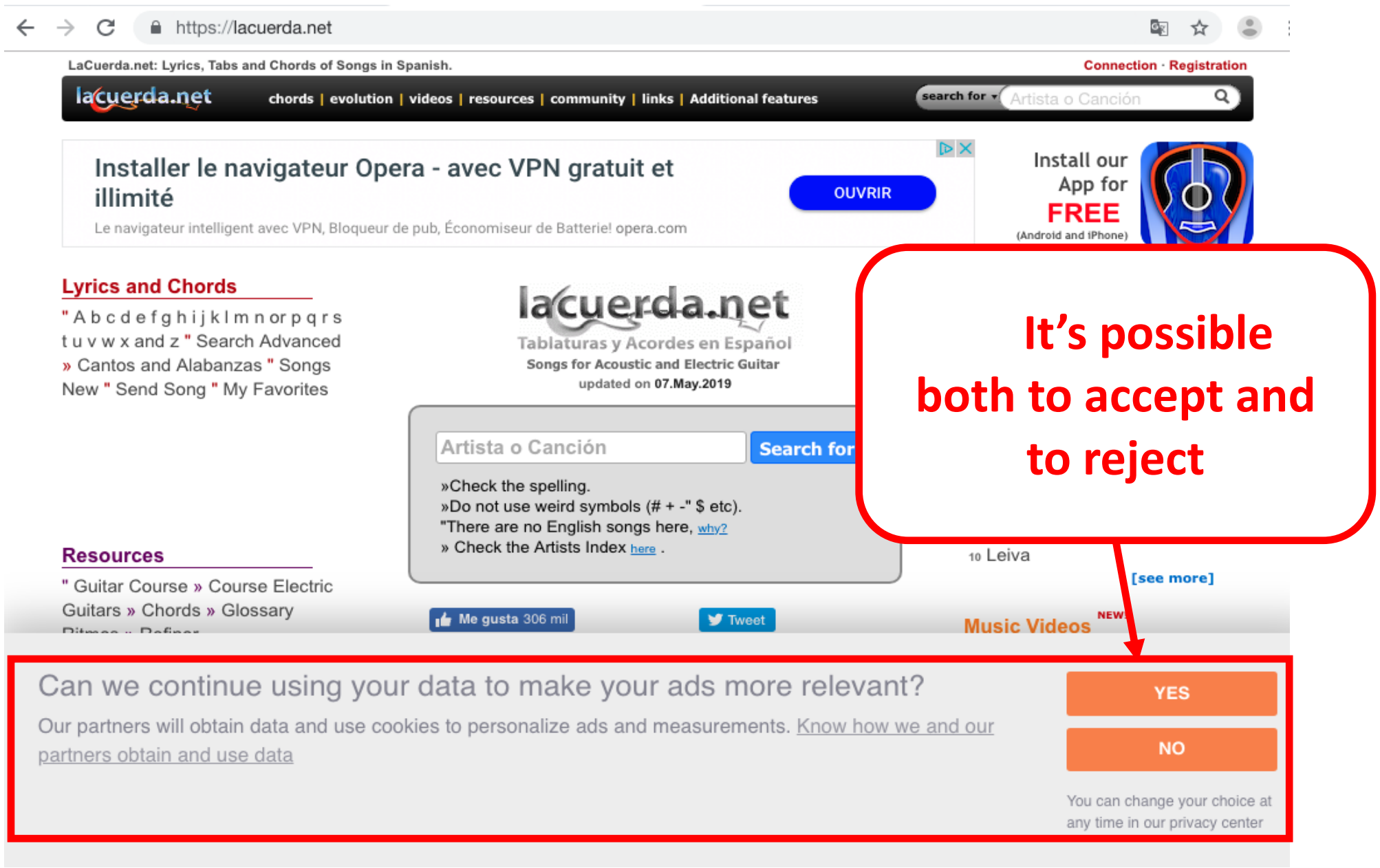

**Figure 25** Compliant example for "Balanced choice" requirement (<https://lacuerda.net/> accessed 17 May 2019)

**How to detect violations?** To detect violations, a human operator needs to evaluate whether a banner gives fair and balanced choices. As of today, it is not possible to verify this requirement automatically because of lack of standards in cookie banner design.

### R14 Post-consent registration

The GDPR mandates in Article 7(1) and Recital 42 that controllers have the obligation to demonstrate that the data subject has consented to processing of his personal data. Further, Article 30 requires that each shall maintain a record of processing activities under its responsibility (which includes consent). These provisions constitute a specification of the principle of accountability, enshrined in Article 5(2) of the GDPR.

This auditable legal obligation entails a *technical* side that is relevant to consider: after a certain user action done via the user interface (like clicking on a button, checking a box, etc.), the user's choice (acceptance or refusal) needs to be "registered" or "stored" in the user's device (a browser in our case). We therefore use the noun "registration" to mean that the consent choice is stored.

Both the need for an auditable consent and an adequate procedure thereto are emphasized by the 29WP and DPAs:

*Consent registration*: both the Spanish, the Greek[170] and the Danish DPAs asserts that consent must be stored for documentation (in case of inspections by DPAs).

*Mechanisms for consent registration*:

- 29WP: an auditable consent can be achieved by keeping a record of the received consent statements, so the controller can show how/when consent was obtained. Consent receipt mechanisms can be helpful in automatically generating such records;
- CNIL[171] (paragraphs 36 and 37): points out that consent (either its acceptance or refusal) should be registered. A tracker may be used with the sole purpose of storing consent or refusal;
- Italian[172] and Irish[173] DPAs: affirm that a special technical cookie is normally used to store and keep track of the acquired consent. The Irish DPA further states that "any record of consent must also be backed up by demonstrable organizational and technical measures that ensure a data subject's expression of consent (or withdrawal) can be effectively acted on";
- Belgium DPA: illustrates that companies must be able to demonstrate that consent was collected by using logs.

Additionally, some DPAs opted to define or comment on the storage period of the user's choice, as shown in Table 17.

**Table 17** DPAs' positioning about the storage period of the user's choice

| DPAs | Positioning about the storage period of the user's choice |
|---|---|
| French | 6 months |
| Spain | consent should be renewed after 24 months |
| Danish | storage period of 5 years |
| Irish | no longer than 6 months, after which the consent request must be renewed |

We include a technical low-level requirement "post-consent registration", as depicted in the requirement box below. Notice that closing a consent banner without a consent being registered as "positive" does not configure a violation of this requirement.

| **Requirement** | **Post-consent registration**<br>Consent should be registered (e.g. stored on a terminal equipment) in a "consent cookie" (or any other browser storage) only after an affirmative action of the user. |
|---|---|
| Violation | A consent registered without any user action. |

**Example.** Figure 24 refers to an example of non-compliant design of the "Post-consent registration" requirement. When accessing the tpi.it website, it is possible to check that the user's consent was registered before the user has made his choice.

[170] cf. Greek DPA (n 42).
[171] cf. CNIL draft recommendation 2020 (n 95).
[172] Italian DPA, FAQs on cookies (2020) <https://www.garanteprivacy.it/web/guest/home/docweb/-/docweb-display/docweb/3585077>
[173] cf. Irish DPA Guidance (n 37).

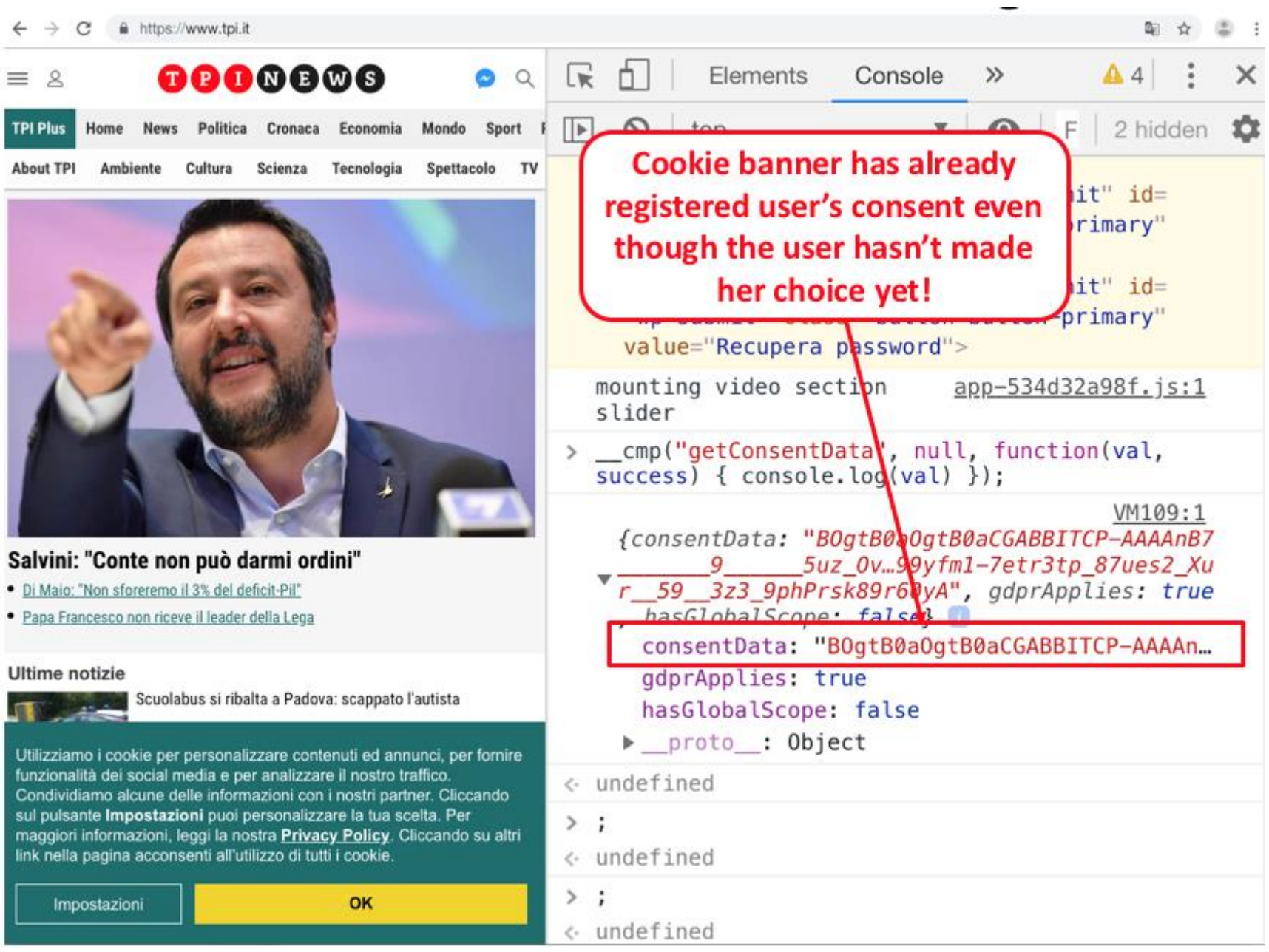

**Figure 24** Violation of the "Post-consent registration" requirement (<www.tpi.it/> accessed 17 May 2019).

**How to detect violations?** Detecting violations is only possible with technical means, but only on websites where it is known how the consent is registered by the publisher (e.g. for websites using the IAB's Transparency and Consent Framework, as demonstrated by Matte et al.).[174]

For the majority of websites, it is not the case and therefore, detecting violations without standardizing the storage of consent is not possible. Two elements would be needed to standardize consent: the data structure for consent storage, and a concrete browser storage (cookie, localStorage, or any other) used to store consent. On websites where storage and structure of consent is known (such as on the IAB TCF), the verification procedure is the following: in a browser with an empty session, open the target webpage. Before doing any interaction with the cookie banner, verify whether the browser storage contains a positive consent.

In order to assure that the violation does not occur, website publishers or other entities that register consent on a website (such as CMP in case of IAB TCF) have to be able to prove that they have rightfully obtained consent after user's interaction with the cookie banner. Technically, the proof of consent could be done through the use of cryptographic primitives, or secure hardware. Additionally, publishers need to record the user's interaction with the banner via screenshots or video recording. As of May 2020, no technical solution exists to handle the problem of proving that consent has been obtained only after the users has interacted with the banner.

### R15 Correct consent registration

Another technical side deriving from the obligation to register consent (Article 7(1)) refers to the fact that the decision made by the user in the banner interface should be identical to the consent that gets registered/stored by the website. If consent is correctly registered, the user will not be pressured to choose again by the same website. To this scope, nagging[175] practices (identified as dark patterns) seems to be related to the functionality of consent dialogs being correctly registered.

Some DPAs express concerns when consent is not registered correctly:

i. The CNIL draft recommendation[176] (paragraph 36): states that as a result of a non-registration of the user's choice, the user will be pressured to accept out of weariness:

174 cf. Matte et al. (n 152).
175 cf. Gray et al. (n 7).
176 CNIL draft recommendation 2020 (n 95).

“Thus, the choice expressed by the user, be it consent or refusal, should be recorded in such a way that the user's consent is not sought again for a certain period of time. Indeed, failure to register the refusal would prevent it from being considered in the long term, in particular during new visits. If the choice that the user has expressed is not registered, he or she would be asked again to consent. This continued pressure would be likely to cause the user to accept out of weariness. Failure to record the refusal to consent could therefore have the consequence of exerting pressure that could influence his or her choice, thus calling into question the freedom of the consent he or she expresses”.

ii. Greek DPA:[177] emphasizes the consequences of an incorrect registration of the user’s choice: “in case trackers are rejected, the user is constantly requested to register a new choice through the perseverance of pop-up windows, whereas, in case trackers are accepted, this choice is maintained for a longer period of time than the choice of rejection”.

iii. Spanish DPA: the users’ preferences may be stored so they are not asked to set them up again every time they visit the relevant page.

In this line, we have devised the technical low-level requirement of “correct consent registration”. Notice that the consent should be *registered both in the user’s browser, and also on the server* of the entity that proves collection of consent.

| | |
|---|---|
| **Requirement** | **Correct consent registration**<br>The registered consent must be identical to the user’s choice in the user interface |
| Violation | A registered consent is different from the user’s choice. |

**Example.** In figure 25, by using Matte et al. Cookie Glasses tool, [178] it is possible to verify whether consent is correctly registered by cookie banners of IAB Europe's Transparency & Consent Framework.

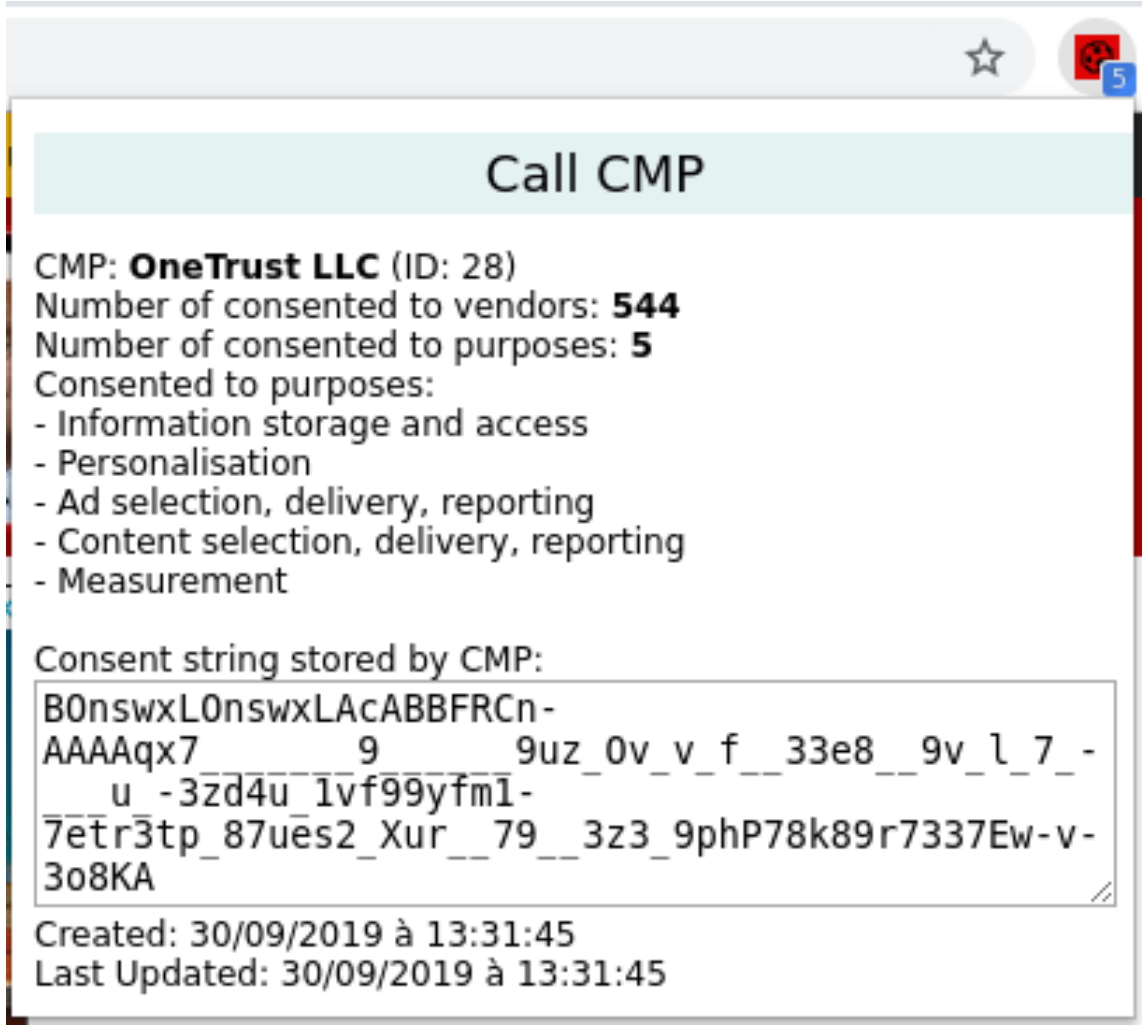

**Figure 25** Browser extension “Cookie Glasses” showing consent registered by cookie banners of IAB Europe's Transparency & Consent Framework

**How to detect violations?** Similar to detecting violations of R14 “Post-consent registration” requirement, assessing this requirement is possible only by a combination of manual and technical means and only on websites where it is known how the consent is registered by the publisher.

Manual verification is needed for the evaluation of the user interface in the banner. For example, a human operator can decide to refuse consent and then compare this choice to the consent registered in the browser. However, since there is no standardized way to structure and store consent, it is not possible to detect violations automatically on all websites.

Matte et al.[179] demonstrated verification of correct consent registration requirement on websites that contain banners of the IAB Europe's Transparency and Consent Framework. Matte et al. used the following procedure: in a browser with an empty session, open the target webpage. After giving a consent on the cookie banner interface, analyze whether the consent given in the user interface of the banner is consistent

[177] cf. Greek DPA (n 42).
[178] cf. Matte et al. (n 153).
[179] cf. Matte et al. (n 152).

with the consent (a) stored in the dedicated browser cookie, or (b) collected from querying the Consent Manager Provider (cookie banner provider in the terminology of IAB Europe TCF).

Notice that technical analysis of stored consent is only possible for consent stored on the "client-side", that is in the user's browser. Website publishers also need to store consent on their own servers, in order to prove consent collection upon request. Server-side storage of consent cannot be verified due to absence of access to such servers.

### 5.4 Readable and accessible

The requirements explained in this section refer to the consent request (i.e. how consent should be collected by the data controller). We derived the requirement "readable and accessible" consent request from our analysis of the following provisions: Article 7(2) GDPR, and its further articulation in Recitals 32 and 42 of the GDPR. Herewith we transpose their excerpts for legibility purposes. In the wording of Recital 32, "if the data subject's consent is to be given following a request by electronic means, the request must be clear, concise and not unnecessarily disruptive to the use of the service for which it is provided". Pursuant to Recital 42, "a declaration of consent pre-formulated by the controller should be provided in an intelligible and easily accessible form, using clear and plain language and it should not contain unfair terms". From these three precepts, we derive that the request for consent should be clearly distinguishable from any other matters, in an intelligible and easily accessible form, using clear and plain language. These requirements are shown in Table 18.

The GDPR (Article 7(2)) mandates that a failure to comply with these requisites constitutes an infringement and renders a non-binding consent, which signals the *practical effect* of these validity conditions. These four requisites were mostly elaborated in the 29WP Guidelines on Transparency[180] and relate to how information should be disclosed.[181]-[182] In this work, we adapt the content of these four elements described in these guidelines and apply them as low-level requirements for a valid consent request, as explained in the following subsections.

We included the "readable and accessible" (and the low-level) requirement for a consent request considering two main factors:

i) The reasonable expectations of data subjects (which are, in general, laymen), as evoked by the recent jurisprudence of the European Court of Justice, in Planet49 judgment[183] that reads: "due to the technical complexity of cookies, the asymmetrical information between provider and user and, more generally, the relative lack of knowledge of any average internet user, the average internet user cannot be expected to have a high level of knowledge of the operation of cookies"; [184]
ii) The average user needs specific information to easily determine the consequences of any consent he might give, in an intelligible, clear way, where layered[185] information is amenable.

[180] Article 29 Working Party, "Guidelines on transparency under Regulation 2016/679" (WP260 rev.01, 29 November 2017).
[181] "The GDPR puts several requirements for informed consent in place, predominantly in Article 7(2) and Recital 32. This leads to a higher standard for the clarity and accessibility of the information", cf. 29WP (WP259 rev.01) (n 4) 14.
[182] The 29WP mentions that "transparency requirements in the GDPR apply irrespective of the legal basis for processing and throughout the life cycle of processing", Article 29 Working Party (n 180) 6.
[183] cf. Planet49 Judgment (n 11) para 114.
[184] On the terminology in the area of consumer protection, see Directive 2011/83/EU on consumer rights, ELI: http://data.europa.eu/eli/dir/2011/83/2018-07-01. See, by way of example, the judgments of the following cases: Case C-485/17 *Verbraucherzentrale Berlin eV v Unimatic Vertriebs GmbH* [2018] ECLI:EU:C:2018:642, para 44; Case C-44/17 *Scotch Whisky Association v Michael Klotz* [2018] ECLI:EU:C:2018:415, para 47; Case C-210/96 *Gut Springenheide and Tusky v Oberkreisdirektor des Kreises Steinfurt* [1998] ECLI:EU:C:1998:369, para 31.
[185] The Handbook on European Data Protection Law refers to the "accessibility in an online environment", as follows:

> "The quality of the information is important. Quality of information means that the information's language should be adapted to its foreseeable recipients. Information must be given without jargon, in a clear and plain language that a regular user should be able to understand. Information must also be easily available to the data subject (…). Accessibility and visibility of the information are important elements: the information must be clearly visible and prominent. In an online environment, layered information notices may be a good solution, as these allow data subjects to choose whether to access concise or more extensive versions of information",

European Agency for Fundamental Rights, "Handbook on European Data Protection Law" (2018 edition) (Publications Office of the European Union, 2018) 147.

**Table 18** Derived low-level requirements and their sources

| Requirements | | Sources at low-level requirement | | |
|---|---|---|---|---|
| **High-Level Requirements** | **Low-level Requirements** | **Binding** | **Non-binding** | **Interpretation: Legal (L) or Computer Science (CS)** |
| **Readable and accessible** | **R16 Distinguishable** | 7(2) | 29WP, Recital 42; DPAs: Belgium, Spanish, UK, Irish | - |
| | **R17 Intelligible** | 7(2) | 29WP, Recital 42. DPAs: UK, Spanish, Danish | - |
| | **R18 Accessible** | 7(2) | 29WP, Recital 42; DPAs: UK, Irish, Spanish, Belgium | - |
| | **R19 Clear and plain language** | 7(2) | 29WP, Recital 42; DPAs: UK, Spanish, Danish | - |
| | **R20 No consent wall** | - | DPAs: UK, French | L based on 7(2) and recitals 32, 42 |

### R16 Distinguishable

The requirement "distinguishable", according to Article 7(2) and the 29WP Guidelines on Transparency (WP260 rev.01),[186] means that the consent request should be clearly differentiated from other non-related information, such as contractual provisions or general terms of use, warning boxes, etc.

R16 is related to R3 ("No merging into a contract"), however, R16 is more general (distinguishable from the other matters), while R3 is an instantiation of one such matter – the existence of a contract.

| **Requirement** | **Distinguishable**<br>The consent request should be clearly differentiated from other non-related information. |
|---|---|
| Violation | When the consent request is mixed with other matters, like terms of use, warning boxes, among others. |

**How to detect violations?** An example of violation shown in R3 shows also a violation of this requirement. Both R16 and R3 can be verified either manually or by using NLP tools that analyze structural properties of the text – we detail the discussion on their verification in section 6.

### R17 Intelligible

Intelligible means, in the context of a consent request, that the collection of consent "should be understood by an average member of the intended audience" (WP260 rev.01).

| **Requirement** | **Intelligible**<br>The consent request should be understood by any user. |
|---|---|
| Violation | When the consent request is not understood by average users. |

**How to detect violations?** Intelligible consent is dependent on the understandability of the target audience – composed by users who try to make their choice in the consent banner interface. Therefore, such requirement can only be analyzed by means of user studies, questioning users about their understanding of various types of explanations in cookie banners. We provide a deeper analysis of verification for this requirement in section 6.

### R18 Accessible

The consent request should be easily accessible, which means that "the data subject should not have to seek out" for the settings to customize her preferences; and "it should be immediately apparent to them where and how this information can be accessed" (WP260 rev.01).

Violations of R18 can be noted, for example, whenever there is a decoupled choice, meaning that a consent request and the settings are located far from the primary interaction with the banner, or when it is required unnecessary user effort (such providing different opt-out links) for users to make a choice.

[186] cf. 29WP WP260 rev.01 (n 180).

| Requirement | **Accessible**<br>The consent request should be easily accessible to the data subject. |
|---|---|
| Violation | When the sliders or settings are far from the settings of a banner. |

**How to detect violations?** This requirement is also subjective to the capacities to find information of the target audience, and therefore can only be verified with user studies. We discuss it in detail in section 6.

**R19 Clear and plain language**

The information of a consent request should use clear and plain language. The 29WP (WP260 rev.01) reads that information should be presented:

> "in as simple a manner as possible, avoiding complex sentence and language structures. The information should be concrete and definitive; it should not be phrased in abstract or ambivalent terms or leave room for different interpretations. In particular the purposes of, and legal basis for, processing the personal data should be clear. (…) The information provided to a data subject should not contain overly legalistic, technical or specialist language or terminology".

While R17 is related to R19 (consent request should be intelligible, in the sense of being understood by an average user), however, R19 refers concretely to the language used in the *text* of a consent request. We have observed that the text within cookie banners is generally prone to the existence of manipulative dark patterns. We name a few violations of this requirement for illustration purposes:

1. Questioning the choice of refusing tracking, e.g. "Would you re-consider?", or "Are you sure?",
2. Use positive framing regarding one choice (to accept tracking), while glossing over any potentially negative aspects of that same choice, e.g. "We value your privacy","We care about your privacy","Go to the website",
3. Use of technical language and legal jargon, e.g. "This site uses cookies", instead of "This site collects your data",
4. Use of compliance wording might influence the user towards accepting consent, e.g. mentioning a Data Protection Authority recommendation, or making salient regulations.

| Requirement | **Clear and plain language**<br>Concrete, explicit, clear |
|---|---|
| Violation | Use of positive framing, technical and overly legalistic language, use of compliance wording, questioning the choice of refusal, use of abstract and ambivalent terms. |

**How to detect violations?** This requirement is also subjective to the capacities to find information of the target audience, and therefore can only be verified with user studies. We discuss it in detail in section 6.

In this section 5.4, we do not present examples of compliant examples and violations of requirements R17, R18, and R19 because they are subjective to the perception and even biases of the target audience, which are users.

**R20 No "consent wall"**

Recital 32 of the GDPR states that the consent request should not be unnecessarily disruptive to the use of the service for which it is provided. In our own opinion of this Recital, *unnecessary disruption* to the use of a website/app reflects a common practice that we name "consent walls". *Consent walls* consist of a mechanism that forces the user to make a choice, by blocking access to the website/app until the user expresses her choice regarding consent.

For publishers that provide a free access to the website independently of the user's choice, a consent wall is discouraged because it forces the user to make a choice that does not influence her.

*Differently from tracking wall*, this practice allows the user to make a choice between acceptance and refusal. A consent wall appears to be unnecessarily disruptive to the use of a website.

Notice that "on interpreting "unnecessarily disruptive" consent request: "it may be necessary that a consent request interrupts the user experience to some extent to make that request effective". In line with Leenes,[187] this disruption could merely occur depending on the user's choice, e.g. a certain functionality may be

[187] cf. Leenes (n 82).

lacking, such as a forum if the user does not accept social media cookies, or be replaced by other content, such as behavioral advertisements being replaced by other types of advertisements.

We take the view that if there are other ways to display the overlay without blocking the access to the service, then such banner is preferred to a consent wall. In practical settings, the website should still be accessible even if the user did not respond to the consent request. If there are other ways to show the banner without blocking (disturbing) the access to the service, or disrupting the user experience, then it is preferred to a consent wall. Thus, we argue that consent walls do not configure a valid design for consent mechanisms they are confusing and unnecessarily disruptive of the user experience. Other consent design implementations could be sought while engaging the users.

This requirement has even stronger practical significance with mobile devices. Its small configuration implies that consent walls can be more obvious while users do not consent. Relevantly, the ICO[188] emphasizes the user experience along with the electronic consent request implementation:

> Message boxes such as banners, pop-ups, message bars, header bars or similar techniques might initially seem an easy option for you to achieve compliance. However, you need to consider their implementation carefully, particularly in respect of the implications for the user experience. For example, a message box designed for display on a desktop or laptop web browser can be hard for the user to read or interact with when using a mobile device, meaning that the consents you obtain would be invalid (…) so you need to consider how you go about providing clear and comprehensive information without confusing users or disrupting their experience.

The CNIL (in paragraph 38 of its draft recommendation for the use of trackers) states that in the absence of any manifestation of choice to either accept or reject, no trackers should be written. Even if the statement below does not explicitly discourage the use of consent walls, it seems to be inclined thereto.

> "nothing prohibits the person responsible for the processing operation(s) to provide the user with the possibility of not making any choice and delaying his or her decision, as long as the user is given the choice between acceptance and refusal. The situation in which the user does not express any positive choice must be distinguished from the situation of refusal. In the absence of any manifestation of choice (neither acceptance nor refusal), no trackers requiring consent should be written. The user could then be asked again as long as he or she does not express a choice".

The requirement box summarizes the "No consent wall" requirement.

| **Requirement** | **No consent wall**<br>The website needs to be accessible even if the user did not respond to request for consent.<br>If there are other ways to show the banner without blocking (or disturbing) the access to the service, then it is preferred than a consent wall |
|---|---|
| Violation | "Consent wall" that blocks the service before the user accepts or rejects consent |

**Example.** Figure 26 depicts a consent wall displayed by the website fandom.com. This consent wall allows to accept or reject consent. Figure 27 shows a cookie banner that is compliant with the "No consent wall" requirement on a desktop version of the website but becomes non-compliant on a mobile device because the cookie banner covers the majority of the screen. Moreover, as the cookie banner of LBC website only proposes to accept consent, it is non-compliant with the "Configurable banner" requirement (Unambiguous consent). As a result, the mobile version of the LBC website has a cookie banner that forces the user to accept the data collection and at the same time blocks access to the website, which violates the "No tracking wall" requirement.

[188] cf. UK DPA (n 26) 28.

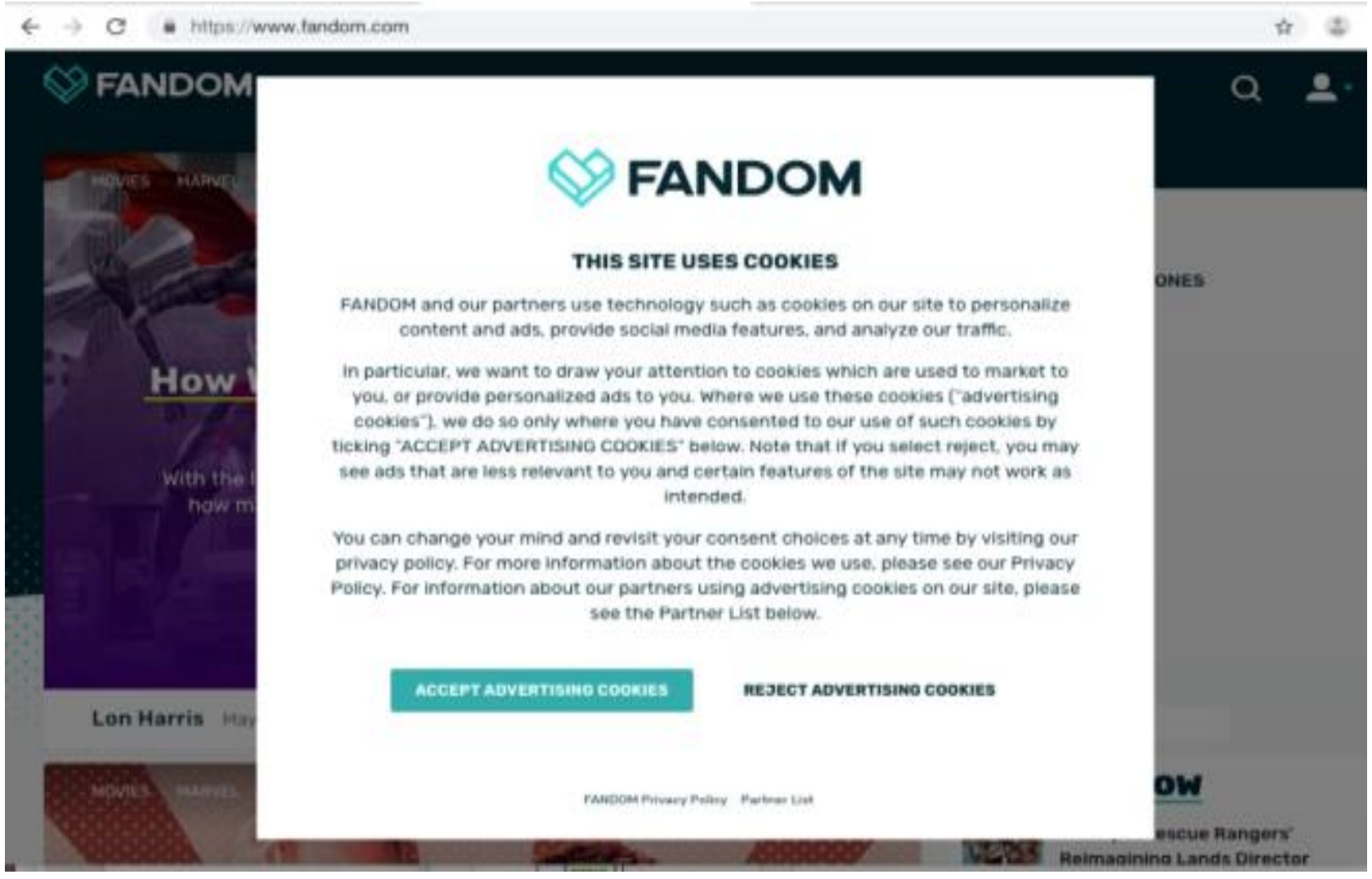

**Figure 26** Violation of the "No consent wall" requirement (<[www.fandom.com/](http://www.fandom.com/)> accessed 17 May 2019).

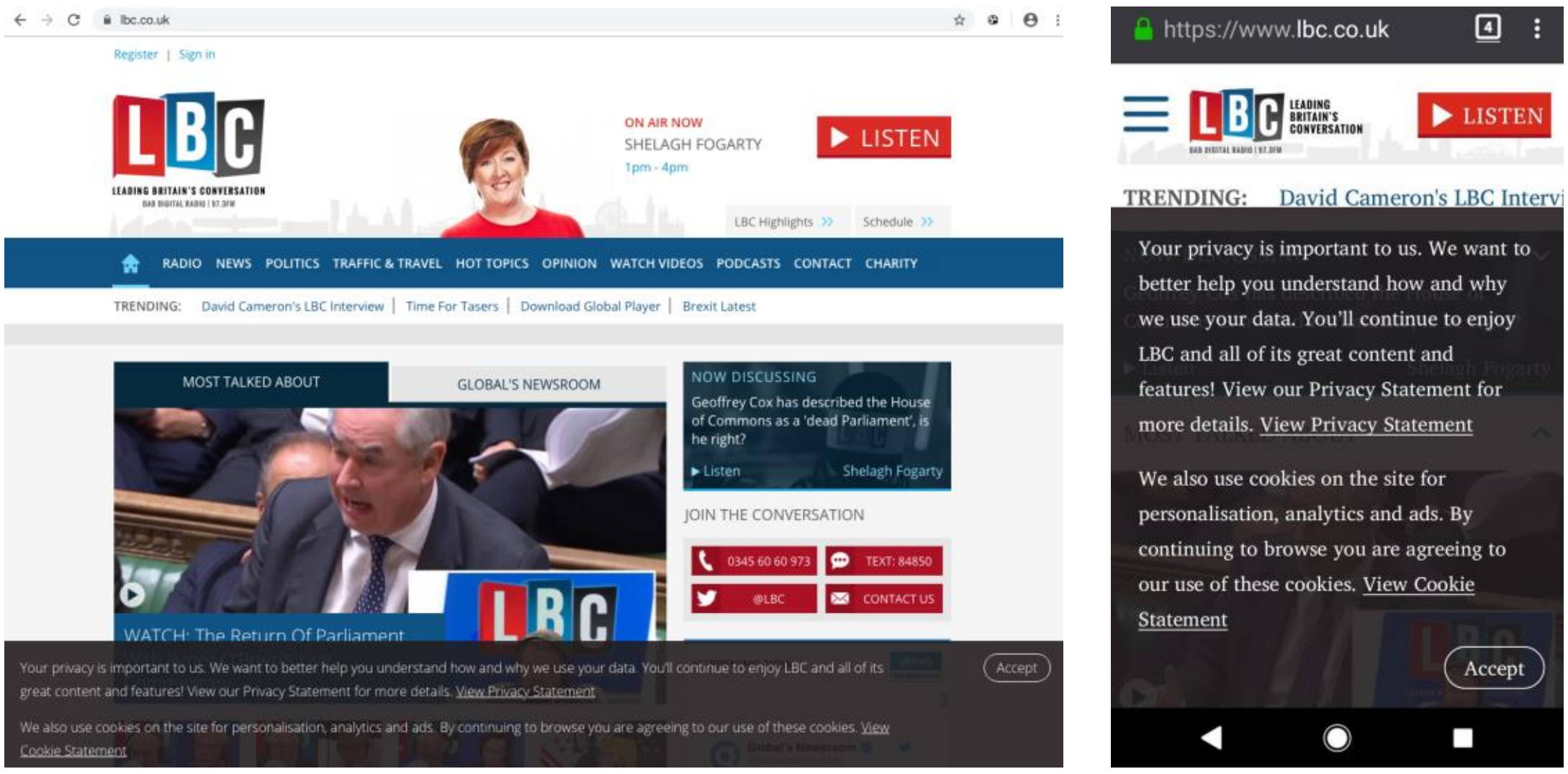

**Figure 27** The Desktop version of the LBC website does not violate the "No consent wall" requirement, however the mobile version of the same website does (<[www.lbc.co.uk/](http://www.lbc.co.uk/)> accessed 25 September 2019).

**How to detect violations?** Detection of such violation is possible manually, by evaluating whether the cookie banner blocks access to the website or not. Currently, technically detecting this violation is challenging because there is no specification that defines which part of the website is a cookie banner. However, techniques based on separation of HTML elements[189] and keyword searches in the cookie banner text could be envisioned. Also, crowd-sourced lists for banners blocking such as the Easylist Cookie List[190] or the Consent-o-Matic tool[191] could be used for that purpose.

The procedure for detection would be the following:

5. Detect whether the element responsible for the cookie banner is displayed;
6. Detect its size relative to the screen size, on different screen sizes (to account for both desktop and mobile)
7. Detect whether the other elements of the page are reachable, or e.g. blocked by an overlay.

We therefore conclude that the technical development of automated tools is not complex, however its accuracy has to be measured at scale to evaluate its effectiveness.

[189] Arunesh Mathur, Gunes Acar, Michael J. Friedman, Elena Lucherini, Jonathan Mayer, Marshini Chetty, Arvind Narayanan. Dark Patterns at Scale: Findings from a Crawl of 11K Shopping Websites. ACM CSCW 2019.

[190] https://fanboy.co.nz/fanboy-cookiemonster.txt

[191] Nouwens, Midas, Ilaria Liccardi, Michael Veale, David R. Karger and Lalana Kagal. "Dark Patterns after the GDPR: Scraping Consent Pop-ups and Demonstrating their Influence." ACM CHI 2020.

## 5.5 Revocable

The GDPR establishes the right of the data subject to withdraw consent in Art. 7(3). We have made the "withdrawal of consent" an additional explicit requirement due to the practical implications of this right.

Primarily, Article 7(3) explicitly states this right as one of the "conditions for consent", or condition for consent validity. Among other provisions, Recital 42 mentions the revocability of consent. The 29WP (WP259 rev.01)[192] confirms that the GDPR gives a prominent place to the withdrawal of consent. The German DPA[193]-[194] also makes salient the requirement for revocability. It states that "anyone using cookies to analyze and track user behavior for advertising purposes or have them analyzed by third parties generally requires the informed, voluntary, prior, active, separate and revocable consent of the user".

Regarding the *easiness* to withdraw consent, Article 7(3) declares that it shall be as easy to withdraw as to give consent. This easiness attribute also relates to withdrawing consent *without detriment*. Easiness and without detriment entail an obligation of the controller at different levels: cost, simplicity of the procedure and finally, service level.

i. *Cost*: free of charge;[195]

ii. *Service level:* without lowering service levels (29WP WP259 rev.01);[196]

iii. *Simplicity of the procedure*: using simple and easily accessible mechanisms, not burdensome.[197] We claim that withdrawal should be done by the *same means* it was obtained in the first place, without the need to ask the user to state the reason for withdrawing consent.[198] We additionally suggest that the "withdrawal tool" should be named appropriately and should be standardized for all environments (including web and mobile). This positioning on the *same means* was already endorsed by the 29WP and other DPAs, as reflected in Table 19.

**Table 19** Positioning of the 29WP and DPAs on the easiness to revoke consent

| Stakeholders | Positioning regarding the easiness of the procedure to revoke consent |
|---|---|
| 29WP (WP259 rev.01)[199] | "the GDPR does not say that giving and withdrawing consent must always be done through the same action. However, when consent is obtained via electronic means through only one mouse-click, swipe, or keystroke, data subjects must, in practice, be able to withdraw that consent *equally as easily*. Where consent is obtained through use of a service-specific user interface (for example, via a website, an app, a log-on account, the interface of an IoT device or by e-mail), there is no doubt a data subject must be able to withdraw consent via *the same* electronic interface, as switching to another interface for the sole reason of withdrawing consent would require undue effort." |
| French DPA[200] | proposes as criteria for easiness: i) the time spent, and ii) number of actions required. |
| Greek DPA[201] | a user "must be able to withdraw his consent in the same manner and with the same feasibility with which he has given it". It refers that revoking should not be cumbersome and illustrates an unlawful practice "following the user's consent or decline, the user is not given any opportunity to change his / her preferences or user preferences may only be changed through his / her web browser settings". |
| Irish DPA[202] | use of an "easy tool, such as a "radio button" on your website which allows users to control which cookies are set and to allow them vary their consent at any time" |

[192] cf. 29WP (WP259 rev.01) (n 4) 21.
[193] cf. German DPA (n 10).
[194] "Since a consent is revocable, a corresponding option for revocation must be implemented. The revocation must be as easy as the granting of consent, Art. 7 (3) sentence 4 GDPR" (our translation), cf. German DPA (n 10).
[195] 29 Opinion 4/2010 (29WP 174) on the European code of conduct of FEDMA for the use of personal data in direct marketing, adopted on 13 July 2010.
[196] ibid.
[197] cf. 29WP Opinion 02/2013 (WP202) (n 132).
[198] cf. 29WP Opinion 4/2010 (WP 174) (n 195).
[199] cf. 29WP (WP259 rev.01) (n 4) 21.
[200] CNIL draft recommendation 2020 (n 95).
[201] Greek DPA, Guidelines on Cookies and Trackers (n 42).
[202] cf. Irish DPA Guidance (n 37).

| | |
|---|---|
| Danish DPA[203] | "when a website has a valid cookie consent solution with a GDPR valid cookie pop-up, the user can simply reject/decline cookies by reopening the cookie pop-up and thus the consent to cookies is withdrawn (hence no cookies are further set in the browser)". |
| Spanish DPA | a system shall be considered easy to use, for example, when users may access easily and permanently to the cookie setup or management system. |

An example of easiness without detriment is the case of a recent decision[204] of the Polish DPA against the company ClickQuickNow referred to a GDPR violation due to the fact that the mechanism for consent withdrawal, involving the use of a link included in the commercial information, did not result in a quick withdrawal. After the link was set up, messages addressed to the user were misleading. Moreover, the company forced stating the reason for withdrawing consent. Furthermore, failure to indicate the reason resulted in discontinuation of the process of withdrawing consent.

The right to withdrawal does not have retroactive effects, meaning that it does not apply for processing that had taken place before withdrawal. Revocation cannot affect nor devalue already conducted research, decisions or processes previously taken on the basis of this data. This reasoning is supported by Article 7(3) of the GDPR that lays down that "the withdrawal of consent shall not affect the lawfulness of processing based on consent before its withdrawal". Moreover, the 29WP (WP187)[205] supports also this view that "withdrawal is exercised for the future, not for the data processing that took place in the past, in the period during which the data was collected legitimately".

Moreover, revocability offers also a possibility for the user to make subsequent changes/configurations to his preferences, at any time. In this line, the 29WP (WP 208)[206] mentions that revocability is "an option for the user to subsequently change a prior preference regarding cookies". In another opinion, the 29WP (WP 259 rev.01)[207] ascertains that "consent is a reversible decision".

In this line, we decompose the revocability requirement in Table 20, explicitly emphasizing the need to communicate withdrawal of consent to all the parties that have previously received it.

**Table 20** Derived low-level requirements and their sources

| **Requirements** | | **Sources at low-level requirement** | | |
|---|---|---|---|---|
| **High-Level Requirements** | **Low-level Requirements** | **Binding** | **Non-binding** | **Interpretation: Legal (L) or Computer Science (CS)** |
| **Revocable** | **R21 Possible to change in the future** | 7(3) | 29WP; DPAs: French, Greek, Irish, Danish, Spanish, German | - |
| | **R22 Delete "consent cookie", communicate to third parties** | - | - | CS |

### R21 Possible to change in the future

Under the revocability requirement, we define the low-level requirement on the possibility to withdraw consent in the future.

| | |
|---|---|
| Requirement | **Possible to change in the future**<br>The website should give an opportunity to withdraw consent after it has been given. The banner should allow the user to change the consent with easiness, by the same means, without detriment |
| Violation | It is not possible to withdraw consent by the same means it was asked;<br>It is cumbersome to revoke – the means of withdrawing are more complex that initial consent;<br>It is rather complex to understand for an average user how to remove cookies, and it is only accessible to the technical experts if other browser storages, such as HTML5 localStorage or cache should be cleaned. Moreover, there are no means to withdraw from browser fingerprinting. |

[203] Danish DPA, "Guide on consent" (n 77).
[204] Polish DPA, "Polish DPA: Withdrawal of consent shall not be impeded" (2019) <https://edpb.europa.eu/news/national-news/2019_en> accessed 7 May 2020.
[205] cf. 29WP (WP187) (n 54) 33.
[206] cf. 29WP (WP208) (n 55) 2.
[207] cf. 29WP (WP 259 rev.01) (n 4) 5.

| | |
|---|---|
| | Revoking poses a delay, while positive consent was instantaneous. |

**Examples.** Figures 28 and 29 show compliant banners to this requirement based on the possibility to change preferences in the future. The banner from the faktor.io website offers users the possibility to review and manage their choices by clicking the fingerprint icon on the bottom right of the screen. This icon is available on every page of the site.

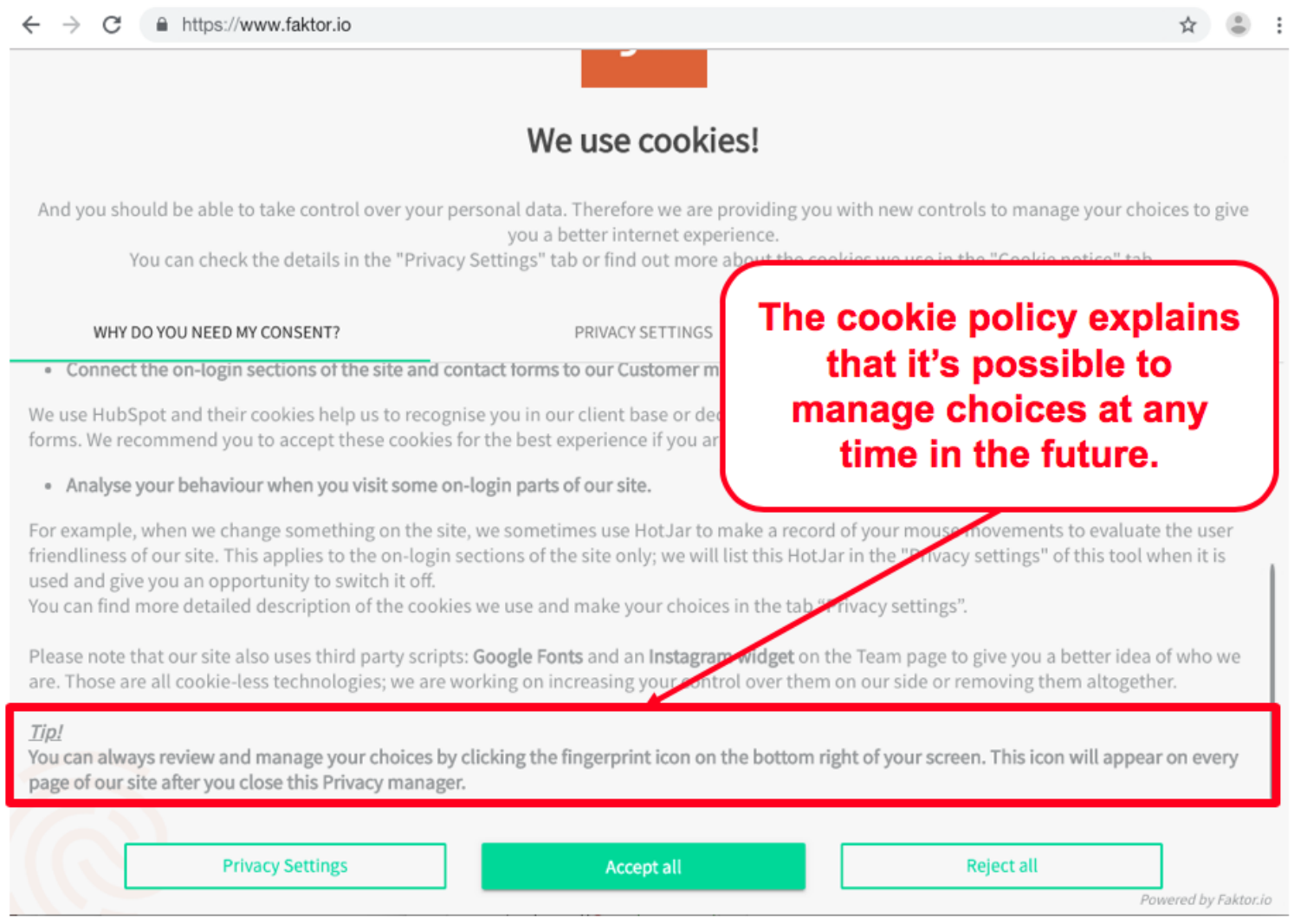

**Figure 28** Compliance with "Possible to change in the future" requirement (<https://www.faktor.io> accessed 17 May 2019).

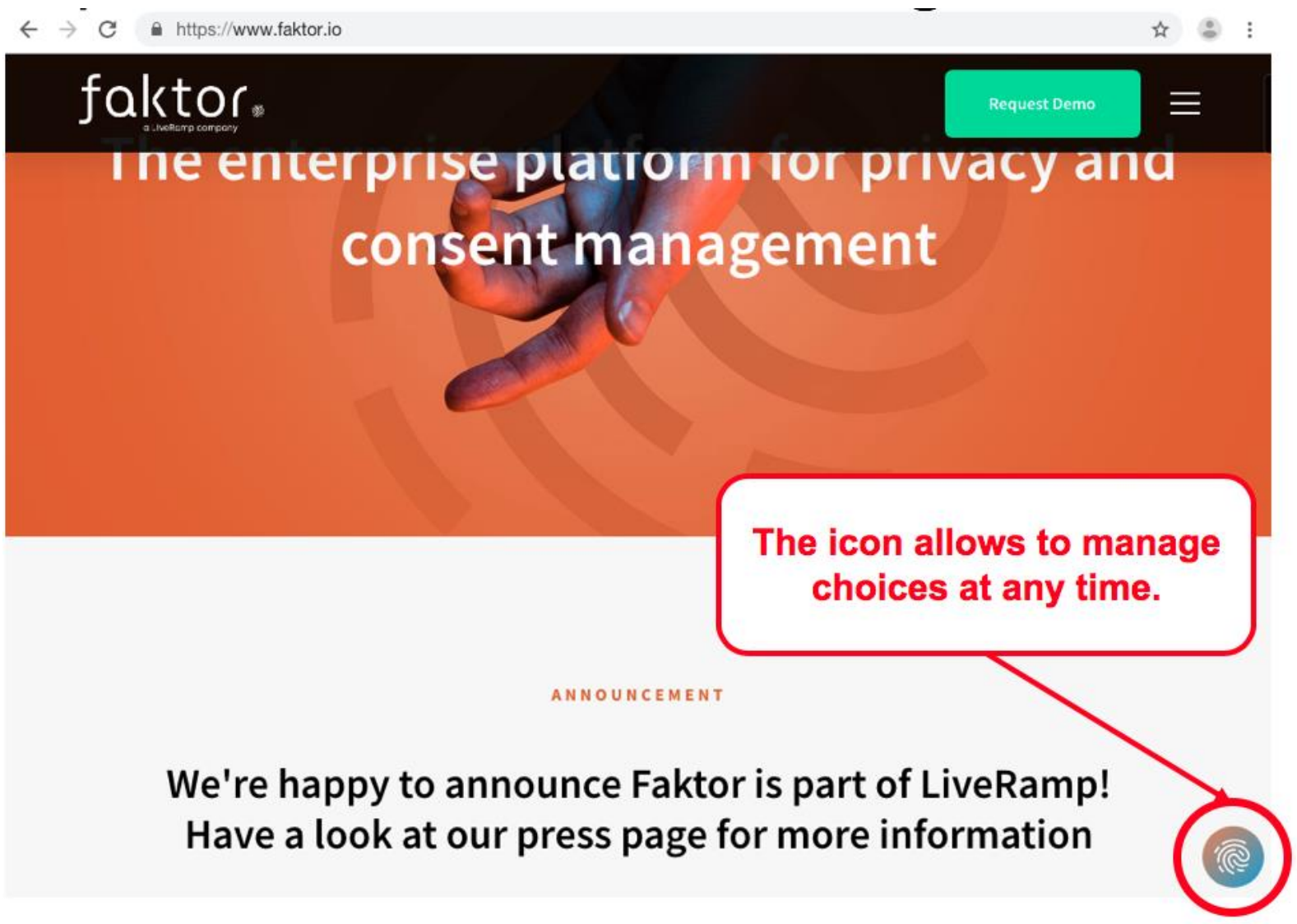

**Figure 29** Compliance with the "Possible to change in the future" requirement (<https://www.faktor.io>  accessed 17 May 2019).

**How to detect violations?** Detection of this violation requires a manual analysis of the banner's interface, by evaluating whether there is a mean to change the consent after it has been given and how easy it is to revoke consent. Only standardized consent design can enable technical means to detect violations.

### R22 Delete “consent cookie” and communicate to third parties

When the user revokes his consent, no BTT cab be further stored/read in the browser. Hence, revoking consent has two technical consequences: blocking and posterior deletion of cookies[208]-[209] in the user’s browser, and as such, data processing will no longer occur. The CNIL[210] states that once the consent is revoked, both the reading and the deposit of new cookies should be blocked. The 29WP (WP 259 rev.01)[211] reasons in the same line,

> As a general rule, if consent is withdrawn, all data processing operations that were based on consent and took place before the withdrawal of consent - and in accordance with the GDPR - remain lawful, however, the controller must stop the processing actions concerned. If there is no other lawful basis justifying the processing (e.g. further storage) of the data, they should be deleted by the controller (art. 17(1)(b) and (3) GDPR. (…) Controllers have an obligation to delete data that was processed on the basis of consent once that consent is withdrawn, assuming that there is no other purpose justifying the continued retention.

Pursuant to the above analysis, we defined the technical low-level requirement that the publisher should delete the registered consent and communicate this withdrawal to all the third parties who have previously received consent.

| | |
|---|---|
| **Requirement** | **Delete “consent cookie” and communicate to third parties**<br>When consent is revoked, the publisher should delete the “consent cookie” and communicate the withdrawal to all the third parties who have previously received consent. |
| Violation | When the “consent cookie” is not deleted, and the publisher does not communicate to third parties that have received the consent |

**Example.** We cannot provide an example for this requirement. As of June 2020, cookie banners rarely give users a way to modify their choice, and when they do, it is still unclear whether this change is actually communicated to a third party.

**How to detect violations?** Detection of such violation is a complex task because it requires checking whether the publisher has communicated the withdrawal of consent to all the third parties who have received it in the first place. As of today, there is no system that would be able to certify this because communication of consent (and of its withdrawal) does not have a standard technical implementation. If consent storage and communication is standardized and is observable in the web browser, technical tools could be devised for complete transparency and verification of this requirement.

# 6 Detection of violations for requirements based on Natural Language Processing and on user perception

This section refers on the mechanisms for detection of violations of requirements that depend on natural language processing (NLP) and user perception of the statements in natural language. We merge these requirements into three groups based on types of techniques that can be used to assess them:

1. Requirements based on the *presence* of information
    - R6 Accessibility of information page
    - R7 Necessary information on BTT
    - R8 Information on consent banner configuration
    - R9 Information on the data controller
    - R10 Information on rights

[208]The European Commission, it its portal, states that “data is deleted unless it can be processed on another legal ground (for example storage requirements or as far as it is a necessity to fulfill the contract”, “What if somebody withdraws their consent?” <https://ec.europa.eu/info/law/law-topic/data-protection/reform/rules-business-and-organisations/legal-grounds-processing-data/grounds-processing/what-if-somebody-withdraws-their-consent_en> accessed 7 May 2020.

[209] It is noticeable that the request for revoking consent does not imply data erasure. For the data to be erased, the data subject needs to exercise this right to erasure. However, revoking consent should imply deletion of data as an immediate consequence.

[210] cf. CNIL (n 36).

[211] cf. 29WP (WP 259 rev.01) (n 4) 22.

2. Requirements that rely on the *distinguishability and structure* of information

- R3 No merging into a contract
- R16 Distinguishable

3. Requirements that can be evaluated based on *user perception* and *understanding*.

- R6 Accessibility of information page
- R17 Intelligible
- R18 Accessible
- R19 Clear and plain language

**Detection of violations for requirements based on the *presence* of information and *distinguishability and structure* of information.**

The detection of violations related to the presence of information can be done manually but can be extremely time-consuming in case the required information is scattered across several pages of a privacy policy text, and privacy policies of included third parties. Libert[212] has measured that the average time to read both a given site's policy and the associated third-party policies exceeds 84 minutes. Additionally, the text of privacy policies is often written with complex linguistic structured. For instance, Sanchez et al.[213] used the Flesch Reading Ease Score (FRES) and the Flesch-Kincaid Reading Level (FKRL), both used by legislators and government agencies, to measure the readability of privacy policies. Libert[214] proposed a formula to compute the average time required by users to read a policy. Therefore, manual analysis of privacy policies should be discouraged and instead automatic means to analyze privacy policies must be considered.

*Natural Language Processing (NLP*) tools have been conceived to analyze privacy policies. For instance, Libert[215] used an approach based on keywords to extract information from privacy policies. Following Brodie et al[216], The Usable Privacy Policy project[217] combines technologies, such as crowd sourcing, to develop browser plug-in technologies to automatically interpret policies for users. Ammar et al.[218] performed a pilot study, thus deriving and collecting a corpus of website privacy policies. This corpus has later been used by Harkous et al.[219] in a more complex deep-learning-based method within the Polisis tool that automatically extracts information flows described in privacy policies. A similar approach has been taken by Zaeem et al.[220] proposing PrivacyCheck browser extension that analyzes privacy policies with data mining. A further recent report compares the results of Polisis and PrivacyCheck on hundreds of privacy policies.[221] For a complete overview of modelling of privacy policies and automatic analysis of privacy policies at scale, a recent survey by Morel and Pardo[222] describes recent tools to analyze privacy policies at scale. Nevertheless, NLP tools have not been developed so far in the context of consent requirements and further research by NLP experts is needed to approach legal requirements presented in this paper.

[212] Cf Libert (n 128)
[213] Iskander Sanchez-Rola, Matteo Dell"Amico, Platon Kotzias, Davide Balzarotti, Leyla Bilge, Pierre-Antoine Vervier, and Igor Santos, "Can I Opt Out Yet? GDPR and the Global Illusion of Cookie Control" (ACM Asia Conference on Computer and Communications Security (AsiaCCS "19), Auckland, New Zealand, 2019).
[214] Cf Libert (n 128).
[215] Cf Libert (n 128)
[216] Carolyn A. Brodie, Clare-Marie Karat, and John Karat. 2006. An empirical study of natural language parsing of privacy policy rules using the SPARCLE policy workbench. In Proceedings of the second symposium on Usable privacy and security (SOUPS '06). Association for Computing Machinery, New York, NY, USA, 8–19.
[217] Wilson Shomir, Florian Schaub, Aswarth Abhilash Dara, Frederick Liu, Sushain Cherivirala, Pedro Giovanni Leon, Mads Schaarup Andersen, Sebastian Zimmeck, Kanthashree Mysore Sathyendra, N. Cameron Russell, Thomas B. Norton, Eduard H. Hovy, Joel R. Reidenberg and Norman M. Sadeh. "The Creation and Analysis of a Website Privacy Policy Corpus." *ACL* (2016).
[218] Waleed Ammar, Shomir Wilson, Norman Sadeh, Noah A. Smith, "Automatic categorization of privacy policies: A pilot study," Tech. Rep.
[219] Harkous, Hamza, Kassem Fawaz, Rémi Lebret, Florian Schaub, Kang G. Shin and Karl Aberer. "Polisis: Automated Analysis and Presentation of Privacy Policies Using Deep Learning." USENIX Security Symposium (2018).
[220] Razieh Nokhbeh Zaeem, Rachel L German, and K Suzanne Barber, "Privacycheck: Automatic summarization of privacy policies using data mining". ACM Transactions on Internet Technology (TOIT), 18(4):53, 2018.
[221] Razieh Nokhbeh Zaeem, Suzanne Barber, "Government Agencies and Companies: a Study Using Privacy Policy Analysis Tools", UTCID Report (2020) <https://identity.utexas.edu/assets/uploads/publications/Comparing-Privacy-Policies-of-Government-Agencies-and-Companies-a-Study-Using-Privacy-Policy-Analysis-Tools.pdf> accessed 18th June 2020.
[222] Victor Morel, Raúl Pardo, "Three dimensions of privacy policies," Working Paper (2019)<https://hal.inria.fr/hal-02267641> accessed 7 May 2020.

To conclude, techniques using NLP and data mining are extensively tested today in order to process and structure privacy policies. Various NLP approaches are usually compared with respect to precision and recall metrics, however this evaluation heavily depends on (a) the “ground truth”, that is how well the underlying corpus is labeled, and (b) the structure of privacy policies that are often not well organized. Nevertheless, these techniques have not been applied to assess legal requirements on consent presented in this work. Further investigation by computer scientists and legal experts are needed to assess whether these requirements are verifiable by automatic means or whether a new standardized format is required to display this information. In the NLP domain, such standardized format is often called by “controlled natural language, template, or pattern”. With such a standardized format, a rather simple algorithm could verify the presence of the different information that is required by valid consent. Additionally, requirements that are based on distinguishability and structure of information need additional set of NLP tools that can answer the question whether consent is bundled with other types of information, such as a contract. We therefore conclude that such requirements can be analyzed manually (however with significant effort), or can be partially analyzed with technical means, whose efficiency is still to be evaluated by the experts.

**Detection of violations for requirements based on *user perception* and *understanding.***
Several requirements for valid consent rely on user understanding and is heavily dependent on the target audience of a dedicated website. For example, requirement R6 “Accessibility of information page” depends on the usability of the website in question, but also on the technical ability of the user to find the information. Requirements R17, R18 and R19 are directly related to understandability of the users, but also on the technical background of users, e.g. when they are presented with statements such as “Can I have some cookies?”, “If you do not allow cookies, website functionality will be diminished” or “Opting in to data collection will enable new and easier functionality“. Such statements are often confusing to users, however in order to quantify whether these statements are intelligible (R17) and whether the language is clear and plain (R19), more structured evaluation of users’ perception and understanding is needed.

*Usable security and privacy* research area has established the standards and techniques in building users’ surveys and interviews in order to evaluate privacy perceptions, understanding and motivations of end users when it comes to privacy settings in online environment. For example, Utz et al.[223] have designed several types of cookie banners and ran a user survey to evaluate how users would interact with them depending on banners’ text, position and provided options. Nouwens et al.[224] have also investigated users’ engagement when placing controls in first- vs second-layer of a cookie banner. However, this research direction is only at its beginning, and more user studies are needed to evaluate whether cookie banners are indeed clear and well-understood by the target end users.

## 7 Discussion on shared consent

In this section, we discuss compliant scenarios related to the possibility of a shared consent. Section 7.1 refers to the shared responsibility between publishers and third parties. Section 7.2 discusses the scenario when consent itself in shared between them.

### 7.1 Shared responsibility between publishers and third parties

According to Article 5(3) of the ePD, when a main website content, fully controlled by the publisher, is setting cookies in his web domain (first-party cookies), then it will be primarily responsible for complying with the requirement to obtain an informed consent. We question if only the website publisher is obliged to display information and collect consent when third-party services are used as processors, or when third-party services act as controllers. The recurrent scenario in which multiple entities are involved in the installation of and access to an information by means of BTT is advanced by the 29WP, the ICO and the CNIL, while referring to BTT as “cookies” or “trackers”:

[223] Christine Utz, Martin Degeling, Sascha Fahl, Florian Schaub, and Thorsten Holz, “Un)informed Consent: Studying GDPR Consent Notices in the Field”, ACM SIGSAC Conference on Computer and Communications Security (CCS’19), 2019, 973-990.
[224] cf. Nouwens et al. (n 191)

- 29WP (WP171)[225] contends that a website publisher that allows third parties to place cookies shares the responsibility for information and consent (joint controllership);
- ICO[226] takes the view that where the website publisher sets third-party cookies, this same publisher and the third party are jointly responsible for ensuring that users are clearly informed about cookies and for obtaining consent. This means they are both determining the purpose and means of the processing of personal data of any user that visits the landing website. In substance, it is considerably more difficult for a third party, which has less direct control on the interface with the user, to achieve this. The ICO further instructs the need to include a contractual obligation into the agreements between publishers and third-parties on the allocation of responsibility to provide information about the third-party cookies and to obtain consent;
- CNIL[227] observes that when several parties may be involved in the user of trackers (e.g. a publisher and an advertising agency), publishers (of mobile sites or applications) are in the best position to inform users of the information on deposited trackers and to collect their consent, because of the *control* they exercise over the consent management interface, and the *direct contact* they have with the user.

Where controllers determine jointly the purposes and means of the processing, they must enter into a joint controllership agreement. Article 26 of the GDPR stipulates that both shall, in a transparent manner, determine their respective responsibilities, including which party provides information and obtains consent from the users.

Lastly, a data processor, in this context, is defined as an entity which installs information and/or has access to information stored on a user device exclusively on behalf of a data controller, without re-using the data collected via the tracker for the processor's own purposes. In such case, the parties must enter into a data processing agreement under Article 28 of the GDPR.

Real-Time Bidding (RTB) scenarios constitute a grey area: in RTB, publishers often have no knowledge about the specific third parties that a data subject might see, as they change between each website request. The CNIL (paragraph 26) states that a publisher does not have to show a banner again every time the list of third parties changes, when changes are not significant. On the other hand, the ICO[228] strongly criticized RTB because of the lack of transparency, due to the fact that publishers cannot have information about all the third parties they're sending user information to, and because of the opaque nature of the industry itself.

### 7.2 Shared consent between publishers and third parties

It is apparent from the case law from the CJEU (in its two decisions of Tele 2 and Deutsche Telekom, that we adapt to this context) that consent can be shared among publishers, insofar the processing operations pursues the same purposes, and that the user was informed thereof, as analyzed in section 5.3.1 (in point ii. consent "not required per publisher"). From these legal sources, we reason that if consent is collected in a lawful way, consent can be shared. Note that this observation is not explicitly prohibited by the law-maker.

From practical side, however, such reasoning raises questions about shared responsibility of the data controllers and, most importantly, implies reliance and trust *on the way* consent was collected by either other publishers or providers of third party content. Below we foresee a practical scenario of a website, in which third-party content a priori does not require consent, but merged with BTT that requires consent.

**Example of shared consent.** Imagine a user visiting two hypothetical websites: search.com and info.com. Figure depicts user's browser and its interaction with the server of search.com:
1) the user visits a website search.com, where a cookie named SID of search.com is placed in the user's browser. This cookie is used for advertising purposes, and hence requires consent. Let's assume a valid consent was collected by search.com before placing of cookies in the user's browser.
2) the user visits the website info.com, and it contains a customized search engine from search.com. Therefore, while visiting the website info.com, the user's browser automatically sends a request to search.com to fetch the needed functional content, i.e. the customized search engine. Upon this request, the browser also automatically attaches the cookie SID of search.com. Therefore, search.com receives its advertising cookie SID when the user visits the website info.com.

[225] Article 29 Working Party, (WP 171) (n 33).
[226] cf. UK DPA (n 26).
[227] cf. CNIL (n 36).
[228] UK DPA, "Update report into adtech and real time bidding" (2020) <https://ico.org.uk/media/about-the-ico/documents/2615156/adtech-real-time-bidding-report-201906.pdf>

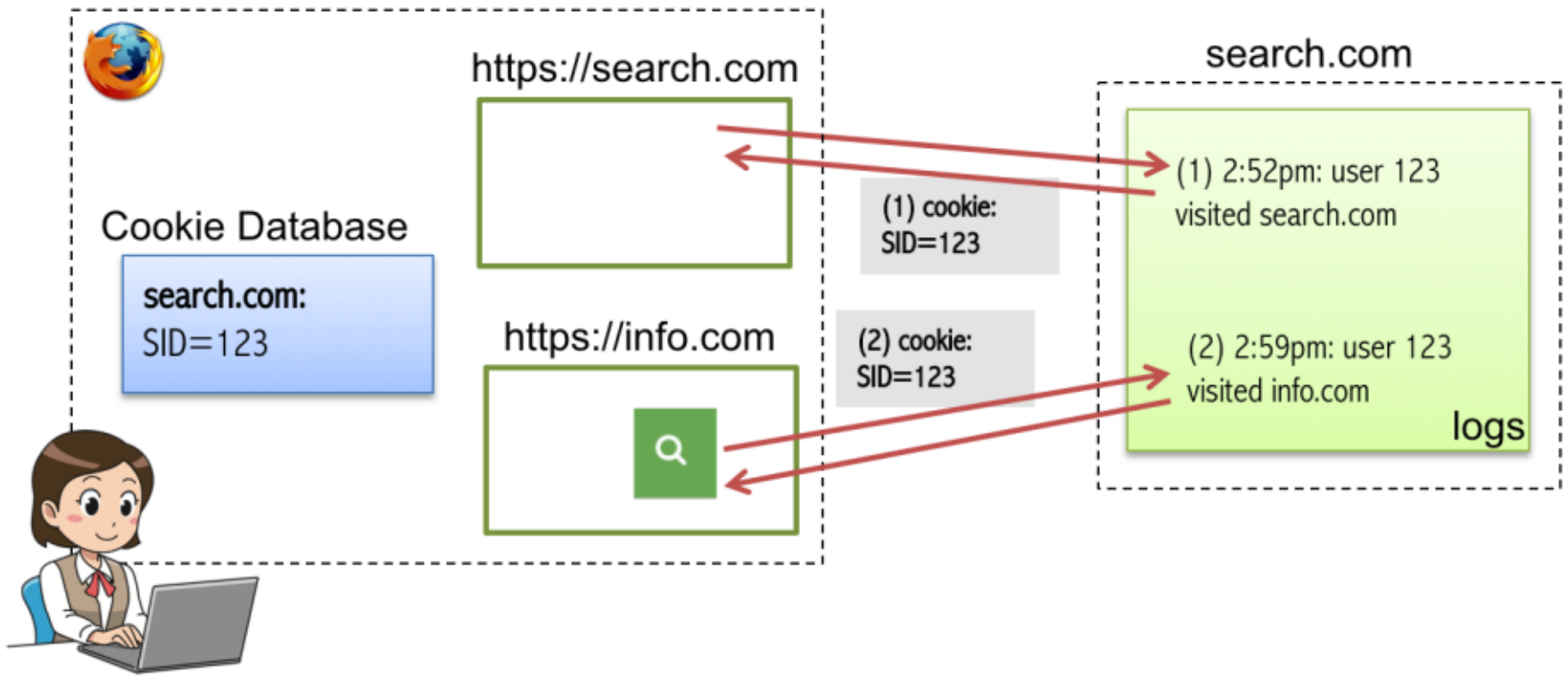

**Figure 30** Example of shared consent.

In this hypothetical scenario, let us analyze how info.com can be compliant with the requirements of a valid consent:

- The publisher of the website info.com decides to collect its own consent for the search.com's advertising cookie SID. To be compliant, the consent should be collected before cookies are sent (see R2 "Prior to sending an identifier" requirement).
- This practice, however, prevents the loading of website's functional content before consent is given (the customized search engine from search.com is not loaded before consent is given), and hence violates requirement R4 "No tracking walls". The publisher of infor.com, therefore cannot collect a valid consent by itself without violating legal requirements and thus, has to rely on the consent collected by search.com.
- The publisher of info.com relies on the consent already collected by search.com, hence info.com has to place full trust in how search.com has collected consent. This scenario will have practical and legal consequences. If the consent is not obtained in a valid way through search.com, then the website info.com will become jointly responsible for the non-compliant consent collection.

Therefore, when a third party merges content that does not require consent (e.g. functional content, such as customized search engine of search.com) with the content that requires consent (e.g. advertising cookies of search.com), this forces the website publisher to rely on the consent collected by the third party.

**Conclusion: only a negative consent can be shared.** We expand our discussion towards scenarios where a publisher relies on other publishers, previously visited by the user, for the collection of valid consent, or on third parties (as in section 7.1). If the user gives a positive consent (i.e. allows at least one type of data processing for at least one purpose), then if the consent collection has violated at least one of the requirements on valid consent (see Table 6), then the publisher can also be claimed responsible for such unlawful consent collection. This triggers a heavy responsibility burden on the publisher side, because he has no control over all the publishers or third parties on the way in which they collect consent (also, websites are very dynamic and quickly change over time, hence even if a publisher has verified consent collection in the past, such evidence might not hold upon a consequent visit to the same website). We underline that such model is not sustainable and very hard to manage for the publishers.

Nevertheless, consent can be shared if the user gives a negative consent (if the user refused all types of data processing for all purposes). In this case, the publisher can safely rely on this consent collected by other parties. Even in the case of an invalid consent in which the user gives a positive consent, but the collected consent is registered as a negative consent, the publisher would be complaint: he would respect a negative consent, and hence would not process any data (processing less data than allowed by the user's consent is always valid).

### 7.3 Summary on shared consent

Given the concerns raised in section 7.1, we believe that the legislator, when updating the EU ePrivacy framework, should clarify that **content not requiring consent must not be merged or served with BTTs that require consent**. Otherwise that publisher either violates one of the requirements on valid consent or is forced to rely on the way consent was collected by other parties. As we discussed in 7.2, relying on consent collected by other parties is not sustainable in practice and therefore puts a publisher in a weak and at the same time liable position for consent collection. Shared consent should be acceptable in practice only when the consent is negative, however a positive shared consent places again a publisher in a complex and liable position at the same time.

# 8 Discussion on the ePrivacy Regulation

This section presents a short evolution of the proposal of the ePrivacy Regulation (ePR) (section 8.1) and discusses whether it includes the proposed requirements for a valid consent (section 8.2).

## 8.1 Brief evolution

The ePR was first introduced by the European Commission in 2017 intended to replace the existing ePrivacy Directive 2002/58, as well as updating the current ePrivacy framework in the EU. The European Parliament adopted its position in October of the same year, but a stalemate still resides at the Council of the EU (in the Working Party on Telecommunications and Information Society - WP TELE). The text has been submitted through iterations of the seven EU presidencies (the holder rotates every six months) which failed to find a compromise between Member States. A Progress Report issued by the Council (Finnish Presidency doc. 14054/19[229] of 18 November 2019) noted that the Regulation continues to divide Member States, and many amendments have been suggested and debated so far and a compromise was not found.

The Croatian Presidency of the Council presented a revised proposal[230] in March 2020 and is still under discussion with the aim of getting a common position agreed. Before the revised Regulation can take effect, it will need to pass through trialogue negotiations among the Parliament, Council and the Commission, after which a compulsory grace period of a maximum of two years will apply to allow EU Member States to implement the Regulation. The Council Presidency also indicated that it is currently reflecting on additional revisions and intends to issue an additional document to be discussed during these meetings – a document which we are not knowledgeable as of today. In order to avoid the legal uncertainty created by some foundational elements of the current proposal, our analysis of the discussions in the Council of the EU specifically and is based on the consolidated text circulated by the Croatian Presidency, 5979/20, of 21 February 2020.

## 8.2 Requirements for valid consent in the recent draft proposal of the ePrivacy Regulation

**Tracking walls.** The European Parliament's draft of 26 October 2017(Article 8(1)(1)(b) and recital 22) called for an explicit ban on tracking walls for the first time (we state it as a requirement R4). However, Recital 21 of the Finnish draft of the ePR proposal of 2019 addressed indirectly the case of legitimizing tracking walls for advertising purposes. This indirect indication reveals that it is a topic of political controversy between the stakeholders. These draft signals that consent is valid (freely given) when the processing related to a service the user requested has advertising purposes. The Recital reads,

> "[I]n some cases the use of cookies may also be necessary for providing a service, requested by the end-user, such as services provided to safeguard freedom of expression and information including for journalistic purposes, such as online newspaper or other press publications (…), that is wholly or mainly financed by advertising provided that, in addition, the end-user has been provided with clear, precise and user-friendly information about the purposes of cookies or similar techniques and has accepted such use".

In the event the upcoming draft of the ePR becomes enforced thereby legitimizing the use of tracking walls, it would impact our results regarding requirement "no tracking wall" (R4, explained in section 5.2).

**Grounds to the processing and storage of BTT.** The ePrivacy Regulation might impact the scope of the analysis carried out in this work, that is when a Browser-based Tracking Technology requires or is exempted of consent, summarized in Table 5 of Section 4.1. Specifically, Article 8 of the draft proposal of the ePR points to the following grounds for storage of and access to information from the from end-users' terminal equipment. We compare these with the purposes needing and exempted of consent that we analyzed in Table 5 of Section 4.1:

[229] Proposal for a Regulation on Privacy and Electronic Communications (2019) <https://data.consilium.europa.eu/doc/document/ST-14068-2019-INIT/en/pdf> accessed on 19 June 2020.
[230] Proposal for a Regulation on Privacy and Electronic Communications (2020) <https://www.parlament.gv.at/PAKT/EU/XXVII/EU/01/51/EU_15125/imfname_10966469.pdf> accessed on 19 June 2020.

a) **communication purposes** (such as load balancing BTT). As we have analyzed, communication purposes would be exempted of consent.
b) **consent**;
c) **necessary for providing a service requested by the end-user**. Herewith we transpose the reasoning upheld in Table 5 about which purposes are necessary for providing a service requested by a user. The following purposes are considered to be necessary for the provision of a service (and therefore exempted of consent, requiring another legal basis): users input, local analytics/measurement, user security for service requested by the user, social media functionality requested by the user, authentication that is session-based, customization that is short-termed. However, other purposes are not considered necessary for providing a service to the end-user and would be subject to consent, according to the observations held in Table 5. These are:
   - advertising
   - measurement/non-local analytics
   - user security service not requested by the user
   - social media service not requested by the user
   - authentication that is persistent
   - customization that is persistent
d) **necessary for audience measuring**, provided that such measurement is carried out by the provider of the information society service requested by the end-user or by a third party, or by third parties jointly. This means that either local and local analytics might be used without consent (and hence subject to another legal basis). In this regard, if such rational will be enforced in the ePR, this would impact our analysis concerning our distinction between local and non-local analytics referred to in Table 5.
e) **necessary for emergency communication**;
**f)** **necessary for the legitimate interests pursued by a service provider**, except when such interest is overridden by the interests or fundamental rights and freedoms of the end-user, such as i. when the end-user is a child; ii. profiling; iii. special categories of personal data.

# 9 Related work

In this section, we give the reader a summary of the current context on legal compliance to BTT. Notably, consent for BTT deployment has been analyzed through different prisms that we regard in this paper: audits to websites in order to promote responsible behavior of web publishers; through guidance policy from stakeholders; through enforcement actions and decisions of the Court of Justice and DPAs; and finally, through legal scholarship literature. For readability issues, this section is divided in four parts. Section 9.1 refers to relevant audits on websites. Section 9.2 considers DPAs guidance on the elements for a valid consent. Section 9.3 explains some of the issued decisions related to valid consent. Section 9.4 shows the related work on consent analysis portrayed by legal scholarship and automatic auditing of websites by computer scientists. Table 21 presents a summary of the audits, guidelines and enforcement actions related to consent to BTT that will be presented with further detail in the following subsections.

**Table 21** Summary of the audits, guidelines and enforcement actions related to consent to consent to BTT

| Audits on websites | Guidelines | Enforcement actions |
|---|---|---|
| 29WP, 2015 | 29WP/EDPB, 2020 | Planet49 Judgment of the CJEU, 2019 |
| EDPS inspection, 2019 | EDPS, 2016 | French DPA decision, 2018 |
| Bavarian Audit, 2019 | UK DPA, 2019 | Spanish DPA decisions, 2019 |
| Dutch Check, 2019 | German DPA, 2019 | Belgium DPA decision, 2019 |
| Irish Sweep, 2020 | Finnish DPA, 2019 | |
| Greek Sweep, 2020 | Spanish DPA, 2019 | |
| | French DPA, 2020 | |
| | Greek DPA, 2020 | |
| | Belgian DPA, 2020 | |
| | Irish DPA, 2020 | |

## 9.1 Recent audits on websites

Recurrent audits on websites are aimed at information-gathering assessment of the state of cookie (and related technologies) consent compliance at scale. Such auditing initiatives reveal the current playing level field of websites that still struggle to comply with the GDPR and ePrivacy rules on consent. In contrast, our paper analyzes legal documents to define more precise requirements for these banners.

Table 22 shows the audits of consent banners performed by stakeholders and the related requirements for consent banners.

**Table 22** Stakeholder's audits on websites per sector and the related requirements

| Stakeholders | Tested websites on the use of cookies and related technologies | Related requirements |
| --- | --- | --- |
| 29WP, 2015 | 478 websites in the e-commerce, media and public sectors across 8 Member States | Unambiguous, revocable, informed |
| EDPS inspection, 2019 | websites of major EU institutions and bodies, e.g. European Council, Council of the EU, Commission, CJEU, Europol, European Banking Authority, EDPS, EDPB, 2018 International Conference of Data Protection and Privacy Commissioners (ICDPPC 2018) | Prior, informed |
| Bavarian Audit, 2019 | 40 Bavarian providers (online stores, media companies, insurance companies, banks, sports teams | Informed, freely given, unambiguous |
| Dutch Check, 2019 | 175 websites of web shops, municipalities and media | Unambiguous, freely given |
| Irish Sweep, 2020 | 38 Irish-based websites of sectors, including media and publishing, the retail sector, restaurants and food ordering services, insurance, sport and leisure and the public sector | Unambiguous, freely given, revocable, prior |
| Greek Sweep, 2020 | audit of the use of cookies by popular Greek websites | Non-specified |

**29WP Cookie Sweep Combined Analysis Report, 2015**.[231] This sweep included 478 websites in the e-commerce, media and public sectors across 8 Member States. Both the automated scan and manual review provide the results, thusly: 74% of studied websites displayed banners, 54% thereof did not request user's consent but were merely informative. 70% of the 16,555 cookies stored were third party cookies. More than half of the third party cookies were set by just 25 third-party domains. The sweep showed that a banner was a popular method of informing visitors on the use of BTT in addition to a link in the header or footer to more information. Only 16% of sites offered a configurable banner. The majority relied on browser settings or an opt-out tool provided on a third-party site (e.g. a third-party advertising site). Amongst those sites which set the highest number of cookies, most had taken some steps to inform users about the use of cookies through a banner which was either permanent (requiring an active click from the user within the banner), a banner which disappears on the next user click anywhere on the page or timed to disappear after a certain length of time. *In comparison to our work*, three requirements were analyzed: unambiguous, revocable, and informed.

**EDPS inspection, 2019**.[232] This inspection was carried out on the websites of major EU institutions and bodies, e.g. the shared website of the European Council and the Council of the EU, the Commission, the Court of Justice of the EU, Europol and the European Banking Authority. The EDPS also inspected the website of the European Data Protection Board (EDPB), the 2018 International Conference of Data Protection and Privacy Commissioners (ICDPPC 2018) and the EDPS website itself. The EDPS developed a tool that automatically collects information on personal data processed by websites via the use of cookies, web beacons, page elements loaded from third parties and security of encrypted connections. The inspection revealed that several of the websites were not compliant with the Regulation nor with the ePD. One of the issues encountered was third-party tracking without prior consent. Other issues encountered included the use of trackers for web analytics without visitors' prior consent. *In comparison to our work*, two requirements were analyzed: prior and informed.

**Bavarian State Office for Data Protection Supervision Audit, 2019**.[233] This audit found that forty Bavarian providers (online stores, media companies, insurance companies, banks, sports teams, etc.) use trackers, but only a quarter of the websites inform users about the use of these tools. The remaining providers either did not inform users at all or only informed them insufficiently about the use of tracking tools as part of their privacy policies. Regarding the use of cookie banners, 20% of websites failed to ask users to consent to the use of cookies. Consent obtained were either not given in advance, they were given uninformed, or there was a lack of voluntariness. *In comparison to our work*, three requirements were analyzed: informed, freely given and unambiguous.

[231]Article 29 Working Party, "Cookie sweep combined analysis – Report" (WP229, 3 February 2015).
[232]European Data Protection Supervisor, "EDPS flags data protection issues on EU institutions' websites" (2019) <https://edps.europa.eu/sites/edp/files/edpsweb_press_releases/edps-2019-04-website_inspections_en.pdf> accessed 7 May 2020.
[233] Bavarian DPA, "Safe on the Internet – Data Protection Check on Digital services" (our translation) (2019) <www.lda.bayern.de/media/sid_ergebnis_2019.pdf> accessed 7 May 2020.

**Dutch DPA Check, 2019**.[234] This DPA carried out a check on approximately 175 websites of web shops, municipalities and media, etc. to determine whether they meet the requirements for placing tracking cookies. Some violations were detected, such as preselected boxes and tracking walls. All checked websites are not compliant. The organizations behind these websites have received a letter from the AP calling on them to adjust their working methods accordingly. *In comparison to our work*, two requirements were analyzed: unambiguous and freely given.

**Irish DPA, 2020.**[235] This DPA conducted a sweep between August 2019 and December 2019 on a selection of 40 websites across a range of sectors, including media and publishing, the retail sector, restaurants and food ordering services, insurance, sport and leisure and the public sector to check compliance with the ePD and the GDPR on the use of cookies and other tracking technologies. The 38 respondents signaled either that they were aware of non-compliant practices with the existing rules, or that they had identified improvements that they could make to their websites in order to demonstrate compliance; some detected practices were: implied consent, non-necessary cookies set on landing. A lack of tools for users to vary or withdraw their consent choices, badly designed—or potentially even deliberately misleading—cookie banners and consent-management tools, and prechecked boxes. *In comparison to our work*, two requirements were analyzed: unambiguous, freely given, revocable, and prior.

**Greek DPA, 2020**. In February 2020, this authority performed a sweeping audit of the use of cookies by popular Greek websites, in which the Authority found that non-compliance with the GDPR was widespread. So far, we could not have access to the content of the sweep.

### 9.2 Guidance on a valid consent for BTT

The analysis of the requirements for a valid consent is contained in the guidelines of the 29WP and the EDPS. Other DPAs provide guidance on obtaining consent specifically for trackers. In this section, we give a brief account of the significant aspects of the most comprehensive guidelines. In general, there is broad agreement that a consent banner must contain information on the BTT used, either directly in the banner or via corresponding links. However, the requirements of the authorities vary with regard to the concrete form in which the user may provide consent, ranging from allowing any action from the user, to opinions requiring a concrete banner design configuration. We perform an in-depth comparative analysis of the current state of the art DPA guidelines to the low-level requirements proposed in this paper in Table 7 (see page 15).

**EDPS Guidelines on the protection of personal data processed through web services provided by EU institutions, 2016**.[236] While these Guidelines are in principle aimed at the EU institutions, anyone or any organization interested in data protection and web services might find them useful. The main topics covered in these Guidelines that are useful for this paper are: the use of cookies, scripts and any other tools to be stored or executed on the user terminal device; server-side processing of personal data and the wider issue of tracking.

**29WP Guidance on cookies**. Per the 29WP guidelines, the following guidance documents were observed in our study, for they interpret closely the consent requirements in respect of cookies and BTT and were quoted alongside this paper.

- Opinion 2/2010 on online behavioral advertising, (WP171, June 2010);
- Opinion 15/2011 on the definition of consent (WP187, July 2011);
- Opinion 04/2012 on Cookie Consent Exemption (WP194, June 2012);
- Working Document 02/2013 providing guidance on obtaining consent for cookies (WP208, October 2013);
- Opinion 9/2014 on the application of Directive 2002/58/EC to device fingerprinting (WP224, November 2014).

[234] Dutch DPA, "Many websites incorrectly request permission to place tracking cookies" (2019) (n 99).
[235] Irish DPA, "Sweep conducted between August 2019 and December 2019" (2020) <https://www.dataprotection.ie/sites/default/files/uploads/2020-04/Data%20Protection%20Commission%20cookies%20sweep%20REVISED%2015%20April%202020%20v.01.pdf
[236] cf. EDPS Guidelines (n 35).

- Guidelines on consent under Regulation 2016/679 (WP259 rev.01, April 2018), which was extended by Guidelines 05/2020 on consent under Regulation 2016/679 (May 2020);

**French DPA Guidelines on cookies and trackers, 2019**.[237] The CNIL published guidelines on cookies and trackers. Hereby we consider the most relevant points related to our work on consent for cookies. *Consent*. Continuing to browse a website after its cookie banner is displayed will no longer be considered to be valid consent for cookie use.
*Auditable*: Operators using trackers have to be able to prove that they have obtained affirmative consent from the user, at all times.
*Scope*: The guidelines apply to all types of operations involving cookies and trackers on any type of device, including smart phones, computers, connected vehicles and any other object connected to a telecommunications network open to the public.
*Cookie Wall:* The user should not suffer any major inconvenience if they refuse to give or withdraw their consent. The practice of blocking access to a website or a mobile application unless consent is provided does not comply with the GDPR.
*Revocable*: Users should be able to withdraw their consent at any time. User-friendly solutions must therefore be implemented to allow users to withdraw their consent as easily as they have given it.
*Operator´s Roles and Responsibilities*: An operator using cookies and trackers is considered to be a controller and is therefore fully responsible for obtaining valid consent.
In 2020 the CNIL launched a public consultation on a draft recommendation on cookies and other trackers[238] in order to adapt the GDPR rules to trackers and consent. As main takeaways, this document proposes designs of cookie banners; enunciates best practices for legal compliance, practical arrangements for implementation, and examples of how to comply with the applicable rules. It suggests means to define specific purposes for processing; proposes neutral design interfaces and design patterns to avoid misleading design practices; provides examples of proof of consent; and finally advocates for the development of standardized interfaces operating in the same way and using a standardized vocabulary to make it easier for users to understand when navigating from one site to another.

**UK DPA Guidance on the rules on use of cookies and similar Technologies, 2019**.[239] The ICO updated its guidance on the use of cookies and other similar technologies. Some of the key points to note from the guidance are herewith described. *Cookie walls* may not comply with the cookie consent requirements and it states these as inappropriate if the use of a cookie wall is intended to require, or influence, users to agree to their personal data being used as a condition of accessing its service, as a user has no genuine choice but to accept cookies. The authority clarifies that implied consent conveyed through statements such as "by continuing to use this website you are agreeing to cookies", pre-ticked boxes or any equivalents, such as sliders defaulted to "on", cannot be used for non-essential cookies. Consent mechanisms incorporating a "more information" section, rather than as part of the initial banner are also deemed non-compliant on the basis that they do not allow users to make a choice before non-essential cookies are set. On the types of cookies, the ICO enunciates that advertising and analytics cookies are not "strictly necessary" and are subjected to consent rules.

**German DPA Guidance, 2019**. The German DPA published the "Guidelines for Telemedia Providers"[240] and Frequented Asked Questions (FAQ)[241] about web tracking and cookie banners. According to the guidance, a cookie banner is only necessary if cookies are set through the website that require data protection consent; if a website only sets cookies for which the site operator does not require consent, the guidance considers the banner avoidable. In this guidance, consent is needed when a web service uses web services on its website that analyze the user across several domains, e.g. social media plugins, advertising networks or analysis tools such as Google Analytics. The regulator alerts that consent to the use of cookies must not be preselected and does not consider the opt-out procedure to be sufficient. The authority published also a note[242] on the use of cookies and cookie banners – "what must be done with consent (ECJ ruling "Planet49")?".

[237] cf. CNIL Guidelines (n 36).
[238] CNIL draft recommendation 2020 (n 95).
[239] cf. ICO Guidance (n 26).
[240] cf. German DPA Guidelines (n 10).
[241] German DPA, "FAQ about Cookies and Tracking" (2019) <www.baden-wuerttemberg.datenschutz.de/wp-content/uploads/2019/04/FAQ-zu-Cookies-und-Tracking.pdf> accessed 7 May 2020.
[242] cf. German DPA (10).

**Finnish DPA Guidance on Confidential Communications, 2019**.[243] The NCSC-FI at Traficom mentioned in the guidelines that consent can be requested by using any preferred method (e.g. browser/application setting or pop-up window) as long as it is not requested by using a pre-ticked checkbox. The use of cookies and the related practices must also be indicated on a website in such a manner that a user can obtain additional information about them.

**Spanish DPA Guide on the Use of Cookies, 2019**.[244] The AEPD published new Guidelines on the Use of Cookies and similar technologies, which were prepared in collaboration with different organizations in the marketing and online advertising industries (e.g. Adigital, IAB Spain, etc.). The Guidelines provide factors for categories of cookies:

- *Who manages cookies* (proprietary or third-party);
- *Purpose* (technical, customization, analytical, and behavioral advertising); and
- *Duration* (session or persistent).

The AEPD provides the following examples of actions that could be considered an affirmative action: the use of the scroll bar, insofar as the information on cookies is visible without using it; clicking on any link contained in the site other than those in the second layer of information on cookies or the privacy policy link; on devices such as mobile phones or tablets, by swiping the initial screen and accessing the content. Even if the Planet49 Judgment[245] ruled otherwise, the AEPD Guidelines state:

> For the action of continuing browsing to be deemed a valid consent, the information notice must be displayed in a clearly visible place, so that due to its shape, color, size or location, it can be secured that the notice has not gone unnoticed to the user. Additionally, it will be necessary, for the consent to be deemed granted, that the user performs an action that can be qualified as a clear affirmative action. For instance, a clear affirmative action may be considered to browse to a different section of the website (other than the second layer of information on cookies or the privacy policy), to slide the scroll bar, closing the first layer notice or clicking on any content of the service. The mere fact of viewing the screen, moving the mouse or pressing the keyboard cannot be considered an acceptance.[246]

**Greek Data Protection Authority, 2020**.[247] This authority issues in February 2020 its Guidelines on the use of internet cookies and trackers, following the completion of a sweeping audit of the use of cookies by popular Greek websites, in which the Authority found that non-compliance with the GDPR was widespread. The practical guidance provides specific recommendations on as well as practices that should be avoided.

The **Irish DPA**, 2020. [248] This authority released in April 2020 a comprehensive guidance note on cookies and other tracking technologies. The main takeaways are: implied consent is not valid; the use of pre-checked boxes and sliders set to "on" as default are non-compliant. A consent banner must not obscure the text of the privacy or cookie notice. It explains that a website operator must take accessibility into account in designing interfaces. Uniquely, the Guidance advises users should not be "nudged" into accepting trackers and should be given the opportunity to consent on a granular basis. The Guidance also says that there should be given equal prominence between "accept" and "reject" buttons.

### 9.3 Enforcement actions of consent by the Court of Justice of the European Union (CJEU) and by DPAs

In this section, we show the enforceable decisions in connection to consent requirements for BTT referred in judgements of the Court of Justice of the EU (CJEU) and administrative decisions issued by DPAs. Table 23 shows the enforcement actions by the CJEU and DPAs in relation to the analyzed requirements.

**Table 23** Enforcement actions by the CJEU and DPAs in relation to the analyzed requirements

| **Enforcement actions** | **Requirements** |
|---|---|
| Planet49 Judgment of the CJEU, 2019 | Specific, informed, unambiguous |

[243] Finnish DPA Guidance (n 13).
[244] cf. Spanish DPA Guide (n 82).
[245] cf. Planet49 Judgment (n 11).
[246] Author's translation from the Spanish version.
[247] Greek DPA, Guidelines on Cookies and Trackers (n 42).
[248] cf. Irish DPA Guidance (n 37).

| | |
|---|---|
| French DPA decision, 2018 | |
| Spanish DPA decisions, 2019 | Unambiguous (configurable banner), informed, prior, revocable |
| Belgium DPA decision, 2019 | Informed, revocable, unambiguous |

**Planet49 Judgment of the CJEU, 2019.**[249] On October of 2019, the CJEU decided that the consent which a website user must give to the storage of and access to cookies on his or her equipment is not validly constituted by way of a prechecked checkbox which that user must deselect to refuse his or her consent. The Court notes that consent must be specific so that the fact that a user selects the button to participate in a promotional lottery is not sufficient for it to be concluded that the user validly gave his or her consent to the storage of cookies. Furthermore, according to the Court, the information that the service provider must give to a user includes the duration of the operation of cookies and whether or not third parties may have access to those cookies. The German Federal Court of Justice (BGH) followed the CJEU's preliminary ruling in the "Planet 49 case" which determined that the request for consent by a preselected tick box constitutes an "unreasonable disadvantage to the user".[250]

**Spanish DPA decisions, 2019**. This DPA[251] in October of 2019 fined Vueling for failing to provide a compliant consent banner. The poorly constructed banner did not provide a configuration panel that allows the user to delete cookies in a granular way. It was considered that the information was insufficient for the intended purpose of allowing users to configure preferences in a granular or selective form.[252] This DPA also fined IKEA[253] for placing cookies before users clicked the only option in the banner: the “OK” button. Users were prompted with a cookie banner stating that “IKEA website uses cookies that make browsing much easier. More information about cookies”. Initially, users were instructed to block cookies through browser settings, also including “strictly necessary” cookies like e.g. shopping cart cookies rendering the website basically impossible to use. It did not identify the purposes of the different cookies used, nor informed about the possibility of setting the usage preferences of the cookies. It did not provide a link to the panel or cookie configuration system enabled to select them in granular form. It did not include a specific button or mechanism for rejecting all cookies. The warning that “If you do not change your browser settings, we will understand that you agree to receive all cookies from the IKEA website” breaches consent requirements. It did not report on how to revoke the consent given.

**Belgium DPA decision, 2019.** On December 2019, this DPA imposed a fine of € 15,000 on an SME operating a legal information website for their noncompliant cookie management and privacy policy. It found that their privacy policy lacked transparency and infringed the rules on information to be provided. In particular, it provided insufficient information about the cookies deployed on the website (e.g. the list of cookies used, their purpose, the identity of third parties concerned, and the lifespan of the cookies) and did not properly identify the controller. Moreover, the cookie policy was only available in English, whereas the website targeted Dutch and French-speaking readers. The website did not obtain opt-in consent for certain types of cookies used, including first-party analytics cookies. There was no easy way for users to withdraw consent.

**Decisions and complaints against the IAB Transparency and Consent Framework (TCF).** The TCF of the IAB Europe implements consent solutions for parties in the digital advertising chain. Herewith we report the French DPA decision. The CNIL, in 2018, sued an advertisement company Vectaury using the IAB framework, invoking a lack of informed, free, specific and unambiguous consent.[254] For the CNIL, the consent text was not clear enough regarding the final use of collected data, and the formulation may lead users to incorrectly assume that refusing consent prevents a free access to the website or lead to more intrusive advertisement. It was also noted that pre-ticking consent-related checkboxes was not compliant with the Recital 32 of the GDPR. It was required the list of recipients of users' data to appear immediately

[249] cf. Planet49 Judgment (n 11).

[250] German Federal Court of Justice for consent to telephone advertising and cookie storage (2020) <https://www.bundesgerichtshof.de/SharedDocs/Pressemitteilungen/DE/2020/2020067.html?nn=10690868> accessed 18th June 2020.

[251] cf. Spanish DPA decision (n 155).

[252] EDPB press release, “The Spanish Data Protection Authority fined the company Vueling for the cookie policy used on its website with 30,000 euros” (2019) <https://edpb.europa.eu/news/national-news/2019/spanish-data-protection-authority-fined-company-vueling-cookie-policy-used_en> accessed 7 May 2020.

[253] Spanish DPA decision, “Procedimiento PS/00127/2019” (2019) <www.aepd.es/resoluciones/PS-00127-2019_ORI.pdf> accessed 7 May 2020.

[254] French DPA, “Decision n° MED 2018-042 of October 30th, 2018 enforcement notice against the company Vectaury” (2018) <www.legifrance.gouv.fr/affichCnil.do?id=CNILTEXT000037594451> accessed 7 May 2020.

when consent text is displayed. In April 2019 a formal complaint[255] was filed against the IAB for a tracking wall on its own website that forces visitors to consent if they want to access the website.

## 9.4 Related work on consent and consent banners

In the following, we outline academic work on consent banners. All of these works focus on measuring or detecting legal violations in cookie banners from a technical point of view. We extensively discuss the legal analysis of detected violations in previous works. Table 24 displays a comparison summary between related works and ours regarding the requirements for consent banners.

**Table 24** Comparison summary between related works and ours on the requirements for consent banners

| Works | Tested websites | Tested high and low-level requirements | Comments |
|---|---|---|---|
| Carpineto et al. | 17k Italian public administration websites | Prior (R1 prior to storing an identifier) | Violations found on websites where a banner is not displayed |
| Traverso et al. | 100 Italian websites | Prior (R1 prior to storing an identifier), unambiguous (R15 correct consent registration) | Correct consent registration was tested via confronting the number of trackers after acceptance and after refusal of consent. |
| Trevisan et al. | 36k EU websites | Prior (R1 prior to storing an identifier) | |
| Van Eijk et al. | 1500 websites from 18 countries (EU, USA, Canada) | - | They test the presence of a banner on EU websites and simulating user's visit from different EU, USA and Canada. |
| Degeling et al. | 6,500 EU websites | Specific, informed, unambiguous (R12 configurable banner) | |
| Sanchez-Rola et al. | 2,000 websites | Prior, unambiguous (R11 affirmative action design, R12 configurable banner, R15 correct consent registration), readable and accessible (R19 using clear and plain language, R20 no consent wall), revocable (R22 delete consent cookie) | |
| Libert et al. | 180 news websites | Prior (R1 prior to storing an identifier) | All third-party cookies were considered, independently whether they require consent or not |
| Utz et al. | 5,000 websites | Free (R4 no tracking wall), unambiguous (R12 configurable banner, R13 balanced choice, R11 affirmative action design) | Also studies the influence of design on users choice. |
| Matte et al. | 23k EU websites | Unambiguous (R11 affirmative action design, R12 configurable banner, R14 post-consent registration, R15 correct consent registration) | No pre-ticked boxes |
| Nouwens et al. | 10k websites | Unambiguous (R13 balanced choices, R11 affirmative action design), free (R4 no tracking wall), informed (high level requirement) | No pre-ticked boxes. Also studies the influence of design on users' choice. |
| Leenes and Kosta | 100 Dutch websites | Free (R4 no tracking walls), unambiguous (R12 configurable banner) | |

[255] Brave, "Complaint against IAB Europe's "cookie wall"(2019)<https://brave.com/wp-content/uploads/2019/04/3-April-2019-complaint-to-Data-Protection-Commission-of-Ireland-regarding-IAB-Europe-cookie-wall-and-consent-guidance.pdf> accessed on 19 June 2020.

| | | | |
|---|---|---|---|
| Matte, Santos et al. | 575 advertisers registered in IAB Europe TCF | - | Study of purposes used in cookie banners of IAB Europe TCF |

Carpineto et al.[256] developed a tool to automatically check the legal compliance of cookie banners in Italian Public Administration websites in 2016. The authors used language-dependent text analysis methods and detected cookies based on lists of known trackers. In this study, the only criteria for non-compliance is whether the website uses tracking cookies however does not display a cookie banner. By automatically analyzing Italian Public Administration websites, authors identified 1,140 non-compliant websites placing tracking cookies in the user's browser.

Traverso et al.[257] measured the impact of the ePD's cookie policy on web tracking on 100 Italian websites. Visiting the same website before and after giving consent by clicking on the accept button of the cookie banner, they measured the difference in the number of included trackers. Their results are alarming: there were few differences between both scenarios. In the no-consent-given scenario, they found an average of 29.5 trackers per webpage, none of them containing 0 tracker, and half of them containing more than 16.

Trevisan et al.[258] built an automatic tool "CookieCheck" to check violations of the ePD in 36 000 popular websites popular in the European Union (plus 4 extra-EU countries) in early 2017. Using a list-and heuristic-based tracking cookie detection method, they tested whether websites requested consent before installing cookies. They found that 49% of websites installed profiling cookies before user consent, a number raising to 74% when considering any third-party cookie. On a smaller set of 241 websites from 3 European countries, they observed that 80.5% of those installing tracking cookies did not regard the user's consent. Interestingly, they observed no significant difference in the number of installed tracking cookies between desktop and mobile browsers.

Van Eijk et al.[259] studied cookie banners after the GDPR came in force in 2018. Leveraging a crowd-sourced list of known banners, they automatically detected cookie banners on 40.2% of European Union websites. Accessing websites from different countries using VPNs, they found that the provenance of the user has not so much impact as the expected audience of a website regarding the prevalence of banners. They also observed important variations between websites of different top-level domains.

Degeling et al.[260] performed a study comparing the information presented to users of 6,500 EU websites before and after the GDPR, focusing on the changes in privacy policies and information presented to users. In particular, the authors studied characteristics of 31 cookie banner libraries by installing them locally. They observed a 6% increase in cookie banners adoption by website pre- and post-GDPR. They have identified the following categories within existing implementations of consent notices:

- "*No option notices*" to simply inform the user that the website uses cookies and if the user continues to use the website, they agree to this use;
- "*Confirmation-only banners"* displays button with an affirmative text, such as "OK", or "I agree", through which, by clicking on it expresses the user's consent;
- "*Binary notices"* provide users with a button to accept and another to reject the use of all cookies on the website;
- "*Category-based notices*" assembles the cookies used by the website into categories. Users can allow or disallow cookies of each category individually by (un)checking a settings menu or toggling an "on–off" switch;
- "*Vendor-based notices*" allow visitors to accept or decline cookies for each third-party service used by the website (conceding more fine-grained control). They originate from third-party libraries, as the IAB Europe’s Transparency and Consent Framework, which refers to its advertising partners as “vendors”.

[256] Claudio Carpineto, Davide Lo Re, Giovanni Romano, "Automatic assessment of website compliance to the European cookie law with CooLCheck" (Proceedings of the 2016 ACM on Workshop on Privacy in the Electronic Society, ACM, Vienna, Austria, 2016) 135-138.

[257] Stefano Traverso, Martino Trevisan, Leonardo Giannantoni, Marco Mellia, Hassan Metwalley less, "Benchmark and comparison of tracker-blockers: Should you trust them?" (Network Traffic Measurement and Analysis Conference, Dublin, Ireland, 2017) 1-9.

[258] Martino Trevisan, Stefano Traverso, Eleonora Bassi and Marco Mellia, "4 Years of EU Cookie Law: Results and Lessons Learned" (Proceedings on Privacy Enhancing Technologies, Issue 2, 2019) 126-145.

[259] Rob Van Eijk, H. Asghari, Philipp Winter, Arvind Narayanan, "The Impact of User Location on Cookie Notices (Inside and Outside of the European Union)" (Workshop on Technology and Consumer Protection (ConPro '19), San Francisco, CA, 2019).

[260] Martin Degeling et al. (n 127).

Sanchez Rola et al.[261] performed a wide manual evaluation of tracking in 2,000 websites, inside and outside of the EU. The aim was to measure how easy it is to opt-out from tracking if the user desires to do so and assessing whether it is possible at all. Their results show that tracking is prevalent, happens mostly without user’s consent, and opt-out is difficult. They note whether banners respect many requirements relevant to our work: whether they offer a way to refuse tracking, whether consent is set automatically when visiting the website, whether tracking happens despite a refusal of consent, and whether the consent cookie is deleted upon refusal.

Concerning tracking, Libert et al.,[262] in a factsheet for the press, studied the impact of the GDPR on the amount of third-party content and cookies on news websites. On about 180 European news sites, they observe a 22% drop in the number of third-party cookies before (April 2018) and after (July 2018) the GDPR, but only 2% drop in third-party content.

Another prominent work related to ours is the research from Utz et al.[263] The authors ran a number of studies, gathering ~5,000 of cookie notices from leading websites to compile a snapshot (derived from a random sub-sample of 1,000) of the different cookie consent mechanisms. They also worked with a German ecommerce website over a period of four months to study how more than 82,000 unique visitors to the site interacted with various cookie consent designs. The authors reached the following findings significant to our paper:

- Cookie consent notices do not offer a choice to the users; they are placed at the bottom of the screen (58%); not blocking the interaction with the website (93%); and offering no options other than a confirmation button that does not do anything (86%);
- The more choices offered in a cookie notice, the more likely visitors were to decline the use of cookies;
- A majority also try to nudge users towards consenting (57%) — such as by using “dark pattern” techniques like using a color to highlight the “agree” button (which if clicked accepts privacy-unfriendly defaults) vs displaying a much less visible link to “more options” so that pro-privacy choices are buried off screen;
  Mentioning cookies in a consent notice decreases the chance that users allow cookie use.

Matte et al.[264] found several plausible violations of both the GDPR and the ePD in the implementations of cookie banners by actors using IAB Europe’s Transparency and Consent Framework (TCF). They automatically and semi-automatically detected four suspected GDPR and ePD violations on more than 1400 websites using this framework (found among 23k websites) to display cookie banners and found at least one violation in more than half of them. Considered violations are positive consent registered before any user action, no option to refuse consent, registered consent not respecting user’s decision, and pre-ticked boxes. Their work includes an analysis by a co-author which is an expert in law as to why considered violations can be considered legal violations.

Nouwens et al.[265] detected dark patterns in about 700 websites using IAB Europe’s TCF, and found that only 11.8% of banners meet minimal legal requirements: unambiguous consent, accepting being as easy as rejecting consent, no pre-ticked boxes. They also measure how some design choices in banners affect users’ decision on consent.

From a legal perspective, both studies by Kosta[266] and Leenes[267] on a regulatory approach towards cookies were prominent to our analysis of the legal and technical side of consent requirements for BTT. Of particular relevance to our work is the study performed by Leenes and Kosta,[268] in which the authors examined manually the practices of 100 Dutch websites with regard to cookie consent mechanisms. They found that most of these websites do not respect the ePD. Those researchers defined a four-tier classification of consent implementation from the analyzed banners:

- explicit agreement to all cookies used on the site, without possibility to opt out;
- implicit agreement to all cookies used on the site, i.e. banners whose button's text is not a response to a question regarding the user's consent;
- coerced agreements to all cookies, i.e. "cookiewalls", when users cannot access the website without accepting tracking cookies;
- detailed choice/consent of cookies, i.e. banners containing a "settings" button.

[261] Iskander Sanchez-Rola, Matteo Dell’Amico, Platon Kotzias, Davide Balzarotti, Leyla Bilge, Pierre-Antoine Vervier, and Igor Santos, “Can I Opt Out Yet? GDPR and the Global Illusion of Cookie Control” (ACM Asia Conference on Computer and Communications Security (AsiaCCS ‘19), Auckland, New Zealand, 2019).
[262] Timothy Libert, Lucas Graves and Rasmus Kleis Nielsen, “Changes in third-party content on European news websites after GDPR” (Reuters Institute for the Study of Journalism Reports: Factsheet, Reuters Institute for the Study of Journalism, 2018).
[263] Cf Christine Utz et al. (n 223).
[264] cf. Matte et al. (n 152).
[265] cf. Nouwens et al. (n 191).
[266] cf. Kosta (n 27).
[267] cf. Leenes (n 82).
[268] Ronald Leenes, Eleni Kosta, “Taming the Cookie Monster with Dutch Law - A Tale of Regulatory Failure” (2015) *Computer Law & Security Review*, Volume 31, Issue 3, 317-335.

Among the 100 sites studied, they found 25 banners of the 1st type, 54 of the 2nd one, none of the 3rd one and 6 of the last one. 87% of visited websites installed cookies "of various type" on first page load, i.e. irrespective of the choice of the user.

Finally, Matte, Santos et al.[269] analyzed the purposes of data processing defined in IAB Europe's Transparency and Consent Framework (TCF). They derive from their legal analysis that most of the purposes defined in this framework cannot be exempted of consent. Measuring purposes that advertisers registered in the TCF declare to use, they observe that hundreds of advertisers rely on a legal basis that could not be considered compliant under the GDPR. While their paper focused on the legal requirements for purposes, our work analyzes design requirement in banners.

## 10 Conclusion

In this paper, we have performed an interdisciplinary analysis of how consent banners are supposed to be implemented to be fully compliant with the European data protection rules on consent. As a result, we defined **22 operational** fine-grained requirements for consent banners to level-set current practices of websites. For each requirement, we assessed if verification by technical means (with computer science tools), manual means (by a human operator) or based on user studies (with surveys and interviews) is needed to assess compliance with valid consent (see Table 6). We identified hurdles related to the assessment of requirements with technical means, described advances made so far in the corresponding technical areas, and identified the need for standardization of consent interfaces and of consent storage and sharing in web applications. Because of the absence of standards for consent, many requirements can be fully assessed manually by a human operator, which is time-consuming and not scalable.

Additionally, our analysis of requirements applies only to cases when browser-based tracking technology (BTT) is used for the purposes that require consent (see Table 5). However, even if computer scientists can detect the presence of a BTT on a given website, in general it is not possible to identify what purpose BTT is used for on a given website (see the discussion on page 18). We therefore believe that legislators should propose standardized and machine-readable means to specify purposes for each BTT that can be further analyzed automatically with technical tools at scale.

Notably, the number of website audits and sweeps, complaints and fines – in response to alleged ePrivacy and GDPR violations when using BBT – have increased over the last year. Very recent guidelines by the EDPB and DPAs have been issued to surpass rogue website practices while requesting and registering consent, as reflected in section 8.

However, we observed regulatory discrepancies regarding the guidelines for a valid consent. Some DPAs move a step further in several issues compared to corresponding guidelines of other DPAs of other Member States of the European Union, thus signifying an increasing trend towards stricter rules concerning online trackers. Thus, the *EBDP may be the most appropriate institution to assure cooperation*, streamlined guidance and consistency of DPA guidelines, leveraging from the collective knowledge of other DPAs. More legal certainty can possibly be given by the upcoming ePrivacy Regulation. On the other hand, the European Commission needs to map inconsistencies and gaps of national laws and work with Member States to further harmonize the implementation of the GDPR.

Some low-level requirements described in this paper (see a full list in Table 6) were not yet discussed at EDPB level (and only some DPAs propose in their guidelines), while few of them have never been considered and result from a cooperation of the authors of this work – legal and computer science experts. We hope that all the low-level requirements presented in this work raise discussion and are included in further updates of the ePrivacy Regulation, and if is not possible, then at least explicitly recommended by the European Data Protection Board.

[269] Matte C, Santos C, Bielova N. (2020), Measuring the usage of purposes and their legal basis by advertisers in the IAB Europe's Transparency and Consent Framework, Annual Privacy Forum (APF) (forthcoming).

**Acknowledgements**

We are grateful to Fabiano Dalpiaz who has given feedback on the section related to natural language processing (NLP), as well as to Frederik Borgesius and Gaëtan Goldberg who have validated some of the requirements during the development of this paper. This work has been partially supported by the ANR JCJC project PrivaWeb (ANR-18-CE39-0008 and by the Inria DATA4US Exploratory Action program